\documentclass[a4paper,11pt]{article}
\pdfoutput=1 

\usepackage{jheppub} 
\usepackage{mathtools}
\usepackage[T1]{fontenc} 
\usepackage{graphicx, amssymb, amsfonts, amsmath, subfigure}
\newcommand{\ba}{\begin{eqnarray}}
\newcommand{\ea}{\end{eqnarray}}

\title{\boldmath On the particle spectrum and the conformal window}

\author[a]{M.P. Lombardo,}
\author[b]{K. Miura,}
\author[c]{T.J. Nunes da Silva,}
\author[c,1]{and E.Pallante\note{Corresponding author.}}

\affiliation[a]{ INFN-Laboratori Nazionali di Frascati, I-00044, Frascati (RM), Italy }
\affiliation[b]{Kobayashi-Maskawa Institute for the Origin of Particles and the Universe (KMI), Nagoya University, 464-8602, Nagoya, Japan}
\affiliation[c]{Van Swinderen Institute, University of Groningen, 9747 AG, The Netherlands}

\emailAdd{mariapaola.lombardo@lnf.infn.it}
\emailAdd{miura@yukawa.kyoto-u.ac.jp}
\emailAdd{t.j.nunes@rug.nl}
\emailAdd{e.pallante@rug.nl}

\abstract{
We study the $SU(3)$ gauge theory with twelve flavours of fermions in the fundamental representation as a prototype of non-Abelian gauge theories inside the conformal window. Guided by the pattern of underlying symmetries, chiral and conformal, we analyze the two-point functions theoretically and on the lattice, and determine the finite size scaling and the infinite volume fermion mass dependence of the would-be hadron masses. 
We show that the spectrum in the Coulomb phase of the system can be described in the context of a universal scaling analysis and we provide the nonperturbative determination of the fermion mass anomalous dimension $\gamma^*=0.235(46)$ at the infrared fixed point. 
We comment on the agreement with the four-loop perturbative prediction for this quantity and we provide a unified description of all existing lattice results for the spectrum of this system, them being in the Coulomb phase or the asymptotically free phase.  
Our results corroborate the view that
the fixed point we are studying is not associated to a physical singularity along  the bare coupling line and estimates of physical observables can be attempted on either side of the fixed point. Finally, we observe the restoration of the $U(1)$ axial symmetry in the two-point functions.
   }
\begin{document} 
\maketitle
\flushbottom
\section{Introduction}
\label{sec:intro}

\begin{figure}[tbp]
\centering
\includegraphics[width=.50\textwidth]{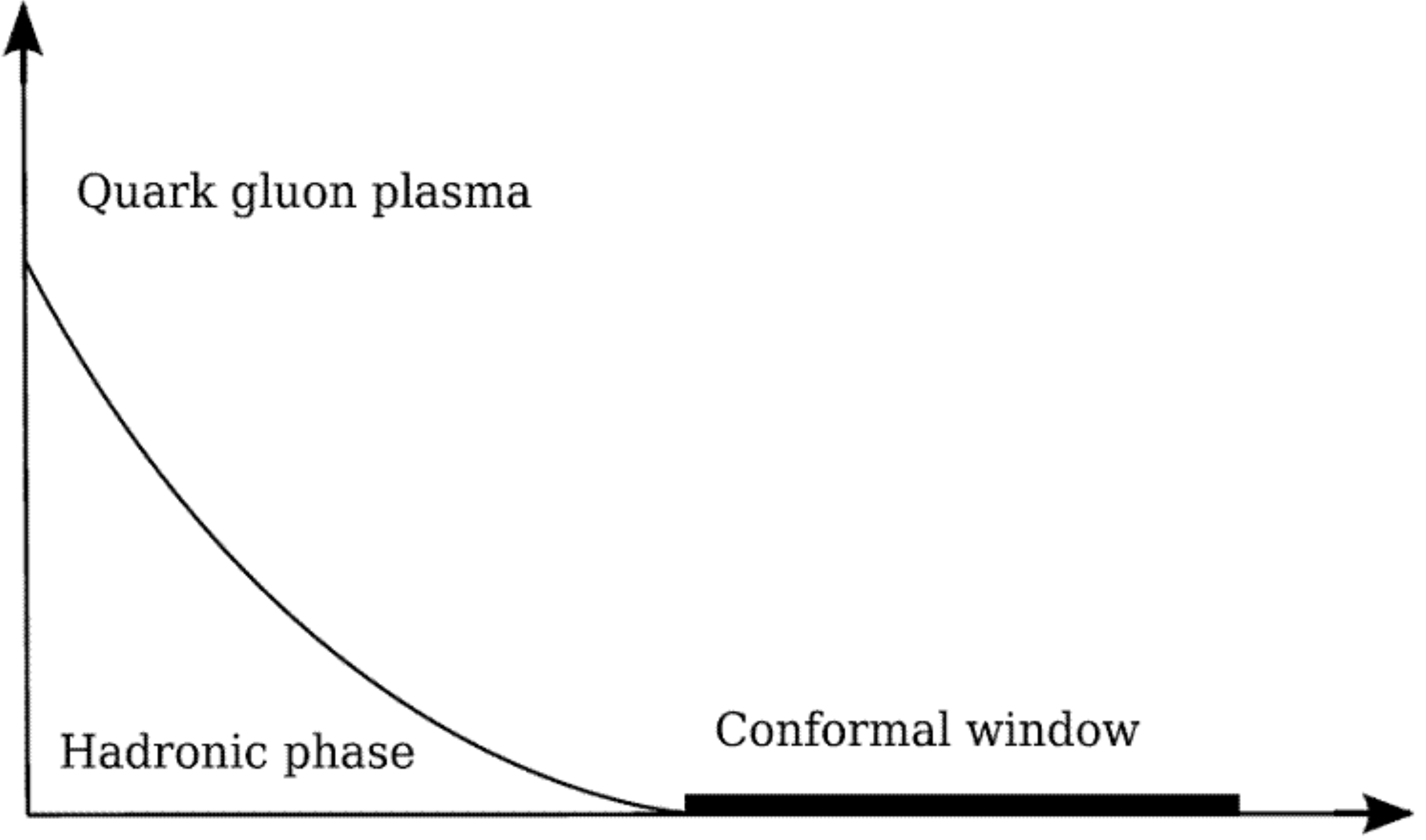}
\caption{\label{fig:phase} The phase diagram in the temperature ($T$) and flavours ($N_f$) plane for a non-Abelian gauge theory with $N_f$ massless Dirac fermions in the fundamental representation. A chiral phase boundary (solid line) separates the hadronic phase from the quark-gluon plasma phase. The endpoint of the chiral phase boundary is where the conformal window opens. }
\end{figure}

This work is devoted to the study of how the particle spectrum of non-Abelian gauge theories changes once they enter the conformal window. Strongly coupled scenarios beyond the standard model provide the phenomenological motivation to study these theories, in particular, the preconformal phase that precedes the conformal window. On the other hand, the properties of these theories inside the conformal window, such as the value of the anomalous dimension of the fermion mass operator along the infrared fixed point (IRFP) line and the ordering of the would-be hadron states can shed light on the dynamics at and just below the lower endpoint of the conformal window. 
Our guiding principle, as in our previous studies of these theories, is the identification of  symmetry patterns; conformal symmetry and chiral symmetry are the key ingredients in this case. Figure~\ref{fig:phase} guides us through the projection on the plane temperature-flavour of the phase diagram of the chosen prototype theory, an $SU(3)$ gauge theory at zero chemical potential with $N_f$ flavours of massless Dirac fermions in the fundamental representation, in other words massless QCD with varying flavour content. 
For small $N_f$ and at zero temperature, QCD has  spontaneously broken chiral symmetry and it is confining. At some finite temperature, the system enters the quark-gluon plasma (QGP) phase, where chiral symmetry is restored and the system deconfines. For increasing $N_f$ the most favoured scenario, first discussed in \cite{Appelquist:1998rb,Miransky:1996pd}\footnote{These works were preceded by the pioneering work in \cite{Banks:1981nn}, where alternative scenarios with no phase transition at $N_f^c$ were considered.} and supported by lattice studies \cite{Deuzeman:2009mh}, identifies the endpoint of the chiral phase boundary with the opening of the conformal window at some critical value $N_f^c$. The conformal window refers to a family of theories that develop an IRFP in the interval $N_f^c\leq N_f\leq N_f^{af}$; at $N_f^{af}$ asymptotic freedom is lost. 

 As long as the theory is renormalizable,  scale invariance at the fixed point has been shown to imply conformal invariance if an energy-momentum tensor $\theta_{\mu\nu}$ exists for which the scale current $s_\mu$ satisfies $s^\mu = x_\nu\theta^{\mu\nu}$  \cite{Callan:1970ze}\footnote{For recent progress and exceptions see \cite{Fortin:2012hn}.}. This is the case for QCD-like theories in the conformal window, as long as nonrenormalizable operators do not play a role. To guarantee scale and conformal invariance, chiral symmetry has to be restored and the theory deconfined in the usual sense. 
 Everywhere in the parameter space of the theory, except at the fixed point, the observables will only show remnants of conformality; these remnants and the realization of exact chiral symmetry determine features of the correlation functions and the spectrum that are quite distinct from those of QCD.  The purpose of this study is to isolate the aforementioned features and identify  the universal scaling properties of the IRFP. To this end, we take the $N_f=12$ system as a prototype of theories inside the conformal window and study both the fermion mass and volume dependence of the would-be hadron masses in a lattice box, and the fermion mass dependence of the spectrum at infinite volume. 
This analysis updates and largely extends  our first study of this system \cite{Deuzeman:2009mh}. Importantly, this is the first study of universal scaling for the complete would-be hadron spectrum\,---\,pseudoscalar, scalar, vector, axial mesons and the nucleon. Inspired by \cite{Cheng:2013xha}, it also provides a unified description of all existing lattice spectrum results for this system. 

The paper is organized as follows. Section~\ref{sec:theo} contains a theoretical premise, where we partly reformulate or adapt to this system existing knowledge for the scaling properties of two-point functions and the spectrum at, or close to, a fixed point. Section~\ref{sec:setup} describes our lattice action and the strategy adopted to compute the spectrum of the $N_f=12$ theory at a fixed lattice gauge coupling.  Section~\ref{sec:spectrum} is entirely devoted to results, in order, the universal scaling and violations of scaling for the would-be hadron spectrum at finite and infinite volume, the fermion mass anomalous dimension at the IRFP, mass ratios, and the degeneracy of chiral and $U(1)_A$ partners. 
We conclude in section~\ref{sec:conclusions}. 

\section{Theoretical premise}
\label{sec:theo}

\begin{figure}[tbp]
\centering
\includegraphics[width=.80\textwidth]{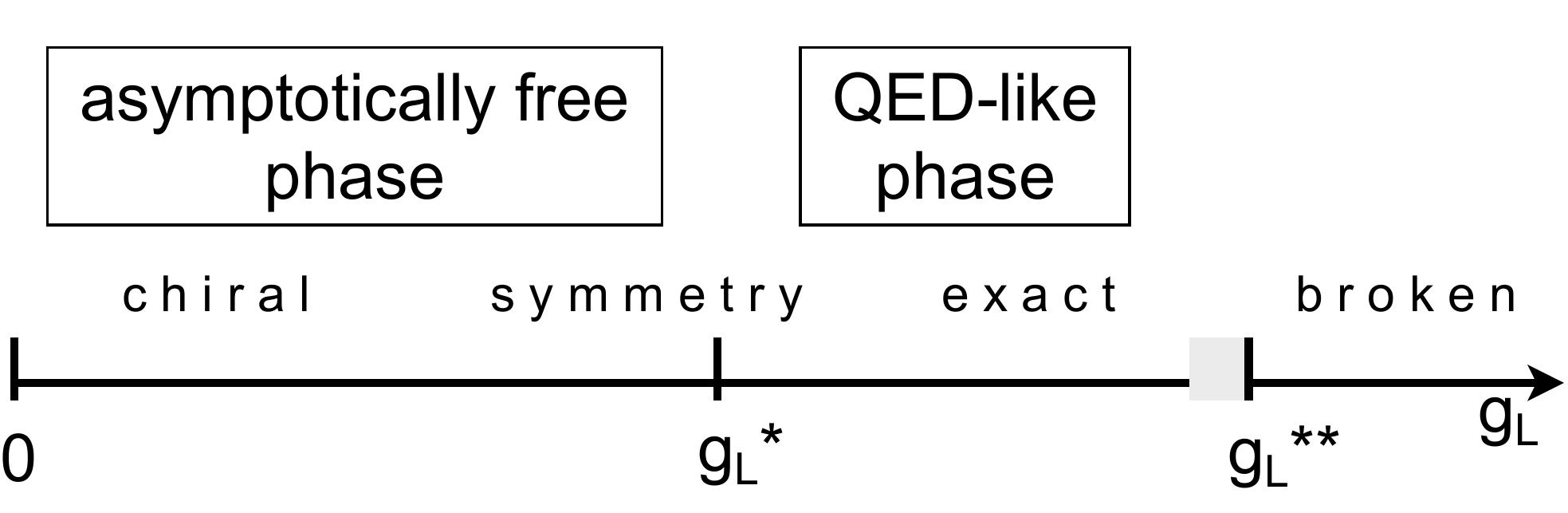}
\caption{\label{fig:gbar} Phases of the many-flavour $SU(N)$ gauge theory formulated on the lattice inside the conformal window. As the bare lattice gauge coupling $g_L$ increases from left to right, the lattice theory encounters an asymptotically free phase (negative $\beta$-function), it crosses the IRFP at some $g_L^*$ and  enters a QED-like phase (positive $\beta$-function); chiral symmetry is exact in all these phases. At $g_L^{**}$ a phase transition occurs to a chirally broken phase. It may be preceded by an exotic phase (grey shaded area) due to the improvement of the lattice action \cite{Deuzeman:2012ee,Cheng:2011ic,daSilva:2012wg}.  }
\end{figure}

As always, symmetries guide our understanding of a physical phenomenon, and we should identify the observables sensitive to those symmetries.   In this case, we are interested in the  two-point functions inside the conformal window, and the relevant symmetries are conformal, chiral and, to a certain extent, the $U(1)$ axial symmetry; the latter will be shortly discussed at the end of this work. Figure~\ref{fig:gbar} describes the pattern of phases for a theory inside the conformal window formulated on the lattice, i.e. a gauge invariant formulation on a discretized spacetime.
At small bare lattice coupling $g_L <g_L^*$ the theory is in the asymptotically free phase, characterized by a negative $\beta$-function. 
The lattice theory then crosses the IRFP at some lattice coupling $g_L^*$ and enters a Coulomb, or QED-like, phase characterized by a positive $\beta$-function. For all $g_L>g_L^*$ asymptotic freedom is lost and the lattice theory has no continuum limit\,---\,the only exception being the possible appearance of a UVFP at stronger coupling. For all $g_L<g_L^{**}$ in figure~\ref{fig:gbar} chiral symmetry is exact. At  $g_L^{**}$ the lattice theory exhibits a zero temperature, i.e. bulk, phase transition to a chirally broken phase. There are indications that the line of bulk phase transitions ends at a finite $\overline{N}_f \gg N_f^{af}$ \cite{deForcrand:2012vh}. 
Finally, the grey shaded area in figure~\ref{fig:gbar} indicates the possible emergence of an exotic phase due to improvement of the lattice action \cite{Deuzeman:2012ee,Cheng:2011ic,daSilva:2012wg}; chiral symmetry is still exact in this phase \cite{Deuzeman:2012ee,daSilva:2012wg}. 

\begin{figure}[tbp]
\centering
\includegraphics[width=.80\textwidth]{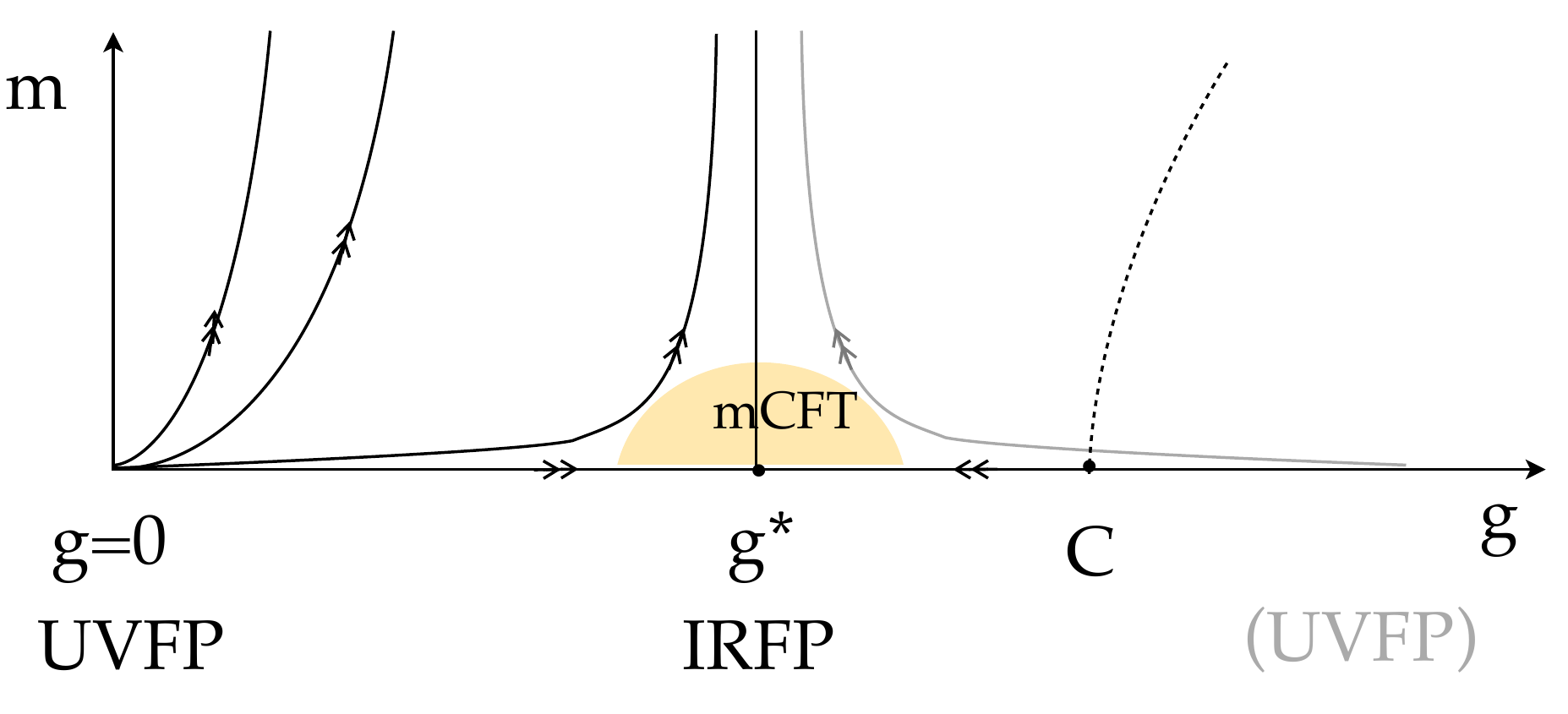}
\caption{\label{fig:RG}  Wilson RG flow and fixed points of non-Abelian $SU(N)$ gauge theories in $d=4$ spacetime inside the conformal window, with fermion mass $m$ and gauge coupling $g$. For the massless theory ($m=0$), the trivial UVFP ($g=0$, asymptotic freedom) becomes unstable towards $g$ perturbations due to quantum corrections and the system flows toward the non-trivial IRFP at  $g^*$, $m=0$.  The fermion mass operator is always relevant and the gauge coupling is irrelevant at the IRFP. 
Mass deformed conformal field theory (mCFT) for $g\simeq g^*$ and $m\to 0$ provides the universal scaling laws for observables at the IRFP. The dashed line is a line of possible initial values $(m,g)$ for the lattice system, and its critical point  $C$ flows into the IRFP. A UVFP at strong coupling may emerge.  }
\end{figure}

In our first study \cite{Deuzeman:2009mh} of the conformal window 
we outlined a strategy based on the physics of phase transitions, in order to characterize the different phases in  figure~\ref{fig:gbar}. 
We highlighted the observables  that are most suitable for probing the emergence of the conformal window and its properties:
the chiral condensate, order parameter of chiral symmetry, the chiral susceptibilities and their ratios, and the would-be hadron masses. 
 This goes beyond the direct probe of the IRFP, it probes the very existence of the conformal window and the nature of the symmetry breaking/restoring patterns. In fact,  the existence of an IRFP might not necessarily imply the existence of a conformal window characterized by a phase transition at some $N_f^c$. 
We also attempted for the first time  in the study of this theory a lattice determination of the fermion mass anomalous dimension, obtaining, with all due caveats and still sizable uncertainties,  a value in good agreement with most of later results; ours were obtained in the QED-like region of figure~\ref{fig:gbar}, and we stress once again the importance of identifying the  phase where observables are measured. 
 In this work we concentrate on the would-be hadron spectrum in the QED-like phase of the $N_f=12$ lattice system, as a specific probe of conformal and chiral symmetries. 
In line with recent work \cite{DelDebbio:2010ze} and renormalization group (RG) theory {\it \`a la} Wilson,
figure \ref{fig:RG} illustrates how the fermion mass term perturbs the RG flow of the continuum massless interacting theory inside the conformal window\,---\,we have assumed that no other couplings are present besides the gauge coupling and the mass itself. 
The mass term is a relevant operator that drives the theory away from the IRFP, while the gauge coupling is irrelevant due to quantum corrections. 
As discussed in \cite{DelDebbio:2010ze}, mass deformed conformal field theory (mCFT) can be used within the basin of attraction of the IRFP  to uncover the universal scaling properties of the correlation functions at infinite and finite volume. 
Away from the IRFP violations of universal scaling will gradually appear, while correlation functions still satisfy all constraints implied by the non spontaneously broken chiral symmetry.
A UVFP at strong coupling may emerge, however no indications of it have been found in preliminary lattice studies of the $N_f=12$ theory \cite{Deuzeman:2010fn}.  
We also note that, at strong coupling, new operators may be promoted from irrelevant to marginal or relevant; if so, the fixed point structure of the theory needs to be reanalyzed in an enlarged space of couplings, possibly complicating the lattice search for new UV/IR fixed points. 

 \subsection{Two-point functions}
\label{sec:CFs}

We are interested in  the two-point  functions made of  currents of the type $J_{M}\sim \bar{q}\Gamma_M q$, with $\Gamma_M = 1, \gamma_5, \gamma_\mu , \gamma_\mu\gamma_5$ for the scalar (S), pseudoscalar (PS), vector (V) and pseudovector (PV) mesons, respectively, and the nucleon correlation function with current $J_N \sim q q q$. In other words, we identify the would-be hadrons  of QCD in order to allow for a direct comparison with the spectrum of theories inside the conformal window.

At the IRFP the theory is massless and interacting, and its two-point functions satisfy universal scaling relations with nonzero anomalous dimensions
\begin{equation}
\langle 0| T\, J_H(x_1)J^\dagger_H (x_2)|0\rangle \sim (x_1-x_2)^{-\Delta_H}~~~~~H=M,\,N
\end{equation}
with $\Delta_H/2$ the scaling dimension of the current $J_H$. More specifically, we are interested in  the Euclidean two-point functions with zero total 3-momentum 
\begin{eqnarray}
\label{eq:Coft}
C_H (t)&=&\int~d^3x\,   
  \langle 0| T\, J_H(t,\vec{x})J^\dagger_H (0,0)|0\rangle \sim t^{-\Delta_H +3} \nonumber\\
&=&\int_0^\infty\,dE\, K(E,t)\, \sigma (E)\, .
\end{eqnarray}
This is the well-known power-law scaling of the correlations $C_H(t)$ at the IRFP.
We have also introduced the representation of $C_H(t)$ in terms of the spectral function $\sigma (E)$ and the kernel $K(E,t)$. Spectral functions are a powerful probe, widely used in the study of the QCD phase diagram at finite temperature and chemical potential; for example, they are a direct probe of the gradual melting of bound states in the quark-gluon plasma close to the critical temperature. 

In the presence of a nonzero fermion mass mCFT provides rigorous results inside the IRFP basin of attraction; these are derived in the next section. Away from the IRFP, known cases, such as QCD at zero temperature or QGP close to its critical temperature, provide a phenomenological insight. 
QCD is confining in the infrared, and free in the ultraviolet. Its complete spectral function $\sigma (E)$ entails the low-energy resonances and the high-energy continuum. Schematically, it is made of  a series of Dirac $\delta$-functions (the propagator poles) and a continuum with a high-energy threshold $\sigma (E) \sim  E^{\Delta_H -4}\left ( A_H\delta (E^2-m_H^2) +C \theta (E -E_0) \right )$, where for simplicity we assumed one resonance (pole) per channel $H$. 
If we take in eq.~(\ref{eq:Coft}) the kernel $K(E,t) = \exp{(-Et)}$, i.e. the Fourier transform of a free boson propagator
 for infinite temporal extent, the two-point function $C_H(t)$ can be exactly computed in terms of upper incomplete $\Gamma$ functions
\begin{equation}
C_H(t) = C_H^{pole}+C_H^{cont}\sim \sum_H m_H^{\Delta_H-3} e^{-m_Ht}+ \frac{C}{t^{\Delta_H-3}}\,\Gamma \left (\Delta_H -3, E_0 t\right )\, .
\label{eq:QCD_CF}
\end{equation}
At large times $t\to\infty$, using $\Gamma (s,x) /(x^{s-1} \exp{(-x)}) \to 1$ as $x\to\infty$, one obtains for the
high-energy continuum contribution 
\begin{equation}
C_H^{cont}(t)\sim \frac{E_0^{\Delta_H-4} }{t} e^{-E_0 t}\, ,~~~~t\to\infty\, .
\label{eq:QCD_CF_infty}
\end{equation}
The high-energy continuum generates an exponentially decaying contribution with time dependent coefficients, and leading $1/t$ behaviour at large times, differently from the low-energy poles.

QGP close to its critical temperature is instead an example of a deconfined, though strongly interacting, theory with restored chiral symmetry;  it is almost analogous to a theory inside the conformal window, except that there is no IRFP.  A realistic description of the QGP two-point functions  is a  rich subject of study that is beyond our analogy. It is here enough  to observe that the system  undergoes a gradual melting of the QCD bound states till their disappearance into a continuum. How gradual is the analogous transition inside the conformal window depends on various factors: the nature of the zero temperature phase transition that opens the conformal window at  $N_f^c$, the strength of the interactions at the IRFP, the quantum numbers of the would-be hadrons. 
\subsubsection{Universal scaling laws at infinite volume}
\label{sec:scaling}

In order to understand the behaviour of the two-point functions $C_H(t)$ and the would-be hadron masses in the surroundings of the IRFP we summarize, partly reformulate and adapt to our case known aspects of the scaling theory at a conformal fixed point.  The scales of the system are the lattice spacing $a$ (that can be thought as the inverse of an ultraviolet momentum cutoff), the characteristic length  $\xi$ that will emerge in the scaling analysis, and the spatial length $L$ of the lattice box; 
we shall consider specific ranges for $a$, $\xi$ and $L$. The couplings are the fermion mass $m$ and the gauge coupling $g$. They have scaling dimensions 
\begin{equation}
[m] = 1+\gamma~~~[g] = -\gamma_g\, ,
\end{equation}
where the scaling dimension is the sum of the canonical dimension and the anomalous dimension, $\gamma$ and $-\gamma_g$, respectively.
At the IRFP, $g=g^*$ and $m=0$, the anomalous dimensions have values $\gamma^*$ and $-\gamma_g^*$, and we introduce the exponent $\delta = (1+\gamma^*)^{-1}$ for later use. For the interacting theory, the coupling $m$ is always relevant in the RG sense ($0<\gamma <2$), while $g$ has $-\gamma_g <0$ due to perturbative quantum corrections and it is thus irrelevant.  
We first consider the lattice system in the infinite volume limit $L\to\infty$ and in the continuum limit $a\to 0$\footnote{The latter limit is realized, in practice, whenever the characteristic length $\xi$, defined later in eq.~(\ref{eq:univ}), is much larger than $a$.}. 
The invariance of the system under a rescaling of coordinates in the presence of a small perturbation of the relevant coupling $m$ at the IRFP provides the universal scaling relations for $C_H(t)$; they are universal in the sense that they do not depend on the microscopic details of the system. 
Under a rescaling of coordinates $x'=x/b$\,---\,where $x$ indicates both space and time\,---\,the form of the correlation is dictated by its scaling dimension $s_H$ 
\begin{equation}
\label{eq:scal}
C_H(t/b; b^{1/\delta}m) = b^{s_H} C_H(t ; m)\, ,
\end{equation}
where $s_H=\Delta_H-3$.
Setting $m=0$, eq.~(\ref{eq:scal}) yields 
$C_H(t/b) = b^{s_H} C_H(t)$, hence $C_H(t)\sim t^{-s_H}$, the scaling form already introduced in eq.~(\ref{eq:Coft}).
Setting $b=\exp{(l)}$ gives a better intuition of the approach to large distances. Since eq.~(\ref{eq:scal}) holds for any $l$, we can choose $l=l^*$ so that $m\exp{(l^*/\delta )} =1$. We can also define a characteristic length $\xi = \exp{(l^*)}$, so that we obtain 
\begin{equation}
\label{eq:univ}
C_H(t) = Z_H\,\xi^{-s_H}\, F\left (\frac{t}{\xi}\right )\, .
\end{equation}
 The coefficient $Z_H$ accounts for the microscopic details of the system and has zero scaling dimension. The second factor carries the scaling dimension $s_H$ of $C_H$, while the third factor is the adimensional universal scaling function that only depends on the ratio $t/\xi$, with zero scaling dimension. In the most general case, the scaling function $F$ will depend upon  all possible products with zero scaling dimension, made of a scale and a coordinate or a relevant coupling. 
The function $F$ is universal in the sense that does not depend on the microscopic details of the system. It depends on the scaling dimension of the operators $H=M,N$, their spin and normalisation. 
 For any nonzero $m$, one can think of $\xi$ as a finite characteristic length $\xi =m^{-\delta}$;  eq.~(\ref{eq:univ}) says that for a change of $m$, the coordinate $t$ changes at the scale $\xi$. 
In the limit $\xi\to\infty$, or equivalently $t\ll\xi$, the system approaches the IRFP and eq.~(\ref{eq:univ}) should reproduce the form $C_H(t)\sim t^{-s_H}$; this constrains  the asymptotic behaviour of the scaling function $F(t/\xi )$.  
 In the opposite limit $t\gg\xi$, the two-point functions decay exponentially, so that 
\begin{eqnarray}
C_H(t) &\sim& t^{-s_H} ~~~~~~~~~~~~~~~~~~~~~~t\ll\xi\nonumber\\
C_H(t) &\sim& \xi^{-s_H}\,f\left (\frac{t}{\xi}\right )\,e^{-t/\xi} ~~~~~t\gg\xi\, .
\label{eq:asymptotic}
\end{eqnarray}
While the time dependence is dictated by the scaling dimension of $C_H$, the numerator in the limit $t\ll\xi$ depends on the spin and normalisation of the operators $H=M,N$. In the opposite limit $t\gg\xi$, 
dimensional reasoning allows for all terms in $f(t/\xi )$ that reproduce the correct scaling at the fixed point, $\xi\to\infty$ or equivalently $t\ll\xi$. Specifically, it allows for all powers $t^{-\alpha}$, with $0\leq \alpha\leq s_H$ and including $\alpha =0$. 
While the limit $t\ll\xi$ is uniquely determined by the conformal fixed point, the details of the limiting behaviour for $t\gg\xi$ depend on the nature of the interactions in the quantum system. Our case is that of composite operators in a deconfined and interacting non-Abelian gauge theory close to a non trivial IRFP; we do not know a priori if the system is weakly or strongly coupled. Without a detailed knowledge of the quantum correlators, we can still infer their most general form close to the fixed point. The presence of a mass threshold, i.e. $\xi$ finite, produces a pole and branch-cuts (i.e., a continuum of critical excitations) in the propagators of the quantum composite operators\footnote{Analogous examples can be found in magnetic systems close to a quantum critical point, where a quasiparticle pole and multiparticle continuum thresholds are present in the system.}. Using for the sake of illustration the schematic form in eq.~(\ref{eq:QCD_CF}) for the continuum part, with threshold $E_0 = n/\xi$, for some $n$, one obtains
\begin{equation}
C_H(t) \sim \xi^{-s_H} e^{-t/\xi} + \frac{C}{t^{s_H}} \Gamma \left (s_H, \frac{nt}{\xi}\right )\, ,
\label{eq:intermediate}
\end{equation}
for  any intermediate time and finite $\xi$. The first term is generated by the would-be hadron pole with residue 
$\xi^{-s_H}$. The second term is a continuum of excitations with energy threshold proportional to $\xi^{-1}$. The relative position of the pole and thresholds depends on the specific theory and interactions. It is always true, however, that all thresholds and the pole merge at the fixed point, where the residue at the pole vanishes.

Note that the would-be hadron pole, the first term in eq.~(\ref{eq:intermediate}), is proper of the deconfined theory close to the IRFP and is not to be identified with the hadron poles of confined QCD.   In fact, it vanishes at the fixed point and the theory smoothly flows to pure Yang Mills at infinite fermion mass, i.e. $\xi\to 0$. This description of theories inside the conformal window does not need, but does not exclude, the occurrence of a phase transition at a finite fermion mass between a deconfined and a confined phase, a scenario proposed in \cite{Ishikawa:2013tua}; note that confinement is always realized in the limit $m\to\infty$, where fermions decouple. 
 
From the practical viewpoint of a lattice study of the system, it is sufficient to observe that the two-point functions $C_H(t)$ in the presence of a fermion mass $m=\xi^{-1/\delta}$ are dominated by a constant times an exponential, i.e.  $\xi^{-s_H} \exp{(-t/\xi )}$, for  $t\gg \xi$ and $\xi\neq 0$, while the time-dependent power-law contributions to 
 $f(t/\xi )$ in eq.~(\ref{eq:asymptotic}) become increasingly important at smaller times and for decreasing masses.
We also observe that the addition of a UV cutoff, i.e. a nonzero lattice spacing $a$, or a finite temporal extent $t\leq T$, do not qualitatively change any of the properties discussed, nor affect the extraction of the would-be hadron mass from the dominant pole contribution. 

The comparison of eq.~(\ref{eq:univ}) with the large euclidean time behaviour $C_H(t)\sim \exp{(-m_H t)}$ for a would-be hadron of mass $m_H$ provides the universal scaling form  
\begin{equation}
\label{eq:scalm}
m_H=c_H\,m^\delta\, ,
\end{equation}
at coupling $g=g^*$
and with  coefficient $c_H$ that depends on the spin of the operator $H=M,N$. 

\subsubsection{Universal scaling laws  at finite volume}
\label{sec:scaling2}

Keeping 
 $g= g^*$, we now consider  the system at finite volume $L$, with $\xi ,\,L\gg a$, and we trade the characteristic scale $\xi$ for the mass $m$, using  $\xi^{-1}=m^\delta$; in this way, the system has effectively one relevant scale $L$, and we can study the scaling of $C_H$  under a change of $L$. 
In full analogy with eq.~(\ref{eq:univ}), the correlation is now\footnote{The large volume limit of the field theory at the IRFP can be treated analogously to the low-temperature perturbation of a lattice spin system at a zero temperature critical point.  } 
\begin{equation}
\label{eq:univL}
C_H(t; m, L) = \tilde{Z}_H\,L^{-s_H}\, F\left (\frac{t}{L} , L m^\delta \right ) \, .
\end{equation}
It is the product of a coefficient $\tilde{Z}_H$, which accounts for the microscopic details, a scaling factor and the universal scaling function $F$ with  zero scaling dimension arguments, made of products of the scale $L$ and a coordinate or a coupling.   
The leading scaling form of the would-be hadron mass in the channel $H$ as a function of  the fermion mass and $L$ now reads
\begin{equation}
\label{eq:scalm_L}
m_H=\tilde{c}_H\frac{1}{L}\,f(x )\,~~~~~~~x=Lm^\delta\, ,
\end{equation}
where the scaling function $f(x)$ depends on the  parameter $x$ with zero scaling dimension. To recover the infinite volume limit of eq.~(\ref{eq:scalm}), one needs $f(x)\sim x$ as $x\to\infty$ and $\tilde{c}_H=c_H$. One has also $f(x)\to const$ as $x\to 0$.
The function $f(x)$ a priori depends on the channel $H=M,N$ through the spin of the corresponding operators. However, once the constant $c_H$ is factored out, its $x\to\infty$ limit is $H$ independent. Moreover, the study of section~\ref{sec:results_scaling} and figure~\ref{fig:Scaling_All} suggests that most of the $H$ dependence is contained in $c_H$ for all $x$. 
 Eqs.~(\ref{eq:scalm}) and (\ref{eq:scalm_L}), important for the scaling analysis of the lattice results, were first derived in \cite{DelDebbio:2010ze}.     
Eq.~(\ref{eq:scalm_L}) says that finding the universal curve followed by $Lm_H/c_H$ as a function of $x$, with $1/\delta = 1+\gamma^*$, is a way to determine the mass anomalous dimension $\gamma^*$ at the IRFP. 
\subsubsection{New operators and corrections to universal scaling}
\label{sec:scalingv}

 The emergence of new marginal or relevant operators can occur in the system at sufficiently strong coupling, e.g., the four fermion operator  can turn from irrelevant to marginal, or relevant. In this case the fixed point structure has to be reconsidered, together with the RG flow towards the fixed point(s) in the enlarged parameter space. This is hardly the case for the IRFP studied here, but it may play a role in the possible emergence of a new ultraviolet fixed point at stronger coupling.  

On the lattice, one can isolate the universal scaling in eq.~(\ref{eq:scalm_L}) by identifying the
 perturbative corrections in the volume $L$ and mass $m$ that produce a deviation from the scaling function $f(x)$ and possible nonperturbative scaling violations. We discuss a few aspects below.
\begin{itemize}
\item[$\bullet$] {\bf perturbative corrections in $L$ and $m$}: close to the fixed point the irrelevant couplings, i.e. $g\neq g^*$ in our case,  generate perturbative corrections made of products of couplings and scales with zero scaling dimension.
 \item[$\bullet$] {\bf nonperturbative corrections}: they should be expected when the scale(s) of the microscopic dynamics becomes comparable to or larger than the characteristic length of the system.  Violations of universal scaling in this context can appear with nonuniversal functions that cannot be factored out of $f(x)$. One example is when the box size $L$ becomes small compared to the Compton wavelength of the would-be hadron $\xi$. Other examples are provided in this context by the occurrence of the bulk phase transition at strong coupling, which changes the underlying symmetries of the system, or the possible appearance of new phases that precede the bulk transition and are induced, e.g., by competing interactions in Symanzik-improved lattice fermion actions on coarse lattices \cite{Deuzeman:2012ee}.
 \end{itemize}
The first type of corrections, perturbative in $L$ and $m$, deserves some discussion. 
They are  induced by the irrelevant coupling $g$, whenever we take $g\neq g^*$. Consider a small variation $\Delta g$; the smaller the scaling dimension $\gamma_g^*$ of $g$, the slower the system will flow back to the IRFP. The leading perturbative corrections to  eqs.~(\ref{eq:univ}) and (\ref{eq:univL})
can be of two types. Firstly, $Z_H$ ($\tilde{Z}_H$) and $m$ (the relevant coupling) should be redefined, in a way that may generically be difficult to compute\footnote{However, a perturbative treatment of the RG equations in the case of an IRFP at weak coupling may provide the leading contributions to these corrections.}.  Secondly, multiplicative corrections to the scaling function $F$ appear. They can be written in terms of products with zero scaling dimension, made of the coupling $\Delta g$ and the relevant scale, thus 
 $1+ \Delta g\xi^{-\gamma_g^*}$, or equivalently $1+\Delta g m^{\delta\gamma_g^*}$, in eqs.~(\ref{eq:univ}) and (\ref{eq:univL}), and $1+  \Delta g L^{-\gamma_g^*}$ in eq.~(\ref{eq:univL}). The first corrections become increasingly unimportant as $\xi\to\infty$ ($m\to 0$), the second ones as $L\to\infty$. 
The same perturbative corrections for $g\neq g^*$ will enter $m_H$ as redefinitions of $c_H$, the fermion mass $m$ itself (equivalently the scaling dimension $\delta$), and multiplicative corrections of the type $1+ \Delta g\xi^{-\gamma_g^*}$  and $1+  \Delta g L^{-\gamma_g^*}$. 
\begin{itemize}
\item The coefficient $\Delta g$, referred to as the fitting parameter $b$ in  section~\ref{sec:spectrum}, 
measures the deviation of the gauge coupling from its value at the fixed point, or, on the lattice, the deviation from $g^*$ along the line of lattice parameters that flows to the IRFP in the continuum limit. It is a nonuniversal parameter that vanishes at the IRFP, possibly changing its sign. In particular, when symmetries allow for a linear dependence on $g-g^*$, we can expect $\Delta g$ (or the parameter $b$) to change sign at the boundary between the two phases of the lattice system, 
 indicating if the latter is located on the strong coupling side of the IRFP, i.e. the QED-like phase on the lattice, or the weak coupling side of the fixed point, i.e. the asymptotically free phase on the lattice. 
We also expect $\Delta g$ to flow to zero in the continuum, i.e. $\beta_L\to\infty$, where the lattice system reaches the IRFP. 
\item The exponent $\delta\gamma_g^*$, entering the perturbative corrections to scaling, is universal and it  is given by the anomalous dimension of the fermion mass $\gamma^*$ and the anomalous dimension of the gauge coupling at the fixed point $-\gamma_g^*$; this tells that the universal scaling function as well as its perturbative corrections contain information on $\gamma^*$. 
\end{itemize}
We will encounter realizations of the perturbative corrections in $m$ and nonperturbative corrections in $L$ in section~\ref{sec:spectrum}.
As a final note, the mixing of the couplings $m$ and $g$ under the RG flow cannot generate in this system a new fixed point {\it \`a la} Wilson-Fischer at $m^*\neq 0$ due to the chirality protection of the fermion mass term, i.e., $dm/d\mu \propto m$ with renormalization scale $\mu$. We study eq.~(\ref{eq:scalm_L}) on the lattice  in section~\ref{sec:results_scaling} and eq.~(\ref{eq:scalm}) in section~\ref{sec:delta}.

\subsection{The fermion mass anomalous dimension at the IRFP}
\label{sec:23}

One aim of this study is to measure $\gamma^*$ nonperturbatively on the lattice, by identifying universal scaling and the pattern of scaling violations in the would-be hadron spectrum.  
Numerous lattice studies have recently reported over a wide variety of evidences that the $N_f=12$ theory  is inside the conformal window, supporting the first results in \cite{Appelquist:2007hu}, based on the flattening of the running gauge coupling, and \cite{Deuzeman:2009mh} that probed the very existence of the conformal window. While the phenomenological interest mainly resides in  theories just below the conformal window, it remains important to determine the pattern and sizes of  observables inside the  window, since they are  directly, and in some cases smoothly, related to their value below the window lower endpoint; one such quantity is the fermion mass anomalous dimension $\gamma^*$ along the IRFP line. Its value determines if the theory is strongly or weakly interacting at the IRFP.
It is null in the free theory, at $N_f=N_f^{af}$, and  
 it  is bounded to be $\gamma^* < 2$ by the unitarity of the conformal theory \cite{Mack:1975je,Ferrara:1974pt}. An appealing conjecture suggests  $\gamma^* = 1$ at the lower endpoint  of the conformal window, $N_f=N_f^c$, so that chiral symmetry breaking is triggered \cite{Appelquist:1996dq}\,---\,for $\gamma^* =1$ the four-fermion operator becomes relevant.
The theoretical question if  $\gamma^* =1$ holds exactly at $N_f^c$ is still open; it is however appealing to think that the exactness is realized \cite{Cohen:1988sq}, and some consequences of this scenario are discussed in section~\ref{sec:delta}. 

\begin{figure}[tbp]
\centering
\includegraphics[width=.60\textwidth]{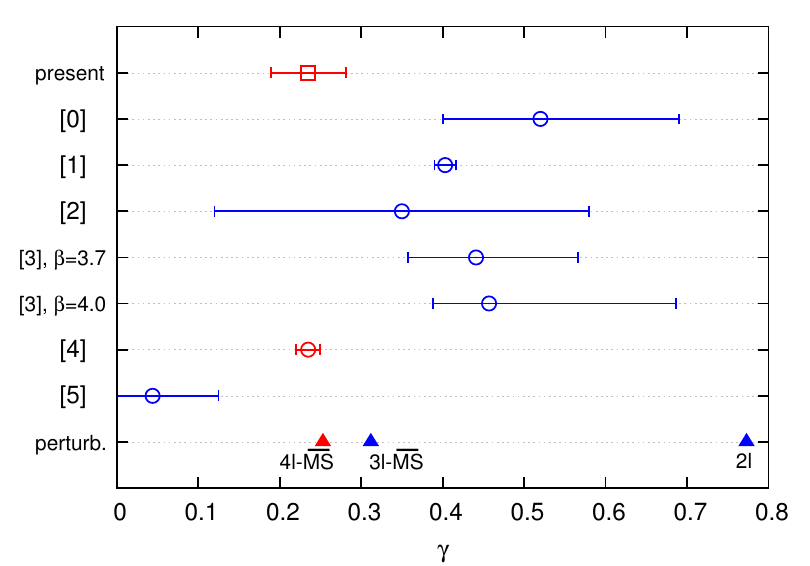}
\caption{ \label{fig:gamma_summary}  Collection of recent lattice determinations of the mass anomalous dimension $\gamma$ for the $N_f=12$ system, and the perturbative prediction of its value at the IRFP to 2, 3 and 4 loops. From top to bottom, the value from this work, $[0]$ from \cite{Deuzeman:2009mh}, $[1]$ from \cite{Appelquist:2010xv}, $[2]$ from \cite{DeGrand:2011cu}, $[3]$ from \cite{Aoki:2012eq}, $[4]$ from \cite{Cheng:2013xha}, $[5]$ from \cite{Itou:2013kaa}, and the perturbative determinations from \cite{vanRitbergen:1997va,Vermaseren:1997fq}. 
}
\end{figure}

The value of $\gamma^*$ can be determined nonperturbatively on the lattice\footnote{It is worth noting that other field-theoretical  techniques, such as conformal bootstrap and variations thereof, are becoming increasingly useful in constraining correlators and anomalous dimensions of operators in strongly coupled field theories.} and it has been computed in perturbation theory at two-loops \cite{Caswell:1974gg}, three and four loops \cite{vanRitbergen:1997va,Vermaseren:1997fq}, see also  \cite{Gracey:1996he,Holdom:2010qs} for studies in the large-$N_f$ limit, \cite{Shrock:2014qea,Ryttov:2012ur} for analyses of renormalization scheme transformations and \cite{Pica:2010xq} for a recent fixed point analysis of classes of gauge theories.
It is thus mandatory to compare the genuinely nonperturbative lattice determination with perturbation theory, and  
in this spirit we analyze the lattice results in section~\ref{sec:spectrum}. Here, we discuss a few relevant aspects that can be inferred from \cite{vanRitbergen:1997va,Vermaseren:1997fq}  and we compute the gauge coupling anomalous dimension $-\gamma_g^*$ to four loops.
By inspection of the four-loop $\beta$-function and the mass anomalous dimension $\gamma$ in  \cite{vanRitbergen:1997va,Vermaseren:1997fq} one observes that: 
\begin{itemize}
\item The $N_f=12$ IRFP coupling in the $\overline{\mbox{MS}}$ scheme moves from $g^2/(4\pi^2)\simeq 0.24$ at two loops to $g^2/(4\pi^2)\simeq 0.15$ at four loops, and the mass anomalous dimension at the IRFP\,---\,a renormalization scheme independent quantity\,---\,moves from $\gamma^*\simeq 0.77$ to $\gamma^*\simeq 0.25$. The latter value provides $\delta\simeq 0.8$; this value will turn out to be in good agreement with our lattice determination. 
\item The IRFP coupling moves towards the origin when going from two to four loops, while the lower endpoint of the conformal window stays around $N_f=8$. 
\item The $N_f$ dependence of the coefficients $\beta_i$ and $\gamma_i$, $i=0,1,2,3$, of the four-loop $\beta$-function and $\gamma$, respectively, deserves some discussion. The four-loop coefficient $\beta_3$ grows rapidly with $N_f$ and it is responsible for the appearance of  a new zero for $N_f\geq 17$, just above the conformal window. Analogously, the four-loop coefficient $\gamma_3$  grows rapidly with $N_f$ and causes a change of sign of the running anomalous dimension at some $g$ for $N_f\geq 8$; importantly, this happens at a coupling $g\gg g^*$ for $N_f=12$, suggesting that perturbation theory may still be reliable for $N_f=12$ at $g=g^*$. 
This is no longer true for $N_f\sim 8$, where $g^*$ is larger and the change of sign occurs for $g<g^*$. 
 We do not further address the problem of the reliability of perturbation theory for $N_f\sim 8$, with or without a truncation to a given order in the loop expansion.  We only note that this is the region where nonperturbative contributions can play a significant role in the disappearance of the conformal window. An instructive comparison between perturbation theory and nonperturbative results can be carried out in the Veneziano limit, see \cite{Shrock:2014qea} for studies in this limit, or with a large-$N_f$ resummation as initially proposed in \cite{Gracey:1996he}. Recent progress in constraining by field-theoretical techniques the correlators of large-$N$ QCD and ${\cal N}=1$ SQCD \cite{Bochicchio:2014vna,Bochicchio:2013eda} should help in the task of  identifying the role of nonperturbative contributions.
\item Finally, we derived the value of the gauge coupling anomalous dimension at the IRFP to two and four loops from \cite{vanRitbergen:1997va} as $-\gamma_g^* = \partial\beta (g)/\partial g\vert_{g=g^*}$. As all other critical exponents, this is a renormalization group invariant quantity. A straightforward calculation gives $\gamma_g^*\vert_{2-loop} \simeq 0.360$ and  $\gamma_g^*\vert_{4-loop} \simeq 0.283$, consistently with the fact that the IRFP moves towards weaker coupling from two to four loops.
\end{itemize}
To summarize, four-loop perturbation theory predicts $\gamma^* \simeq 0.25$, i.e. $\delta \simeq 0.8$, and the universal exponent $\delta\gamma^*\simeq 0.23$ for the $N_f=12$ system at the IRFP. This prediction misses the nonperturbative contribution, and some renormalization scheme dependence can be induced by the truncation of the perturbative expansion. A comparison with the nonperturbative lattice determination of $\gamma^*$ is therefore instructive. Figure~\ref{fig:gamma_summary} collects recent  lattice determinations of $\gamma^*$ and the predictions of perturbation theory, anticipating the result of this work later discussed in section~\ref{sec:spectrum}. 
The most salient feature of figure~\ref{fig:gamma_summary} is the agreement among lattice determinations, and their agreement with the four-loop perturbative prediction, once a universal scaling analysis is carried out.  This is true for \cite{Cheng:2013xha} and this work. Previous pioneering  determinations of the mass anomalous dimension, the very first one in \cite{Deuzeman:2009mh} and the ones in \cite{Appelquist:2010xv,DeGrand:2011cu,Aoki:2012eq,Cheng:2013eu,Itou:2013kaa} were obtained through the analysis of specific $H$ channels in the would-be hadron spectrum or the eigenvalues of the Dirac operator at some bare lattice coupling, without a systematic identification of the universal scaling contributions and violations thereof. Despite this, all determinations are contained in an interval that is well below $\gamma =1$. This shows the stability of the prediction and the fact that the (lattice) system is not largely sensitive to deviations from the IRFP. Importantly, measurements can be done on both sides of the fixed point, being it the asymptotically free side or the QED-like side. In section~\ref{sec:delta} we futher discuss our determination of $\gamma^*$ and the implications of the obtained value. 
\subsection{The Edinburgh plot}
\label{sec:edinburgh}

Besides conformal symmetry, one relevant ingredient characterizing the two-point  functions in the conformal window is restored chiral symmetry. Everywhere in the asymptotically free and the QED-like phase, chiral Ward identities must be fulfilled by the renormalized correlation functions. There is no Goldstone boson, since chiral symmetry is not spontaneously broken, and all chiral partners, scalar and pseudoscalar, vector and axial, must be degenerate in the chiral limit\footnote{Exact chiral symmetry implies the degeneracy of the complete renormalized two-point  functions in the channels that are chiral partners, and the degeneracy of  the corresponding renormalized chiral susceptibilities\,---\,the integrals of the two-point functions.}.

\begin{figure}[tbp]
\centering
\includegraphics[width=.60\textwidth]{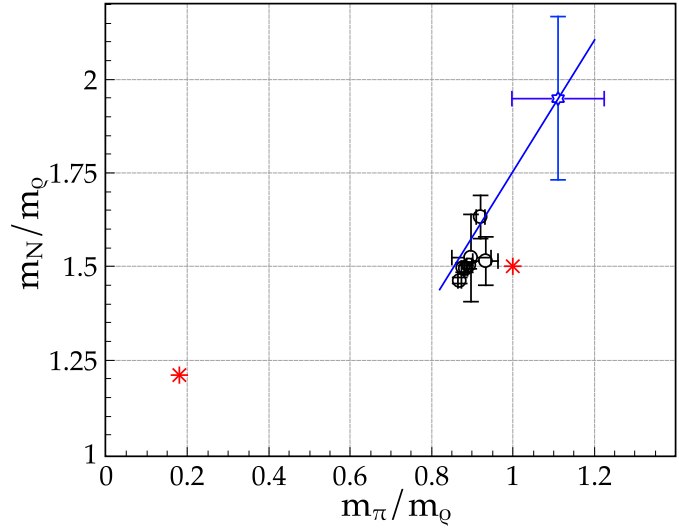}
\caption{ \label{fig:Edplot}  The $N_f=12$ Edinburgh plot for the infinite volume lattice results in table~\ref{tab:masses} and \ref{tab:FVextrap} at $\beta_L=3.9$: the ratio of the nucleon (N) and the vector ($\rho$) mass is shown as a function of the ratio of the pseudoscalar ($\pi$) and the vector mass. We show the  scaling point (blue diamond) with coordinates $(x,y)=(c_\pi /c_\rho , c_N/c_\rho )$, with $c_{\pi ,N}$ from Fit I in table~\ref{tab:delta} and $c_\rho$ in table~\ref{tab:usKMI_PSV} (This work). The superimposed (blue solid) line $y=(c_N/c_\pi ) x$ (error band not shown) entails the perturbative scaling violations derived in section~\ref{sec:spectrum}. 
The coefficients $c_H,\,H=\pi ,\,\rho,\,N$ are a priori $\beta_L$ dependent, so that the solid (blue) line as well as the scaling point  flow to their continuum value, as $\beta_L\to\infty$, cf. section~\ref{sec:results_scaling}.
The QCD physical point (red star, leftmost) and the heavy quark limit (red star, rightmost) are also displayed.}
\end{figure}

The Edinburgh plot, widely used in lattice QCD studies and suggested as a probe of the conformal window in \cite{Deuzeman:2012pv}, is constructed in terms of adimensional ratios of masses, and/or decay constants, and it traditionally offers a powerful way to combine results of lattice calculations performed at different lattice spacings and with different lattice actions. In this case, it also provides a clear visualisation of different mass regimes and distinguishes between QCD and theories inside the conformal window; we use it here to illustrate the behaviour of the $N_f=12$ system, adopted as a prototype of theories inside the conformal window. 
Figure~\ref{fig:Edplot} shows the Edinburgh plot for the $N_f=12$ infinite volume lattice results in table~\ref{tab:masses} for $am > 0.025$ and in table~\ref{tab:FVextrap} for $am\leq 0.025$.
The physical point of QCD (left of figure) corresponds to $m_\pi/m_\rho\simeq 0.18$ and $m_N/m_\rho\simeq 1.21$. At the other side of the figure a useful theoretical limit is the heavy quark mass limit, where all masses of the would-be hadrons are given by the sum of their valence quark masses so that $m_\pi/m_\rho =1$ and $m_N/m_\rho =3/2$.
A QCD scenario would correspond to a curve in figure~\ref{fig:Edplot} that extrapolates to the QCD physical point for decreasing quark masses, i.e., it would join the two red points in figure~\ref{fig:Edplot}. Instead, we observe that the two mass ratios are ``stuck'' at a tiny corner of the plot, away from the heavy-quark limit and the QCD physical point  for a wide range of bare fermion masses $0.01\leq am\leq 0.07$. This is to be expected inside the conformal window; would-be hadron masses scale as $m_H=c_H\,m^\delta$ at the IRFP, ideally producing one point in the Edinburgh plot. Moving away from the IRFP\,---\,in mass, coupling(s) and volume\,---\,produces some scattering  of the data. 
A mild mass dependence of the infinite volume mass ratios is induced by perturbative scaling violations for $g\neq g^*$ and $m\neq 0$. 
Anticipating the results of section~\ref{sec:spectrum}, these are represented by points distributed along the solid line that passes through the scaling point in figure~\ref{fig:Edplot}. All scaling violations still obey the constraints implied by the underlying restored chiral symmetry. 
\section{Numerical setup}
\label{sec:setup}
\subsection{The action}
\label{sec:action}

We have generated configurations of an $SU(3)$ gauge theory with twelve degenerate flavours $N_f$ of staggered fermions in the fundamental representation using a tree level Symanzik improved gauge action 
\begin{equation}
S = -\frac{N_f}{4} \mathrm{Tr}\,{\ln{M(am,U)}} + \sum_{i=0,1}\beta_i(g^2)\sum_{\mathcal{C}\in \mathcal{S}_i} \mathrm{Re}(1-U(\mathcal{C})) 
\end{equation}
where $U(\mathcal{C})$ are the traces of the ordered product of link variables along the closed paths $\mathcal{C}$ divided by the number of colors. $\mathcal{S}_0$ and $\mathcal{S}_1$ contain all the $1\times1$ plaquettes and $1\times 2$ and $2\times 1$ rectangles, respectively. The $SU(3)$ lattice coupling of the unimproved action is given by $\beta = 6/g_L^2$ and the $\beta_i$ are defined in terms of $\beta$ as $\beta_0 = (5/3)\beta$ and $\beta_1 = -(1/12)\beta$. According to the way lattice simulations are performed and reported, we use $\beta_L (= \beta_0)=10/g_L^2$ for the lattice results of this work, and $\beta =6/g_L^2$ for some of the existing lattice results discussed in section~\ref{sec:spectrum}.  

Improvement is extended to the fermionic sector following the Naik prescription. The action of the fermionic sector can be written in terms of the one component staggered fermion field $\chi(x)$ as
\ba
S_F &&= a^4\sum_{x;\mu} \eta_\mu(x)\bar{\chi}(x)\frac{1}{2a}\left\{c_1\left [ U_\mu(x)\chi(x+\mu ) -U^\dagger (x-\mu)\chi(x-\mu)   \right ] \right. \nonumber\\
&&+ c_2\left [ U_\mu (x)U_\mu (x+\mu)U_\mu(x+2\mu)\chi(x+3\mu)   \right . \nonumber\\
&&\left.\left . - U^\dagger_\mu(x-\mu)U^\dagger_\mu(x-2\mu)U^\dagger_\mu(x-3\mu)\chi(x-3\mu)\right ]\right\} \nonumber\\
&&+a^4 m\sum_x\bar{\chi}(x)\chi(x)
\label{eq:naikaction}
\ea
with the phase factor $\eta_\mu(x) = (-1)^{(x_0+x_1\ldots +x_{\mu -1} )}$. Order $a^2$ accuracy at tree level is achieved by using the Naik choice $c_1=9/8$ and $c_2=-1/24$.

This action is the same used in previous studies conducted by our group on $SU(3)$ with $N_f=12$ \cite{Deuzeman:2009mh, Deuzeman:2012ee,  daSilva:2012wg}. In particular, this action corresponds to the choice $D$ of \cite{Deuzeman:2012ee}. The theory under study exhibits a bulk transition separating a region at weak coupling where chiral symmetry is restored from a region at strong coupling where chiral symmetry is broken \cite{Deuzeman:2009mh}, as it is expected for all theories in the conformal window.  For small enough bare fermion masses and with our choice of action, the competition induced by next-to-nearest neighbour interactions in eq.~(\ref  {eq:naikaction}) causes the emergence of an intermediate phase at finite lattice spacing, just before chiral symmetry is broken, as one goes from weak to strong coupling \cite{Deuzeman:2012ee}, see also Fig.~\ref{fig:gbar}. 
 
We generated configurations for a range of bare fermion masses going from $am = 0.01$ to $am = 0.07$ at fixed coupling $\beta_L =10/g_L^2= 3.9$. This choice guarantees that our simulations are carried in the chirally restored region, away from the bulk transition and the exotic phase induced by the improvement. For the heaviest bare masses $am = 0.06$ and $am = 0.07$ we simulated volumes $16^3\times24$ and $24^4$. For bare masses $am = 0.05, 0.025$ and $0.020$ we have simulated volumes $24^4$ and $32^4$. For bare masses $am = 0.01, 0.04$ we have simulated volumes $24^3\times 32$ and $32^4$. In addition, volume $16^3\times 32$ was simulated for bare masses $am = 0.01, 0.02, 0.025$ in order to make it possible for us to obtain infinite volume estimates of the spectrum for these quark masses.

During the runs, the parameters for the acceptance rate have been tuned to yield a good acceptance while keeping a fixed trajectory length $l \approx 0.4$ for all ensembles. Configurations were saved every five simulated trajectories, so that saved configurations are separated by approximately two unit trajectory lengths. Measurements of observables such as the chiral condensate and the average plaquette were conducted on the fly, while the measurements of the particle spectrum were conducted on the saved configurations following the strategy described in the next section. 

\subsection{Strategy for the spectrum measurements}
\label{sec:strategy}

We have measured the two-point functions with currents $J_M\sim \bar{q}\Gamma_M q$ in the scalar, pseudoscalar, vector and axial channels and the nucleon correlation function on the saved ensembles using corner-wall sources with fixed Coulomb gauge. The gauge fixing procedure in traditional QCD is known to reduce contamination from excited states and helps to better isolate the ground state of the system. 
In order to extract the lowest-lying masses we found it useful to construct the meson correlators from quark propagators with different combinations of temporal boundary conditions. This procedure was discussed in \cite{Sasaki:2005ug}, it is extensively used in lattice QCD and, recently, it was explicitly implemented for $SU(3)$ with $N_f=12$ in \cite{Aoki:2012eq}. 
Some caveats are in order inside the conformal window, where 
the two-point function has the form in eqs.~(\ref{eq:univ}) and  (\ref{eq:asymptotic}).
We first summarize the strategy in the case of an exponentially decaying two-point function with constant coefficient, which is realized in all studied cases over a large time interval.  Consider the quantity
\begin{equation}
\label{eq:combined}
\overline{C(t)} = \frac{1}{2}\left[ C_{p.b.c.} (t) + C_{a.b.c.} (t)\right]  
\end{equation}
built from meson correlators with periodic ($C_{p.b.c.}$) and antiperiodic ($C_{a.b.c.}$) temporal boundary conditions on a lattice of temporal extent $T$. It can be shown that the resulting combined correlator $\overline{C(t)}$ has its lattice temporal extent effectively doubled and it can also be written as the periodic correlator ${C_{p.b.c.}(t)}$ with doubled period $2T$ \cite{Sasaki:2005ug}.
In the case of the pseudoscalar meson, the staggered two-point function does not contain a staggered parity partner state.  Taking into consideration a possible constant oscillation term that might appear as a consequence of the wrapping of a quark line around the antiperiodic time boundary, we can then write
\begin{equation}
\overline{C_{PS}(t)} = A \left( e^{-m_\pi t} + e^{-m_\pi (2T - t)}\right) + B(-1)^t
\end{equation}
The constant oscillation term can be removed by using the combination
\begin{equation}
\tilde{C}_{PS}(2t) = \frac{\overline{C_{PS}(2t)}}{2} + \frac{\overline{C_{PS}(2t+1)}}{4} + \frac{\overline{C_{PS}(2t-1)}}{4}
\label{eqn:jtrick}
\end{equation}
so that the final correlator is 
\begin{equation}
\tilde{C}_{PS}(2t) = A \left( e^{-m_\pi 2t} + e^{-m_\pi (2T - 2t)}\right).
\label{pionCorrelator}
\end{equation}
The effective doubling of the temporal extent allows to better isolate the first term in eq.~(\ref{eq:intermediate})
 and enlarges the corresponding effective mass plateau.  

A similar combination of meson correlators with different boundary conditions in the time direction can be performed for the other mesons. We use the PV correlator to extract the masses of the would-be $\rho$ vector meson and the $a_1$ axial meson. The averaged PV correlator can be  written in this case as
\begin{equation}
\overline{C_{PV}(t)} = A_{\rho} \left( e^{-m_\rho t} + e^{-m_\rho (2T - t)}\right) + A_{a_1} (-1)^t \left( e^{-m_{a_1} t} + e^{-m_{a_1} (2T - t)}\right) + B(-1)^t\, ,
\label{eqn:formPVCorr}
\end{equation}
\begin{figure}[tbp]
\centering
\includegraphics[width=.45\textwidth]{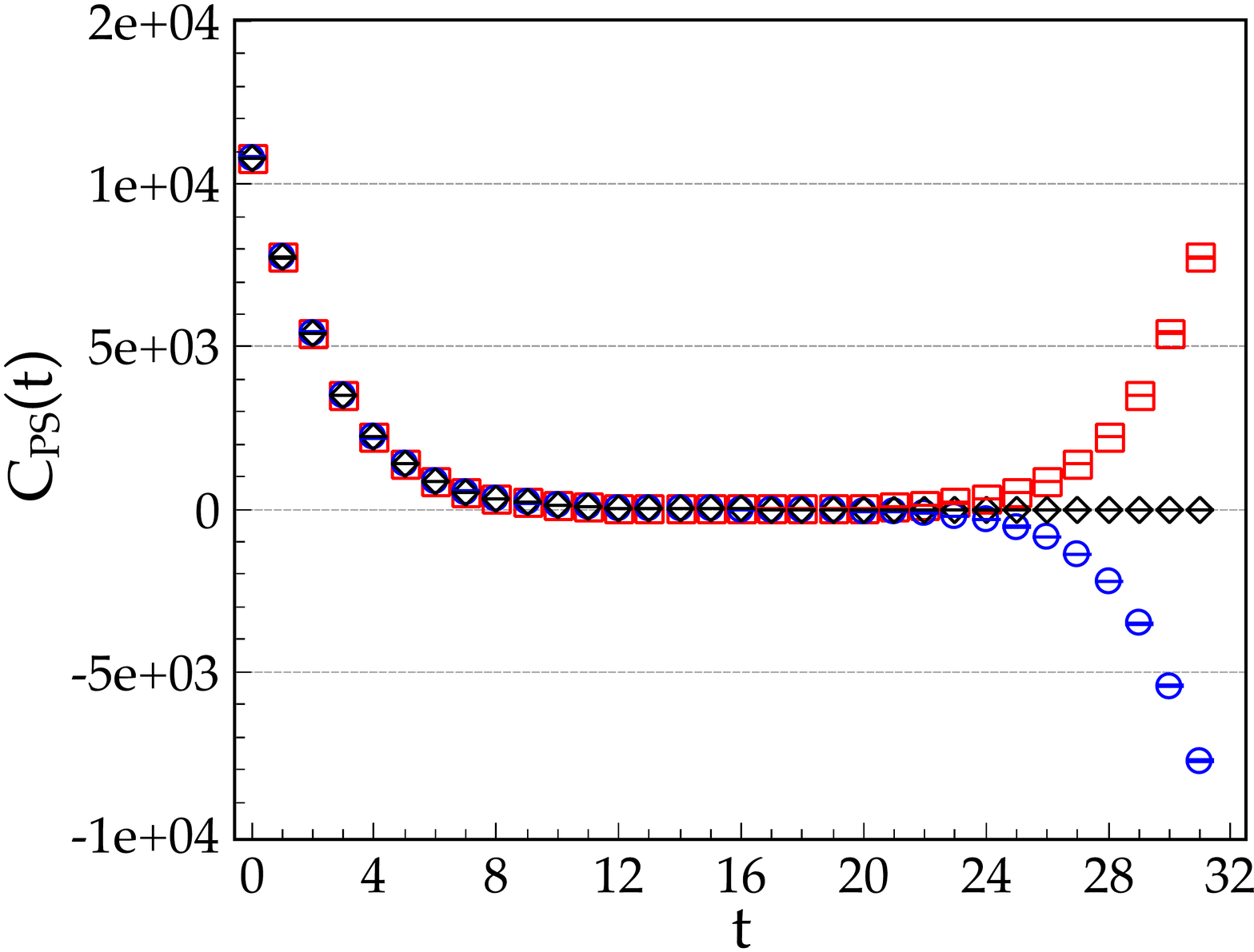}
\hfill
\includegraphics[width=.45\textwidth,origin=c]{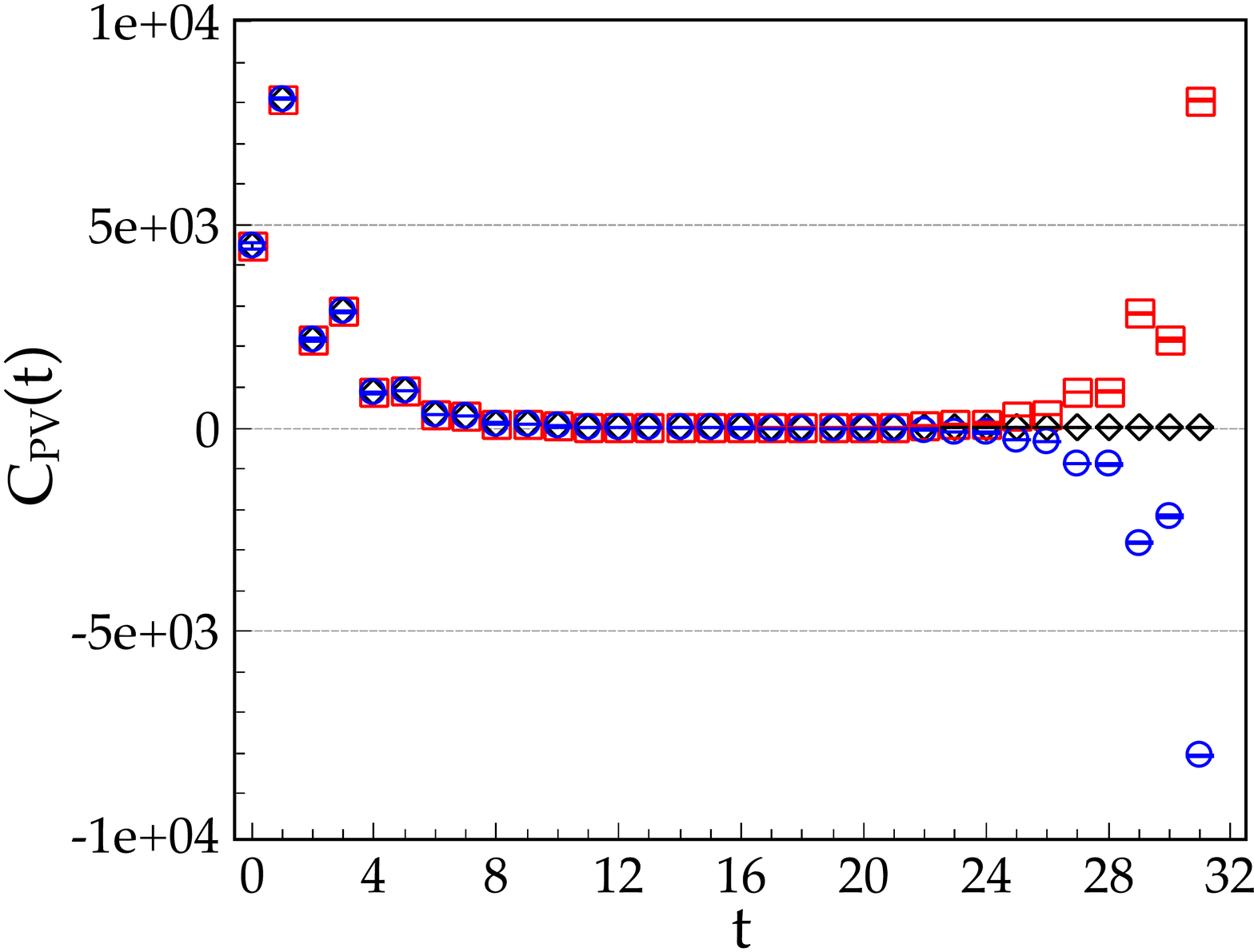}
\caption{ Examples of antiperiodic (blue circles), periodic (red squares) and combined (black diamonds) meson correlators from eq.~(\ref{eq:combined}), obtained from configurations generated with bare quark mass $am = 0.05$ at volume $V=32^4$. (Left) The pseudoscalar correlator, (right) the pseudovector correlator.}
\label{fig:AP-AA-comparisons}
\end{figure}
%
\begin{figure}
\centering
\includegraphics[width=.45\textwidth]{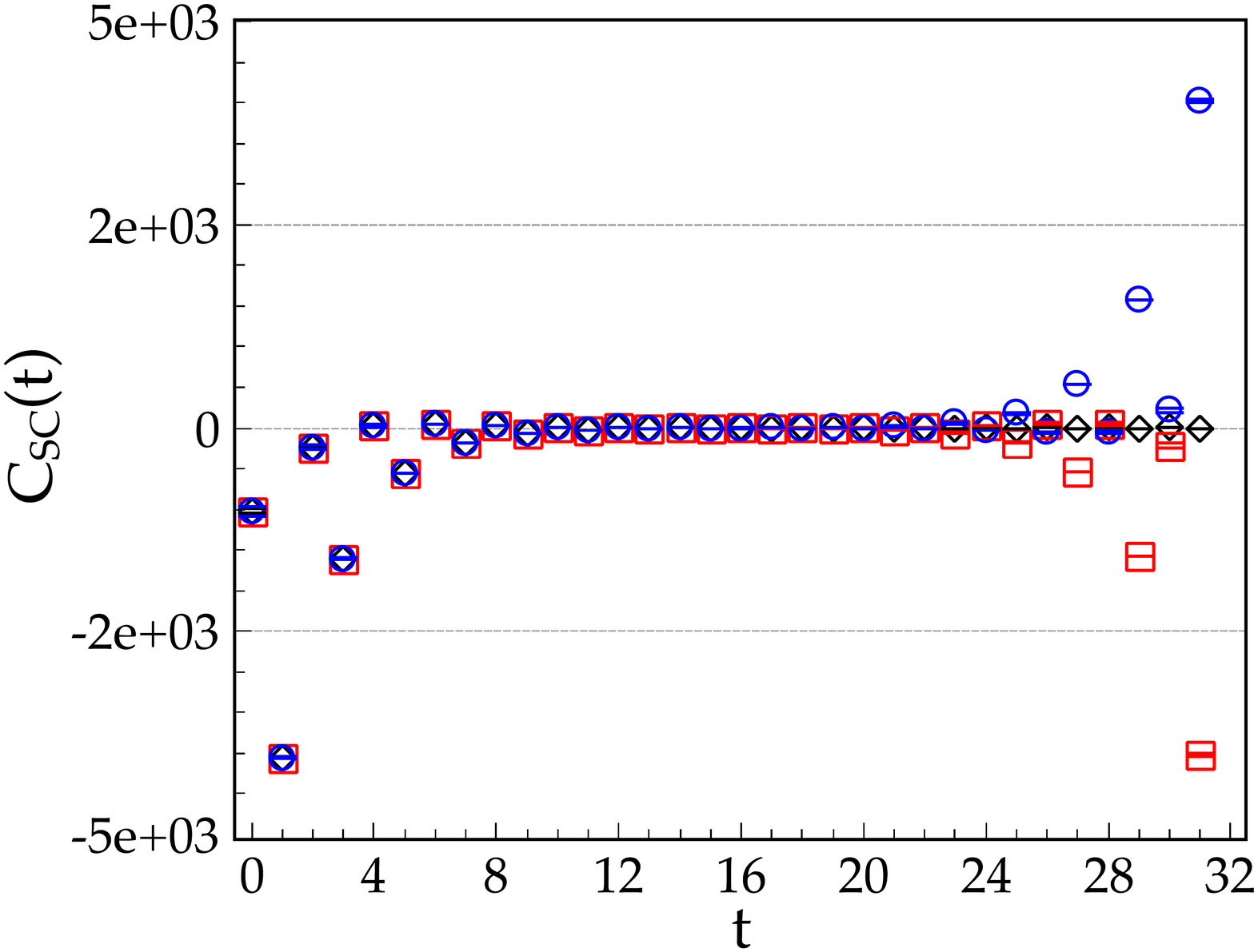}
\hfill
\includegraphics[width=.45\textwidth,origin=c]{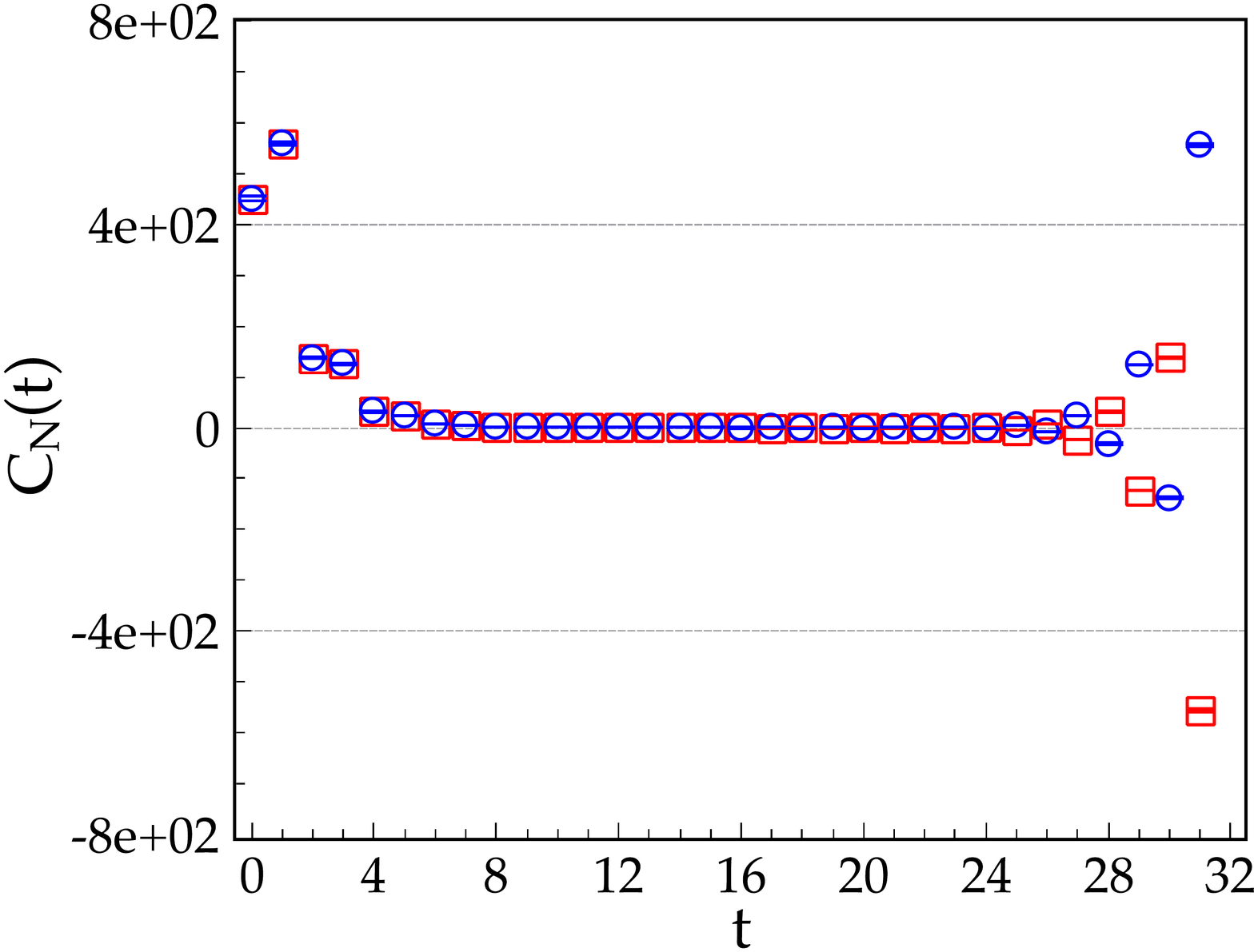}
\caption{ (Left) The periodic (red squares), antiperiodic (blue circles) and combined (black diamonds) scalar correlators, (right) the periodic (red squares) and antiperiodic (blue circles) nucleon correlator for a bare quark mass $am = 0.05$ and volume $V=32^4$.}
\label{fig:nuc-AP-AA-comparisons}
\end{figure}
It is possible to proceed with a similar combination to that of eq.~(\ref{eqn:jtrick}), so that
\begin{equation}
\tilde{C}_{PV}(2t) = \tilde{A}_{\rho1} \left( e^{-m_\rho 2t} + e^{-m_\rho (2T - 2t)}\right) + \tilde{A}_{\rho2} \left( e^{-m_{a1} 2t} + e^{-m_{a1} (2T - 2t)}\right)\, .
\label{vectorCorrelator}
\end{equation}
 In fact, such an approach has been followed in \cite{Aoki:2012eq}, where it was noted that, for the range of volumes and quark masses studied by the authors,  $\tilde{A}_{\rho1} \gg \tilde{A}_{\rho2}$,  and the resulting correlator is well approximated by a single exponential form with coefficient $\tilde{A}_{\rho1}$
\begin{equation}
\tilde{C}_{PV}(2t) \simeq \tilde{A}_{\rho1} \left( e^{-m_\rho 2t} + e^{-m_\rho (2T - 2t)}\right).
\label{vectorCorrelatorApprox}
\end{equation}
We have noticed that while the approximation (\ref{vectorCorrelatorApprox}) holds true for our heavier quark masses, it starts to break down for our lightest quark masses. In addition to that, we are also interested in studying the behaviour of the mass of the axial meson $m_{a_1}$. For these reasons, we fit the PV correlators obtained from our configurations to the complete functional form (\ref{eqn:formPVCorr}) in order to extract both $m_\rho$ and $m_{a1}$. Examples of the initial correlators and the quality of the final combined correlators are shown in figure~\ref{fig:AP-AA-comparisons}.

In similar fashion to eq.~(\ref{eqn:formPVCorr}), we extract the mass $m_\sigma$ of the scalar meson from the averaged $S$ correlator 
\begin{equation}
\overline{C_{S}(t)} = A_{\sigma} \left( e^{-m_\sigma t} + e^{-m_\sigma (2T - t)}\right) + B_{\sigma} (-1)^t \left( e^{-M t} + e^{-M (2T - t)}\right) + B(-1)^t\, .
\label{eqn:formSCCorr}
\end{equation}
Lastly, we built the nucleon correlators from quark and antiquark propagators with antiperiodic boundary conditions in the temporal direction.  These were then fitted with the usual expression for the lowest-lying states of a staggered baryon two-point function containing four parameters
\begin{equation}
C_N (t)= A_N \left( e^{-m_N t} - (-1)^t e^{-m_N(T-t)} \right) + B_N (-1)^t \left( e^{-Mt} - (-1)^t e^{-M(T-t)} \right)
\end{equation}
with parity partner of mass $M$. Examples of these correlation functions are shown in figure~\ref{fig:nuc-AP-AA-comparisons}.
\section{Results}
\label{sec:spectrum}

Table~\ref{tab:masses} collects all lattice measurements of this work.  
\begin{table}[htb]
\centering
\begin{tabular}{|c|c|c|c|c|c|c|c|}
\hline
$am$ & $Volume$  & $am_\pi$ & $am_\rho$ & $am_N$ & $am_\sigma$ & $am_{a_1}$\\
\hline 
$0.01$& $16^3\times 32$  & 0.4421(17)   & 0.516(10)  & 0.867(26)  & 0.4388(15) &  0.5086(36) \\
      & $24^3\times 32$   & 0.2701(44) 	 & 0.3002(61)     & 0.478(9)    & 0.2682(36) & 0.305(30)	   \\
			& $32^4$ & 0.1942(21) 	 & 0.2114(52)    & 0.326(10)    & 0.2057(22) & 0.2295(31)	   \\
\hline
$0.02$& $16^3\times 32$&   0.4480(16)  & 0.499(25)  &  0.846(27) & 0.446(8) & 0.509(14)  \\
      & $24^4$ &   0.3112(34) 	 & 0.3432(85)     & 0.528(14)    	& 0.3297(35) &  0.3903(91) \\
			& $32^4$ &  0.2624(13) 	 & 0.2857(11)    & 0.428(8)    & 0.3117(20) & 0.3464(34)	   \\
\hline
$0.025$& $16^3\times 32$&   0.4511(16)  & 0.5081(46)  & 0.892(31)   & 0.4598(48) &  0.5102(32) \\
	& $24^4$&   0.3447(12)  & 0.374(10) & 0.591(21)  & 0.3916(82) & 0.411(20)  \\
      		& $32^4$ & 0.3087(13) 	 & 0.3332(16)    & 0.530(13)    & 0.3786(49) &	0.4048(55)   \\
\hline
$0.04$& $24^3\times 32$&   0.4236(57)  & 0.4890(60) & 0.726(13)  & 0.543(24) & 0.617(49)  \\
      & $32^4$ & 0.4210(16) 	 & 0.4717(23)    & 0.709(2)  & 0.5359(51) & 0.5783(83) 	   \\
\hline
$0.05$& $24^4$&   0.5020(23)  & 0.5652(76) & 0.851(17)  & 0.6452(26) &  0.703(41) \\
      & $32^4$ & 0.5031(21) 	 & 0.5689(17)    & 0.850(5)    & 0.6463(32) &	0.7097(83)   \\
\hline
$0.06$& $16^3\times 24$&   0.5921(48)  & 0.678(12)  & 1.028(49)  & 0.747(13) &  0.823(40) \\
      & $24^4$ & 0.5881(20) 	 & 0.6700(18)    & 1.003(8)   & 0.746(19) & 0.831(10)	   \\
\hline
$0.07$& $16^3\times 24$&   0.6600(19)  & 0.7663(48)  & 1.116(47)  & 0.831(14) & 0.914(48)  \\
      & $24^4$ &  0.6596(27) 	 & 0.7597(24)    & 1.111(6)    	& 0.831(27)& 0.918(34)  \\
\hline
\end{tabular}
\caption{\label{tab:masses} Masses of the lowest-lying would-be hadrons, the pseudoscalar ($\pi$), the vector ($\rho)$, the scalar ($\sigma$)\,---\,obtained from the quark-line connected part of the isoscalar correlator\,---\,the axial ($a_1$), and the nucleon ($N$) for bare quark masses $am=0.01$ to $0.07$ and lattice coupling $\beta_L=3.9$. The volumes span from $16^3\times 24$ to $32^4$.  }
\end{table} 
Simulations have been done at inverse lattice coupling $\beta_L=3.9$, located in the QED-like region of figure~\ref{fig:gbar}. The same coupling was also used in our first study \cite{Deuzeman:2009mh}. The masses of all would-be hadrons have been measured for a range of bare fermion masses between $am=0.01$ and $am=0.07$, and volumes between $16^3\times 24$ and $32^4$. This section is organized as follows. After comparing the two-point functions with the ones of the free theory, and testing them against eqs.~(\ref{eq:univ}) and (\ref{eq:asymptotic}) in section~\ref{sec:twopoint_r}, we provide in section~\ref{sec:QED} a tool to establish in which phase, QED-like or asymptotically free, the lattice system is. Section~\ref{sec:results_scaling} is dedicated to the spectrum in a finite volume. It establishes the realization of universal scaling for the lattice results according to eq.~(\ref{eq:scalm_L}) and provides a unified description of all available lattice data for the $N_f=12$ system while identifying the pattern of scaling violations on both sides of the IRFP. Section~\ref{sec:FV} treats the extrapolation to infinite volume, needed for the lightest masses. Section~\ref{sec:delta} is dedicated to the spectrum at infinite volume, it established the realization of universal scaling for the lattice results according to eq.~(\ref{eq:scalm}) and identifies the pattern of scaling violations, in particular, for the spin-1 states of this work and those in \cite{Aoki:2012eq}. This analysis leads to the determination of $\gamma^*$ that consistently describes all available lattice data for the spectrum of the $N_f=12$ system at finite and infinite volume. Finally, section~\ref{sec:mass-deg} is a brief discussion of mass ratios and degeneracies of chiral partners, probe of restored chiral symmetry.  
\subsection{Two-point functions}
\label{sec:twopoint_r}

We have analyzed all two-point functions according to eq.~(\ref{eq:univ}) and its asymptotic forms in eq.~(\ref{eq:asymptotic}).
Our results  are easily summarized. For the entire range of masses explored, 
the best fits to $\overline{C(t)}$ (with period $2T$) over the late time range are obtained for the form $a\exp{(-mt)}$, symmetrized on $2T$, with $a$ constant and mass $m$. Time dependent corrections will increasingly be present at small times for decreasing masses, rendering more difficult the determination of the would-be hadron masses. The corrected form $a\exp{(-mt)} +b/t \exp{(-nt)}$, with $b>0$ and $n>m$, ameliorates the fits at smaller times, as expected\,---\,for our lightest mass $am=0.01$, such corrections start to become relevant when considering times $t<10$ and requires $n\gtrsim m$. 

\begin{figure}[tbp]
\centering
\includegraphics[width=.45\textwidth]{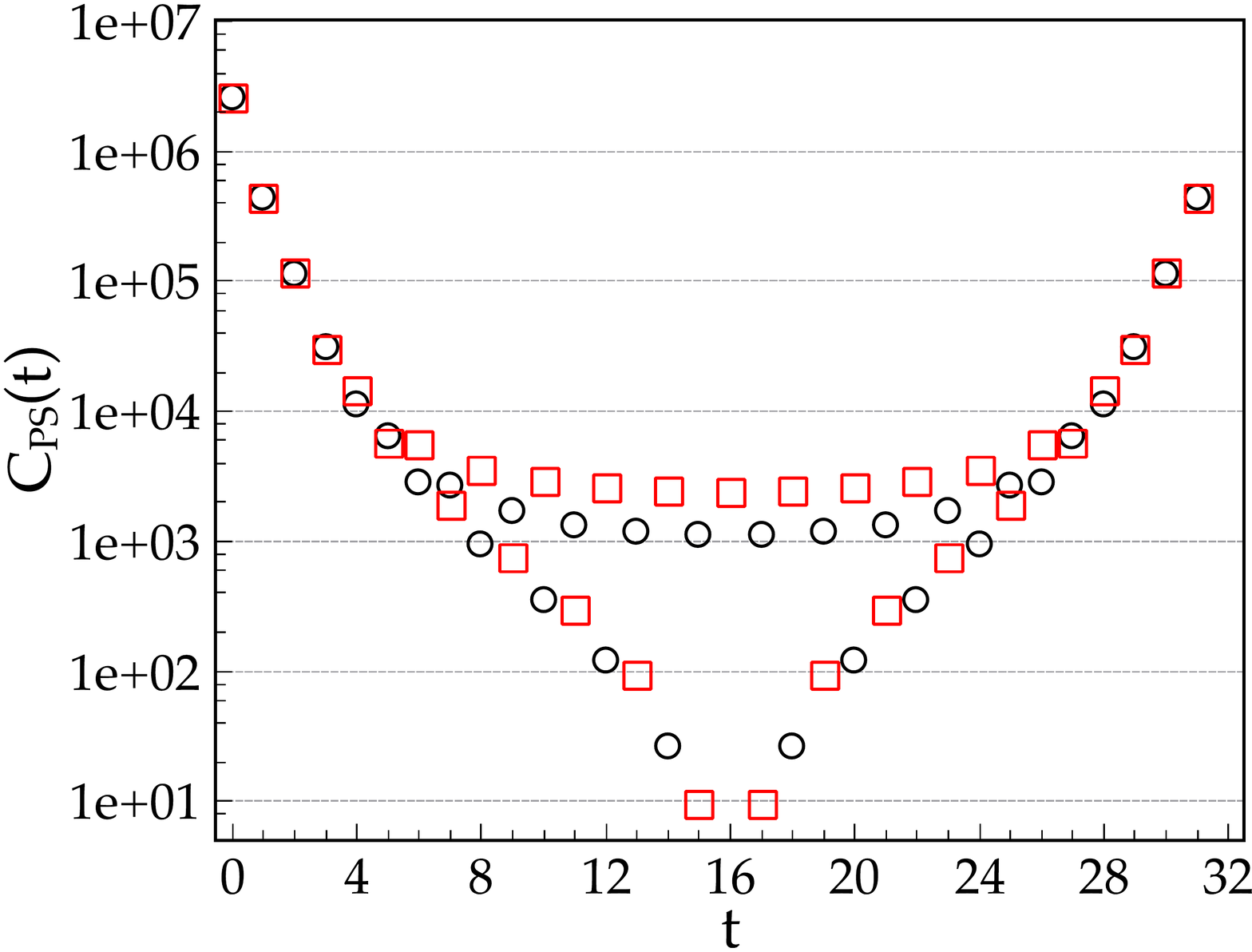}
\hfill
\includegraphics[width=.45\textwidth,origin=c]{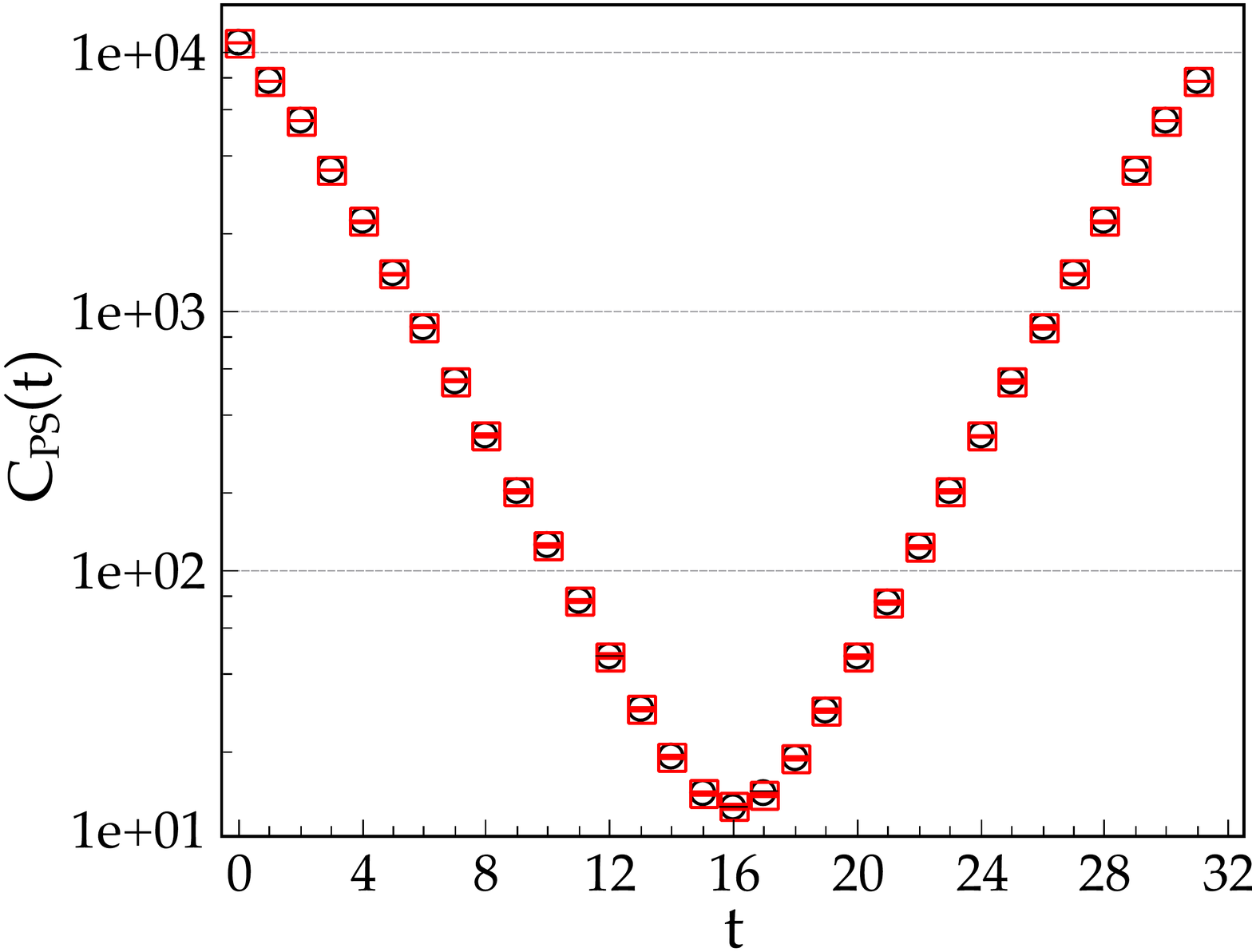}
\caption{ \label{fig:freePPversusAA} (Left) The pseudoscalar two-point function $C_{p.b.c.}$ (logarithmic scale) in the free case with $am=0.05$, made of two free quark propagators with antiperiodic boundary conditions (red squares) and periodic boundary conditions (black circles) in the temporal direction. (Right) The same two-point function at $\beta_L=3.9$.  Vertical axes are rescaled to match.}
\end{figure}

In figure~\ref{fig:freePPversusAA} we compare the two-point functions at $\beta_L=3.9$ with the corresponding ones obtained in the free case, with the same lattice staggered action and the same bare fermion masses; this comparison is useful to clarify how far the studied regime is from the free limit. In fact, if the theory is deconfined one may expect a faster approach to the free limit than it is realised in the confined theory. One useful ingredient in the comparison is that two-point functions built with increasingly free quarks  exhibit increasing sensitivity to the change of boundary conditions, both in the spatial and temporal direction. The zero-momentum free meson two-point functions  reproduce the known analytical form for staggered correlators on even and odd temporal sites \cite{Boyd:1994np}, and indeed figure~\ref{fig:freePPversusAA} (left) shows the significant difference between the standard periodic meson two-point function built with periodic (P) and the one built with antiperiodic (A) temporal boundary conditions on the single free quark propagators. This difference is absent in the two-point functions at $\beta_L=3.9$ in figure~\ref{fig:freePPversusAA} (right), for the same bare fermion mass. 
Note also that exact point-by-point degeneracy of the scalar and pseudoscalar free meson correlators in the chiral limit is only realized at zero lattice spacing, since they involve a sum over the momenta of the fermion and antifermion propagators. In accordance with their analytical form \cite{Boyd:1994np}, we find that pairs of chiral-partner free correlators are exactly degenerate at odd times and increasingly degenerate at even times towards the chiral limit. This comparison confirms that the two-point functions at $\beta_L=3.9$ are  significantly away from the free limit, and well described by an exponential with a constant coefficient $A_H$ over a large time interval.  
\subsection{Would-be hadrons in the QED-like region}
\label{sec:QED}

\begin{figure}[tbp]
\centering
\includegraphics[width=.60\textwidth]{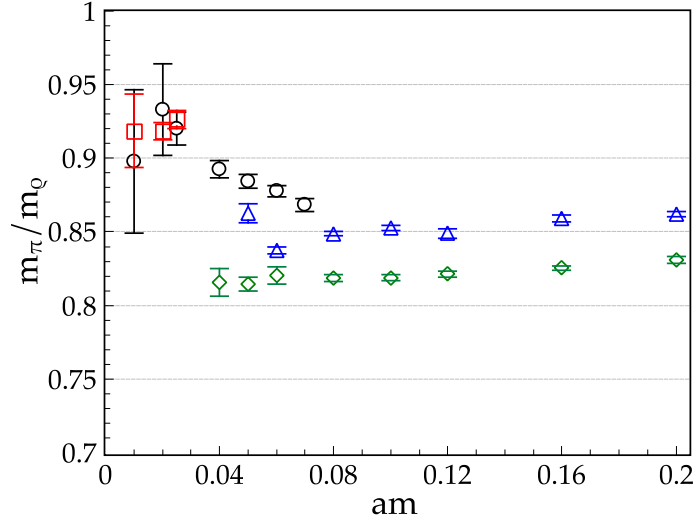}
\caption{ \label{fig:mpimrhoRatio}  The pseudoscalar ($\pi$) to vector ($\rho$) mass ratio as a function of the bare fermion mass.  Data at $\beta_L=3.9$ (left of figure) are at the largest volumes from table \ref{tab:masses} (black circles) and the infinite volume extrapolation from table~\ref{tab:FVextrap} is  shown for the three lightest points (red squares). Data at $\beta_L=4.0$ would draw a line to the right of $\beta_L=3.9$ \cite{Deuzeman:2009mh}.
Data from \cite{Aoki:2012eq} (right of figure) are obtained with a HISQ staggered action, at $\beta =6/g_L^2=3.7$ (green diamonds) and $\beta =4.0$ (blue triangles), $am=0.04$ to 0.2.   }
\end{figure}

 Figure~\ref{fig:mpimrhoRatio} provides a tool to understand which phase of the $N_f=12$ system we are looking at. It shows the mass ratio of the would-be pseudoscalar and vector mesons  as a function of the bare fermion mass; these are the
infinite volume lattice results in table~\ref{tab:masses} for $am >0.025$ and in table~\ref{tab:FVextrap} for $am\leq 0.025$.
  It also provides a direct comparison of our results with those of \cite{Aoki:2012eq}, the latter obtained with a HISQ staggered action at two lattice couplings.  In our initial study \cite{Deuzeman:2009mh}, where more than one lattice coupling\,---\,including $\beta_L=3.9$ and $4.0$\,---\,was considered, we could conclude that our results were located in the QED-like region of the theory, i.e. on the strong coupling and non asymptotically free side of the IRFP, with a positive $\beta$-function. The results in figure~\ref{fig:mpimrhoRatio} update that study at lattice coupling $\beta_L=3.9$; data for the ratio at $\beta_L=4.0$ would be located on a curve with similar slope, to the right of $\beta_L=3.9$. 
The analogous study in \cite{Aoki:2012eq} led the authors conclude that their results are instead located on the weak coupling and asymptotically free side of the IRFP. The same can be inferred from  figure~\ref{fig:mpimrhoRatio}, where the crucial ingredients are the  slopes and the ordering of curves. 
A line of constant physics would lead to a  constant ratio $m_\pi /m_\rho$; a realization of such a line occurs at the IRFP, where the $\beta$-function is zero and universal scaling holds with $m_{\pi\,,\rho} =c_{\pi\,,\rho} m^{\delta}$. Away from the fixed point, a family of curves at different lattice couplings as in figure~\ref{fig:mpimrhoRatio} carries information about the sign of the $\beta$-function. 
For our data, the crossing of a line of constant ratio with the curves at fixed lattice coupling\,---\,$\beta_L=3.9$ and an ideal line for $\beta_L=4.0$ at its right\,---\,implies a positive sign of the $\beta$-function, where to first approximation we assume a constant physical mass between the intersections. The data of \cite{Aoki:2012eq} have the opposite behaviour, and correspond instead to a negative sign of the $\beta$-function. At first sight, the reduced slope of the curves in the latter case would suggest that the data of \cite{Aoki:2012eq} are less affected by violations of scaling and plausibly closer to the IRFP. Another possibility, implied by the results in section~\ref{sec:results_scaling} and in line with \cite{Cheng:2013xha}, is that different mass regimes are covered by the two sets of lattice measurements, both affected in different ways and to different degree by violations of universal scaling. 
Summarizing, the combined set of data in figure~\ref{fig:mpimrhoRatio} nicely covers the region on both sides of the IRFP. This illustrates the fact that observables in this system are actually sensitive to the change of sign of the $\beta$-function and that some clever combination of these observables can be used to locate the IRFP; we are currently investigating a strategy along this line. 
\subsection{The spectrum in a box}
\label{sec:results_scaling}

\begin{figure}[tbp]
\centering
\includegraphics[width=.45\textwidth]{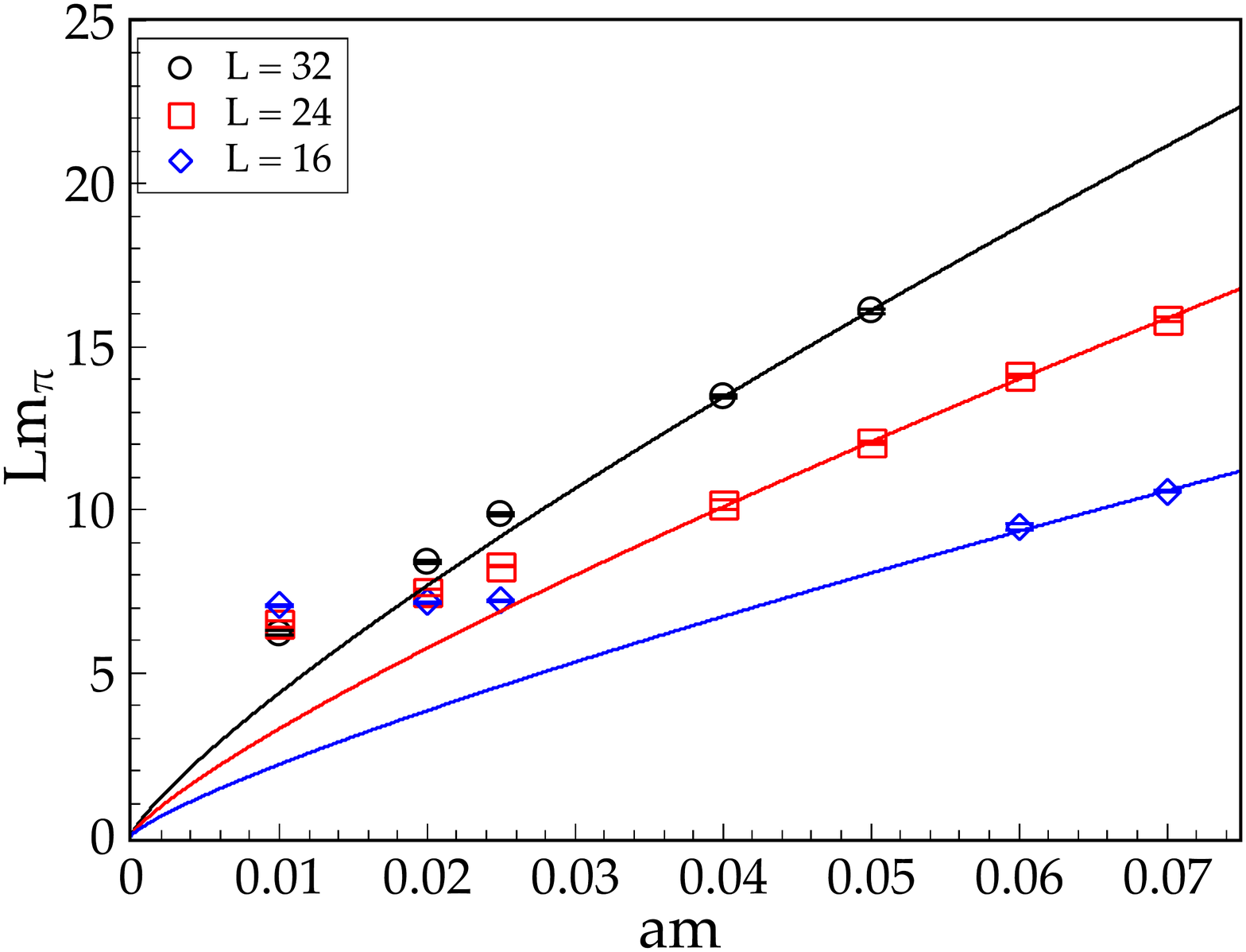}
\includegraphics[width=.45\textwidth,origin=c]{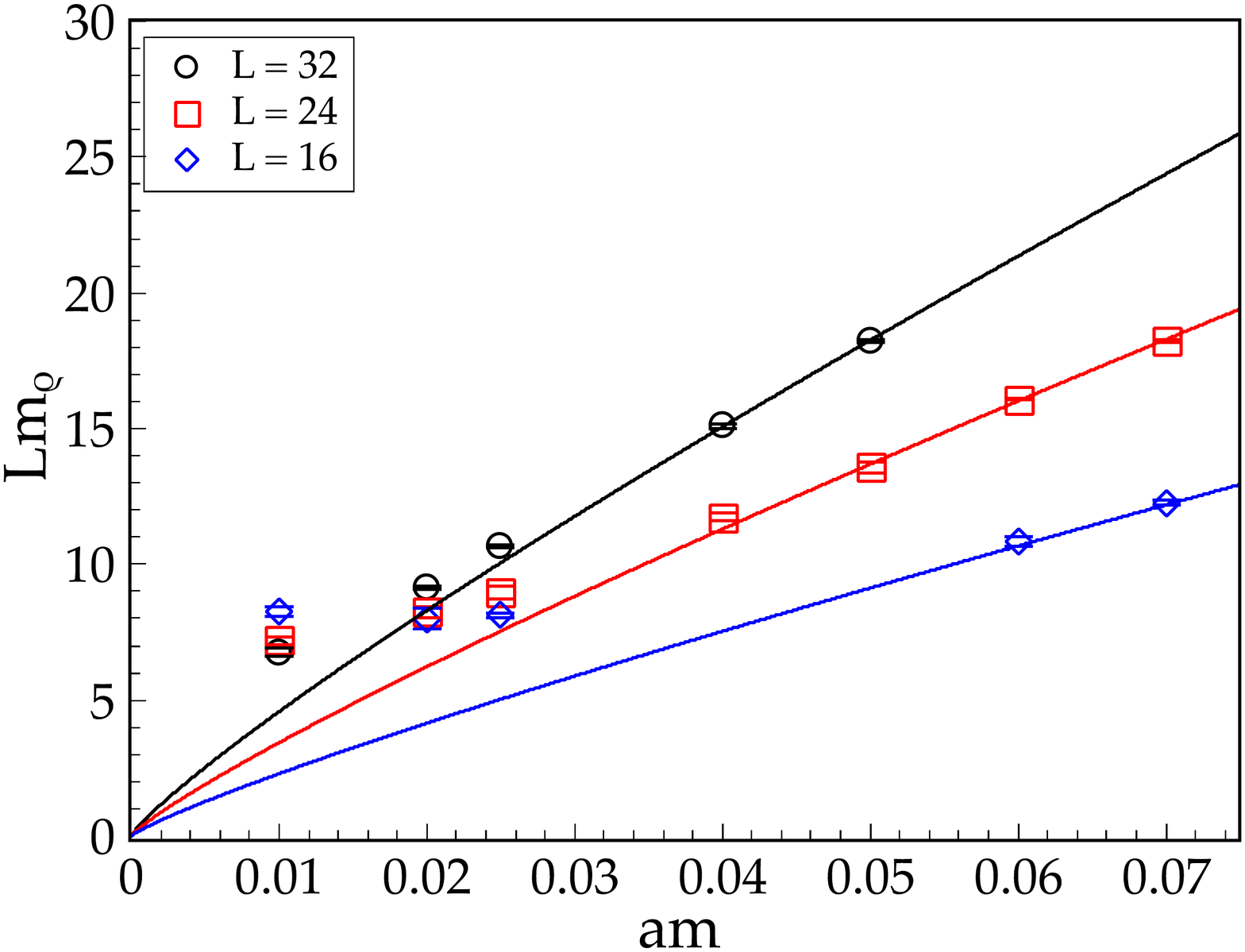}
\caption{ \label{fig:LmHversusam} 
$Lm_H$ for the pseudoscalar (left) and vector (right) would-be hadrons as a function of $am$ and for varying $L$. For sufficiently large volumes and masses the points fall onto a curve (superimposed), which is the power-law best-fit curve obtained at infinite volume (table~\ref{tab:delta}).  } 
\end{figure}

Figure~\ref{fig:LmHversusam} illustrates the pseudoscalar and vector products $Lm_H$ for each given $L$ as a function of the bare fermion mass.  It is clear that  finite volume effects are present at the largest spatial volume for the three lightest bare masses $am=0.01\,,0.02$ and $0.025$.
At the same time, these data offer the interesting option of a finite size scaling study with the aim of  identifying a universal scaling behaviour and define the functional form appropriate to extrapolate these data to infinite volume.
Barring the emergence of new operators, we proceed to identify universal and nonuniversal behaviours in the space of couplings $(g,m)$. 
 A comparison with the superimposed best-fit curves obtained at infinite volume in section~\ref{sec:delta} helps locating the threshold where substantial deviations from a genuine power-law appear in the scaling of $Lm_H$, at fixed $L$.   
These deviations can a priori contain nonasymptotic contributions to the scaling function $f(x)$ of eq.~(\ref{eq:scalm_L}), as well as genuine scaling violations not described by $f(x)$. 
The following analysis is devised to identify these contributions  at small and large $x$.

Anticipating the results of section~\ref{sec:delta}, we note that the infinite volume best fit  to eq.~(\ref{eq:scalm}) for the pseudoscalar,  scalar and nucleon masses gives a critical exponent  $\delta = 0.81$, while the vector and axial states favour a slightly larger exponent $\delta = 0.86$, see table~\ref{tab:delta}. While the difference in the values of $\delta$ for different $H$ channels is in itself an indication of scaling violations, we note that universality appears to be realized in all but the vector channels and $\delta =0.81$ gives a value of $\gamma^*=1/\delta -1$ in agreement, within uncertainties, with the best fit reported in \cite{Cheng:2013xha} and, noticeably, with the four-loop perturbative prediction \cite{vanRitbergen:1997va,Vermaseren:1997fq}; it is thus tempting to conclude that our data are in the universal scaling regime and provide a measure of the mass anomalous dimension at the IRFP. 
The study that follows supports this conclusion. 

\begin{figure}[tbp]
\centering
\includegraphics[width=.45\textwidth]{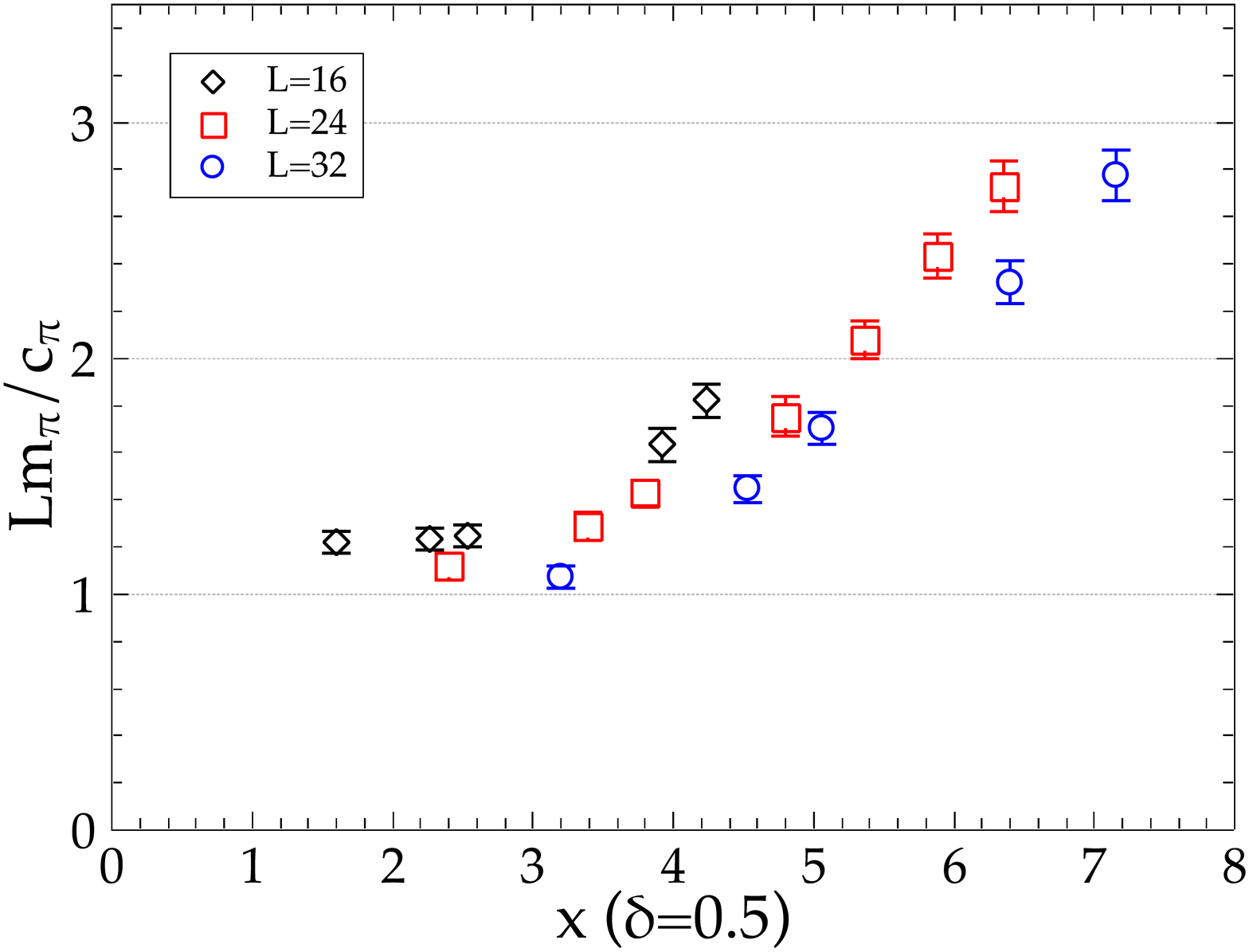}
\includegraphics[width=.45\textwidth,origin=c]{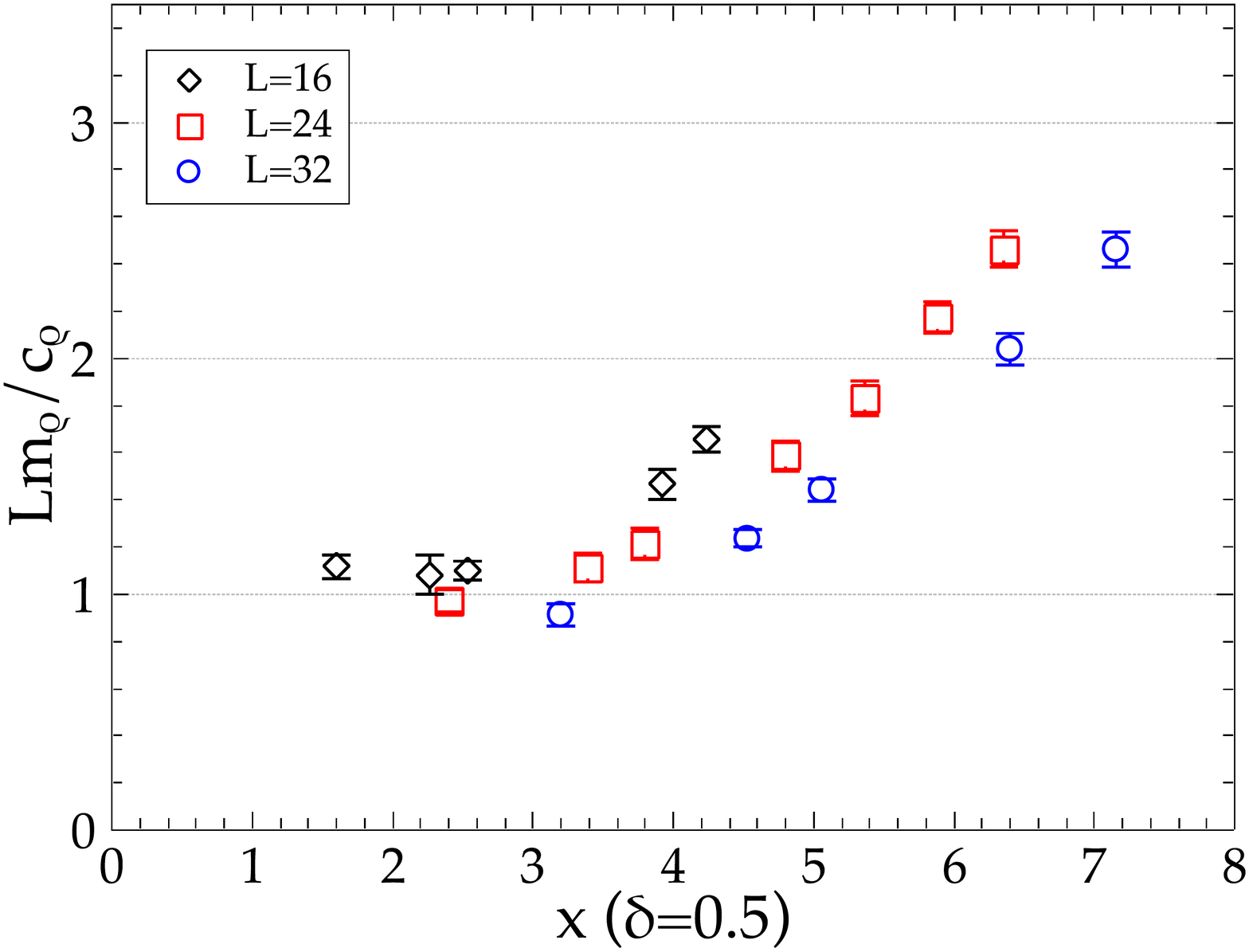}
\vspace{1.cm}
\includegraphics[width=.45\textwidth]{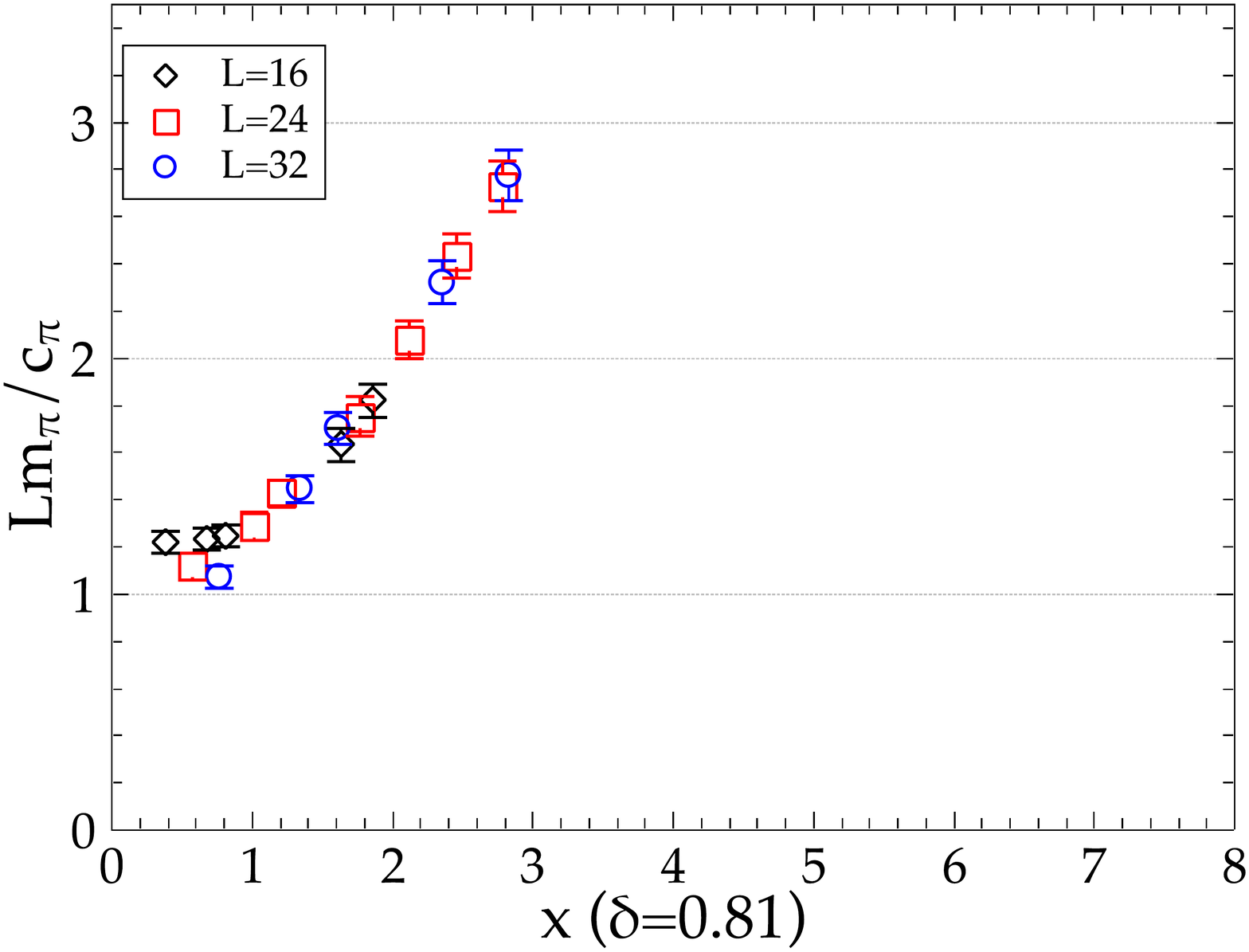}
\includegraphics[width=.45\textwidth,origin=c]{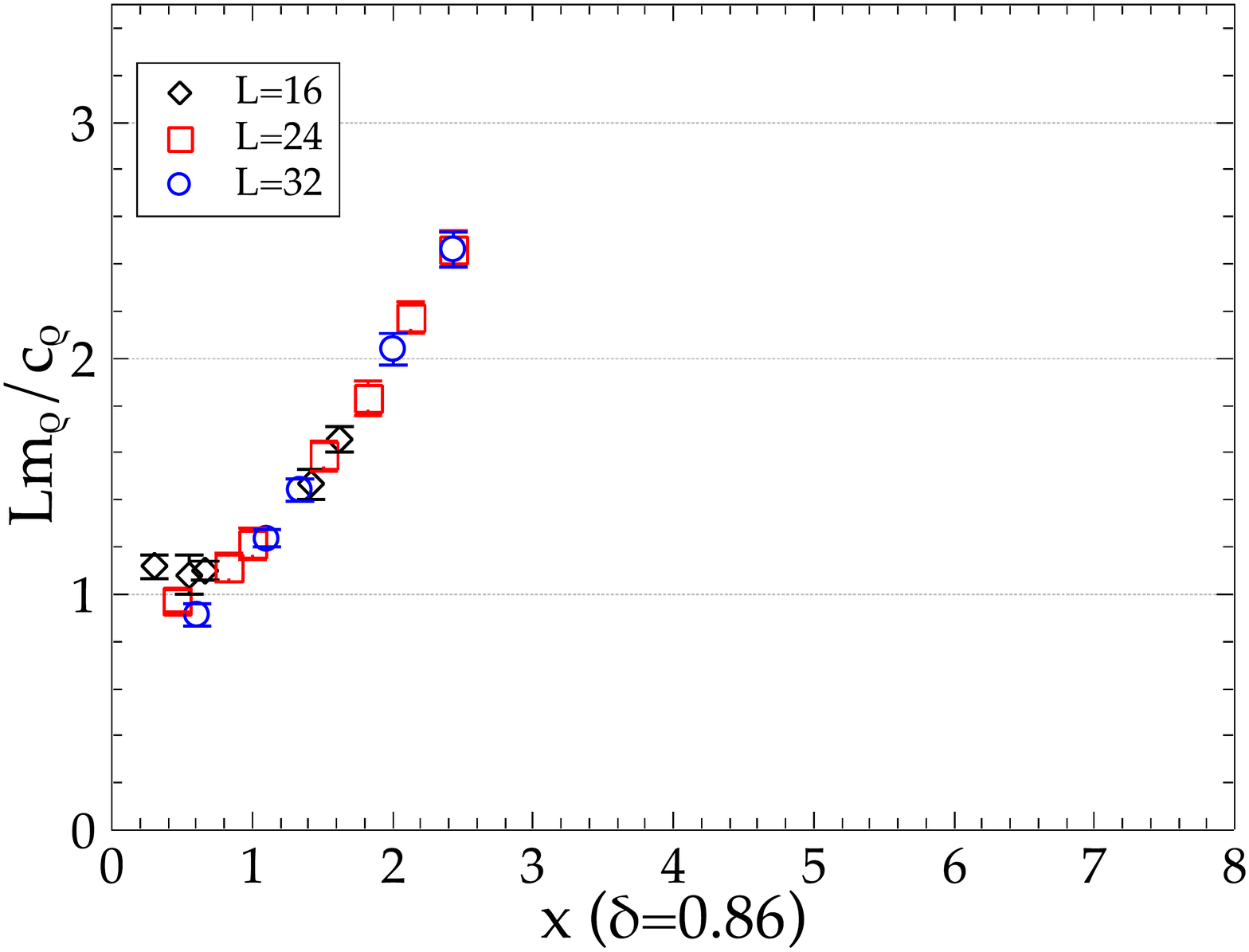}
\vspace{1.cm}
\includegraphics[width=.45\textwidth]{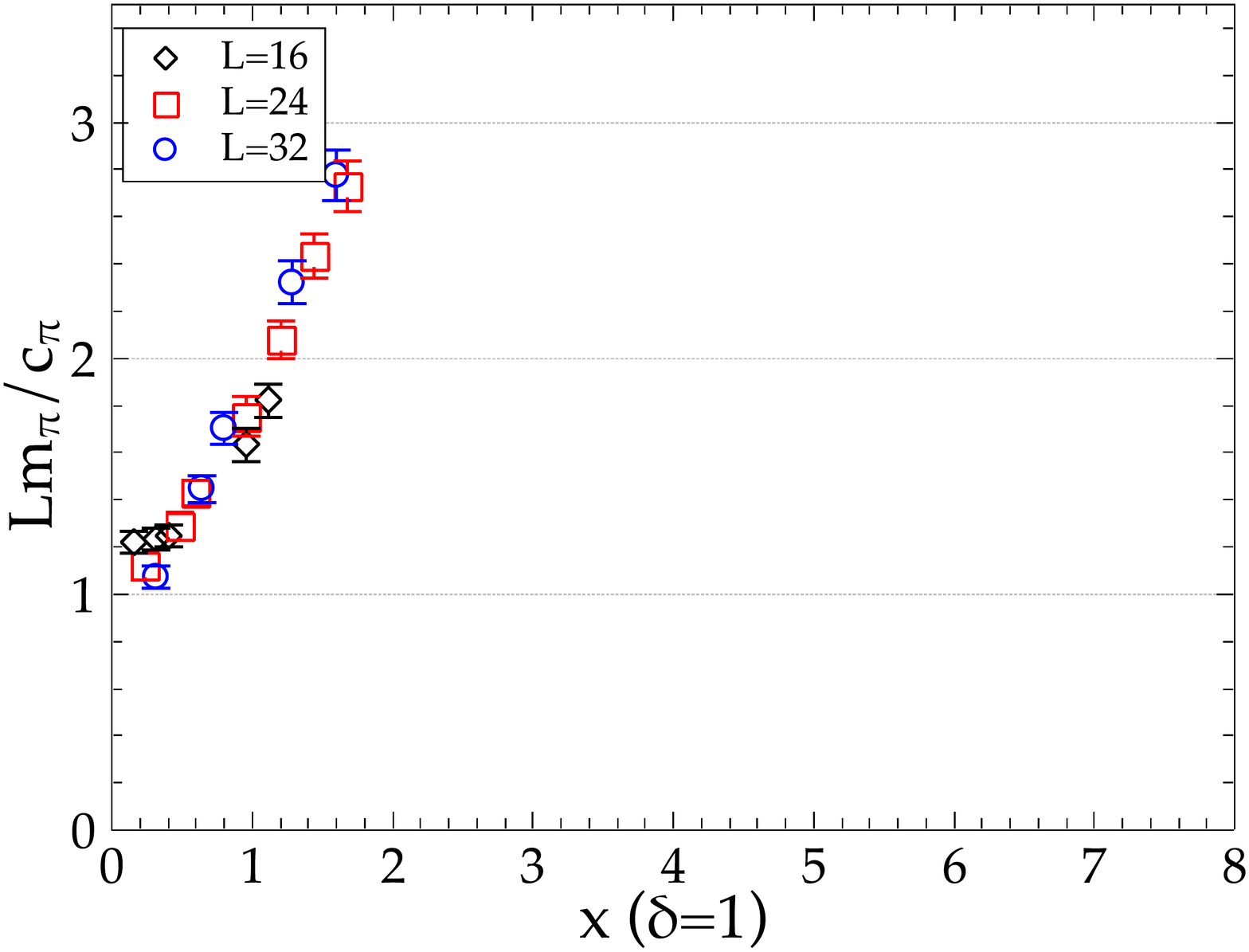}
\includegraphics[width=.45\textwidth,origin=c]{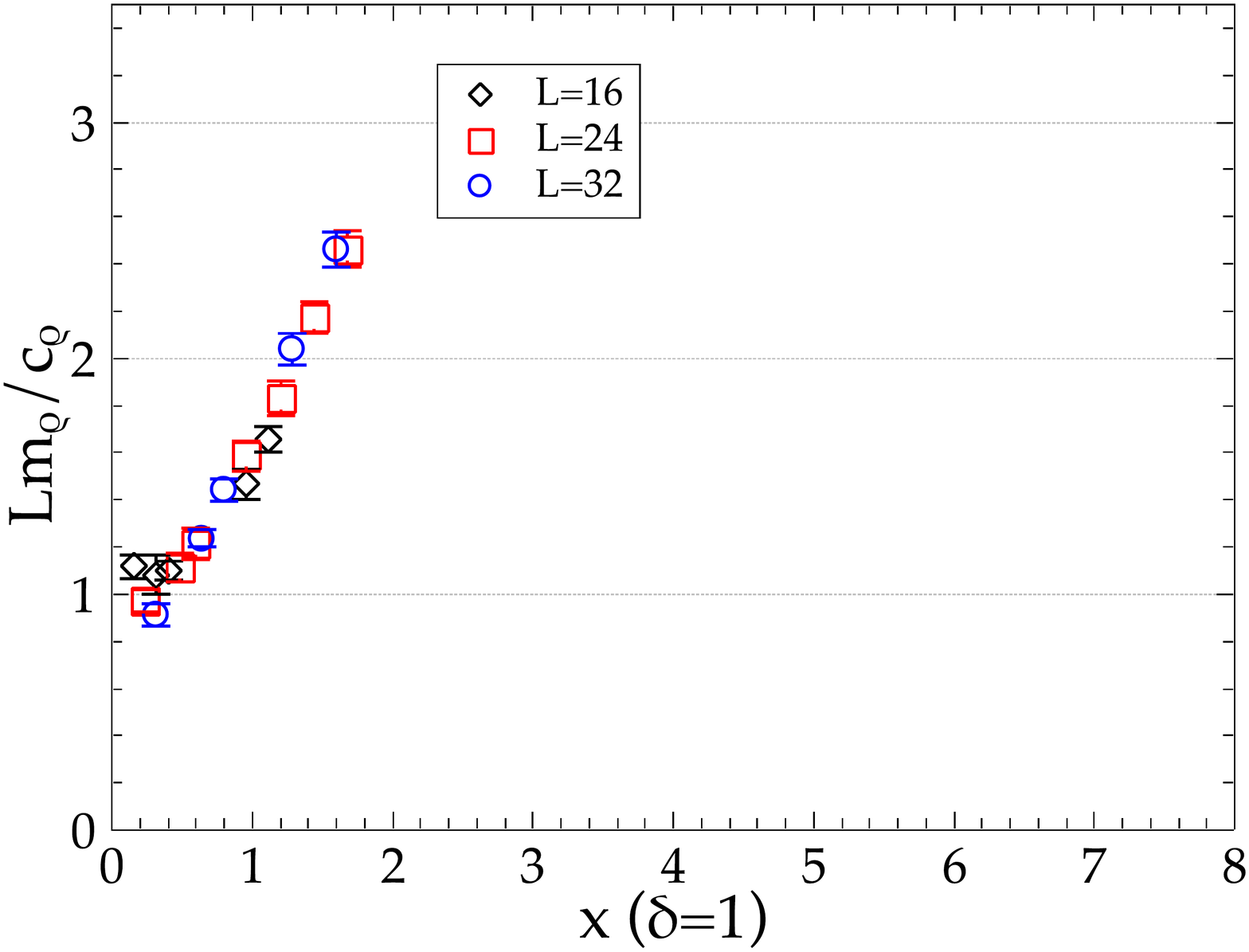}
\caption{ \label{fig:VPSscaling} Ratios $Lm_H/c_H$, $c_H$ from table~\ref{tab:delta}, for the pseudoscalar state (left) and the vector state (right) as a function of the scaling variable $x=Lm^\delta$ for varying $\delta$ on $[0.5,1]$ and $L=16,24,32$. The central figures display the data alignment for the best-fit values of $\delta$ in table~\ref{tab:delta}.}  
\end{figure}

\begin{figure}[tbp]
\centering
\includegraphics[width=.60\textwidth]{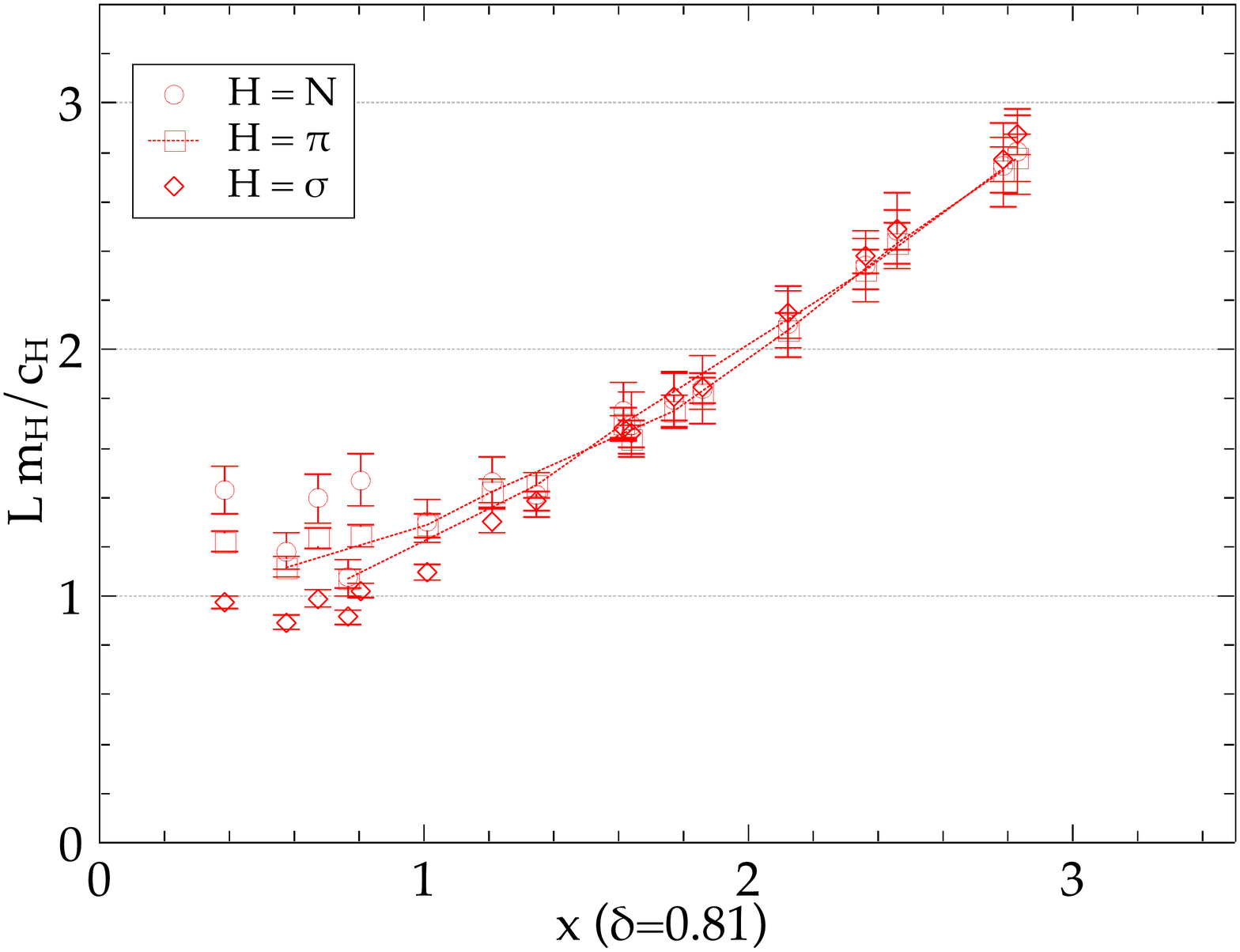}
\vspace{1.cm}
\includegraphics[width=.60\textwidth]{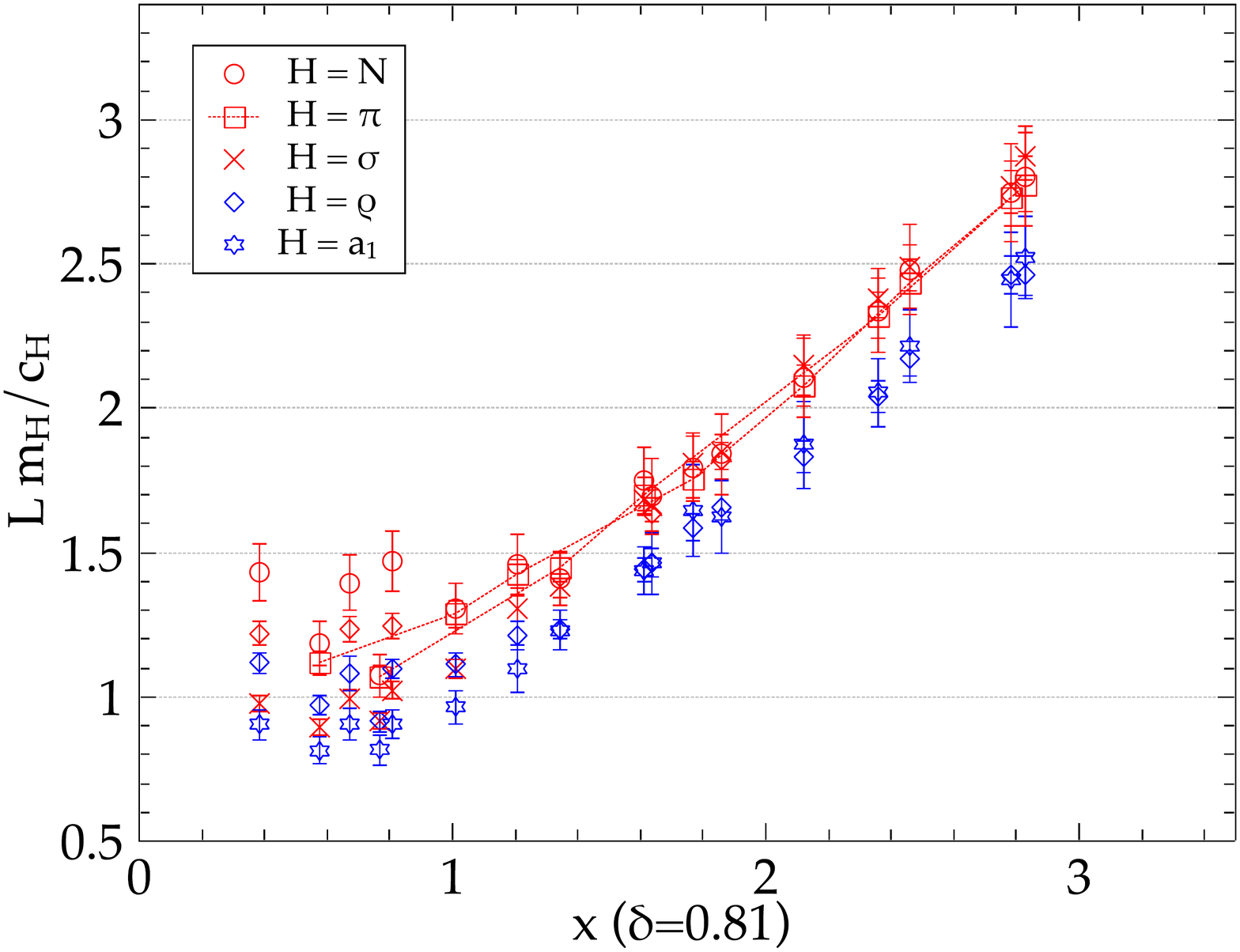}
\caption{ \label{fig:Scaling_All}  $Lm_H/c_H$, $c_H$ from table~\ref{tab:delta}, as a function of $x = Lm^\delta$ with $\delta =0.81$, excluding the vector and axial states (top) and including them (bottom).  } 
\end{figure}

In figure~\ref{fig:VPSscaling} we vary the scaling variable $x$ about the best-fit value for $\delta$ (central figures) and on the range $\delta =[0.5,1]$ to study the $x$ dependence of the ratio $Lm_\pi/c_\pi$ (left) and $Lm_\rho/c_\rho$ (right). 
For $x\gtrsim 1$, the data in the central figures align on a common curve. They increasingly scatter and deviate from it when $\delta$ is moved away from its best-fit value, over the range $0.5$ to 1.  
Figure~\ref{fig:Scaling_All} reports all states for the reference value $\delta=0.81$; 
we observe the universal behaviour of the pseudoscalar, scalar and nucleon states at $x\gtrsim 1$ and the displacement and  slight change  of slope of the vector and axial states. 
For $x\lesssim 1$, the asymptotic behaviour of the universal scaling function $f(x)\to const$ as $x\to 0$,  is corrected by nonperturbative $L$-dependent scaling violations. These are discussed in the next section. 

\begin{figure}[tbp]
\centering
\includegraphics[width=.60\textwidth]{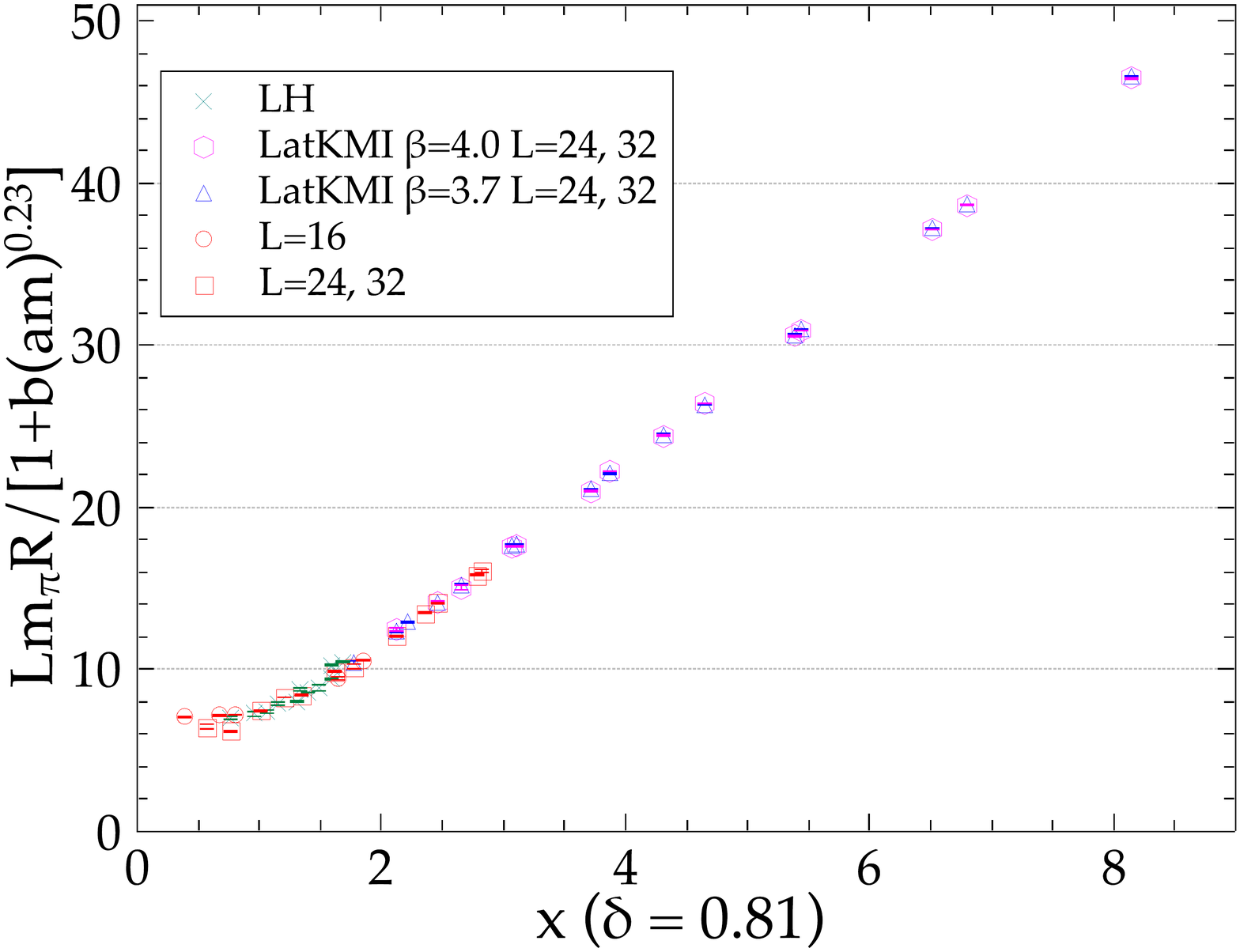}
\caption{ \label{fig:c_noietal} 
Collapse of curves for the rescaled pseudoscalar product $Lm_\pi R/(1+bm^\omega )$, with $\omega = 0.23$ from 4-loop perturbation theory and $R, b$ in table~\ref{tab:noietal}, as a function of the universal scaling variable $x=Lm^\delta$ with $\delta=0.81$ from this work. Data are from \cite{Fodor:2011tu} at $\beta =2.2$ (LH, green crosses), \cite{Aoki:2012eq} at $\beta =4.0$ (LatKMI, magenta hexagons) and $\beta =3.7$ (LatKMI, blue triangles), and from this work for $L=24,32$ (red squares) and $L=16$ (red circles). } 
\end{figure}

Inspired by recent work \cite{Cheng:2013xha}, we now attempt a unified description of the finite volume results of this work and the results obtained for the same system with other lattice actions, at a priori different bare lattice couplings and fermion masses.
In particular, we consider the results  at $\beta =2.2$ in \cite{Fodor:2011tu} and the results  at $\beta =3.7$ and $\beta =4.0$ in \cite{Aoki:2012eq}. We limit this analysis to the pseudoscalar channel, studied in all works, while later in section~\ref{sec:delta} we compare our results and those in \cite{Aoki:2012eq} for the vector state. Figure~\ref{fig:c_noietal} summarises this study, where we show the collapse on a common universal curve of data obtained with different lattice actions and lattice couplings, once perturbative corrections to the universal scaling are divided out.   
\begin{table}[tbp]
\centering
\begin{tabular}{|c|c|c|}
\hline
 &  $R$  & $b$ \\
\hline 
This work &   1 & 0      \\
\hline
LH  &  1 &  0  \\
\hline
LatKMI 3.7  &  $1.054$   & $-0.5435$    \\
\hline
LatKMI 4.0 & $1.193$    & $-0.4926$  \\
\hline
\end{tabular}
\caption{\label{tab:noietal} Values of $R$ and $b$ used in figure~\ref{fig:c_noietal}. }
\end{table} 
The general conclusions of this analysis are in good agreement with the study in \cite{Cheng:2013xha}, while our analysis differs from \cite{Cheng:2013xha} in some details and interpretation of the parameters. We briefly highlight the relevant ingredients and results. 
\begin{itemize}
\item Perturbative corrections to universal scaling are present whenever the system is close to, but not at the fixed point. As pointed out in \cite{Cheng:2013xha}, perturbative corrections due to  $g\neq g^*$ and a finite fermion mass can explain the deviations from universal scaling of many results for the $N_f=12$ system. As explained in section~\ref{sec:scaling2}, these contributions can be parameterized to leading order as multiplicative corrections $1+b\,m^\omega$, with universal exponent $\omega = {\delta\gamma_g^*}$. 
\item   
We find that our best-fits  favour values of $\omega$ lower than $\omega =0.41$ of  \cite{Cheng:2013xha}, though we cannot perform a fully unconstrained fit. It is therefore appealing to consider the value $\omega =0.23$ given by $\delta=0.81$\,---\,our central value in good agreement with the 4-loop prediction within uncertainties\,---\,and the 4-loop prediction $\gamma_g^* \simeq 0.283$. 
\item  The values of $b$ in table~\ref{tab:noietal} and the analysis of the vector state in table~\ref{tab:usKMI_PSV}  
 corroborate the interpretation of $\Delta g$ in section~\ref{sec:theo}. We observe that all data are sufficiently close to universal scaling. Our data are located in the QED-like phase of the system and do not show corrections to scaling at these light masses, thus $b=0$. The data from \cite{Fodor:2011tu} are close to our data, and again they seem not to be sensitive to corrections to scaling at the light masses they consider; the only difference is that their data are located away from the asymptotic linear form of $f(x)$ and we cannot clearly discriminate to which phase they pertain. 
The results from \cite{Aoki:2012eq} are located on the other\,---\,i.e. asymptotically free\,---\,side of the fixed point, thus showing $b<0$. We expect the data of \cite{Cheng:2013xha} to be located between ours  and the data from \cite{Aoki:2012eq} on the asymptotically free side, with $b<0$, in agreement with their analysis. In section~\ref{sec:delta} we provide an example of a positive $b$ for our system in the QED-like phase, needed to successfully describe perturbative corrections to scaling for the vector and axial states. Thus, the parameter $b$ changes sign at the boundary between the two phases of the lattice system and we expect it to flow to zero in the continuum, i.e. $\beta_L\to\infty$, where the lattice system reaches the IRFP.  
\item An overall rescaling is the usual procedure to bring together sets of data that follow universal scaling. We perform a rescaling of  $Lm_\pi$ by the factor $R$ in table~\ref{tab:noietal}. $R$ is given by the ratio of the  coefficients $c_H$ entering the infinite volume functional form  $m_H = c_H m^\delta (1+b_Hm^\omega )$ for each given data set and channel $H$. 
Such a rescaling is asymptotically equivalent to a rescaling of the variable $x$, as used in \cite{Cheng:2013xha}, provided it does not  enter the corrections to the universal scaling function. The factor $R$ for LatKMI is derived in section~\ref{sec:delta}. 
$R$ shows a monotonic dependence on the lattice bare coupling $\beta_L$, converging to its universal value as $\beta_L\to\infty$. This statement, as the previous one for $b$, assumes that the lattice system is in the basin of attraction of the IRFP. 
\item Once rescaled by $R$ and once the perturbative mass corrections are divided out, the product $Lm_\pi$ for all the  lattice data in figure~\ref{fig:c_noietal} is described by a universal curve $f(x)$ for all $x$ values, except for the presence of nonperturbative violations of scaling for $L\lesssim \xi$ for some data of this work. 
\end{itemize}

\subsection{Extrapolation to infinite volume}
\label{sec:FV}

For $am=0.04$ to $0.07$, no residual finite volume dependence is left within the estimated uncertainties; we thus take the result at the largest available volume as the infinite volume value for $am=0.04$ to $0.07$.  For the three lightest bare fermion masses, $am=0.01,\,0.02,\, 0.025$ we have instead performed an extrapolation to infinite volume. 
L\"uscher's formula \cite{Luscher:1986pf,Luscher:1985dn} for particles in a box does in principle apply  to any interacting quantum field theory, provided the scattering amplitude of the particles involved is known.  The latter problem is perturbatively solved by chiral perturbation theory ($\chi$PT) for QCD-like theories in the chirally broken phase, predicting the leading order behaviour of the Goldstone boson mass to be \cite{Gasser:1987zq, Colangelo:2002hy} $m_\pi (L) = m_\pi + c\,{\exp{\left (-m_\pi L\right )}}/{(m_\pi L)^{3/2}}$ for $m_\pi L\gg 1$. 
The functional form to be used in our case is what describes the  scaling violations at small $x$ in figures~\ref{fig:VPSscaling} and \ref{fig:Scaling_All}.

\begin{figure}[tbp]
\centering
\includegraphics[width=.60\textwidth]{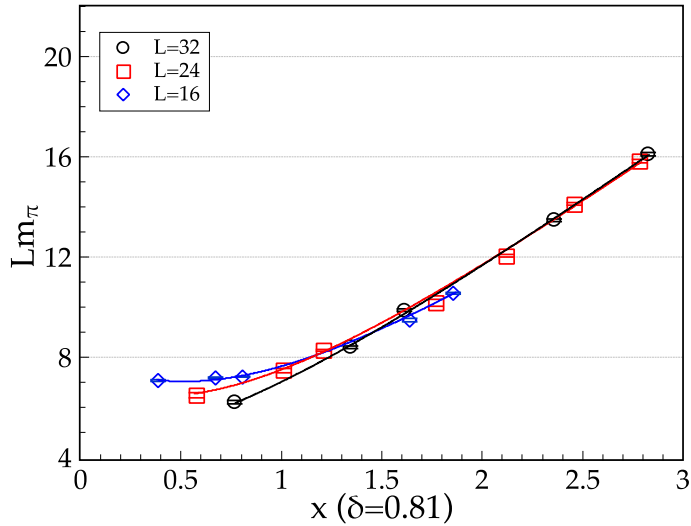}
\caption{ \label{fig:smallx_fit} 
$Lm_\pi$ as a function of the scaling variable $x=Lm^\delta$ with $\delta =0.81$, for $L=16,\,24,\,32$. 
The curves are the best fits to the functional form $F(x,L)=ax+c\exp{(-kx)}$, with $a,c,k$ in table~\ref{tab:smallx_fit}. The analogous figures for $H=S,V,PV,N$ are in figure~\ref{fig:smallx_fit_app} in the appendix.} 
\end{figure}

The small $x$ behaviour of the pseudoscalar would-be hadron is analyzed in figure~\ref{fig:smallx_fit}, for $\delta=0.81$. 
At the smallest values of $x$, $Lm_\pi$  shows an $L$-dependent deviation from a common curve.  $L$ dependent deviations from scaling can be expected whenever the box size $L$ becomes comparable to or smaller than the would-be hadron Compton wavelength $\sim \xi$.
The entire range of $x$ in figure~\ref{fig:smallx_fit} can be described in terms of the universal scaling function $f(x)$, with asymptotics $f(x)\sim x$ as $x\to\infty$ and $f(x)\to const$ as $x\to 0$, and a nonperturbative $L$-dependent violation of scaling at small $x$\,---\,note that perturbative corrections in $L$ of the type $1+\Delta g L^{-\gamma_g}$ would instead multiply the entire scaling function $f(x)$ and modify its behaviour at all $x$. Hence, 
$F(x,L) = ax + g(L)\tilde{f}(x)$ should describe figure~\ref{fig:smallx_fit}, except for the presence of nonlinear universal contributions to $f(x)$ at intermediate $x$. The coefficient $a$ is nothing but $c_H$, $H=\pi$ of table~\ref{tab:delta}, the function $g(L)$ increases for decreasing $L$ according to figure~\ref{fig:smallx_fit}, and $\tilde{f}(x)\to const$ as $x\to 0$. 
Figure ~\ref{fig:smallx_fit} also displays the best-fit curves for the simplified ansatz $F(x,L) = ax + c \exp{(-kx)}$, with best-fit values of  $a,c,k$ in table~\ref{tab:smallx_fit}. 
\begin{table}[tbp]
\centering
\begin{tabular}{|c|c|c|c|}
\hline
 &  $L=16$  & $L=24$ & $L=32$ \\
\hline 
$a$ & $5.21(20) $  & $5.59(10) $ & $5.63(20)$     \\
\hline
$c$ & $8.07(70) $  & $7.1(2.0) $  & $4.7(3.2) $      \\
\hline
$k$ & $1.20(20) $  & $1.31(30)$  &  $1.22(70)$     \\
\hline
$\chi^2/dof$  & $8.5$  & $10$ & $9$ \\
\hline
\end{tabular}
\caption{\label{tab:smallx_fit} Best-fit values of the parameters $a,c,k$ and $\chi^2/dof$ for the fits of $Lm_\pi$ to the functional form $F(x,L)=ax+c\exp{(-kx)}$, with $x=Lm^\delta$ and $\delta =0.81$. }
\end{table} 
Rather than aiming at the optimal $\chi^2/d.o.f$,  the purpose of this example is to illustrate the trend of small volume corrections through effective parameters $c$ and $k$. The latter is quite stable for varying $L$, while, as expected, the parameter $c$ increases with decreasing $L$; a polynomial $g(L) \sim (1-bL)$ perfectly describes the data and provides a volume dependence milder than the universal scaling form $1/L\exp{(-L/\xi )}$ for $m_\pi (L)$ at small $x$. 
At the same time, the shift in the paramater $a$ at $L=16$, as compared to the larger volumes in table~\ref{tab:smallx_fit}, should be attributed to the intermediate $x$ contributions to $f(x)$ that are not captured by the simple ansatz for $F(x,L)$. 

\begin{figure}[tbp]
\centering
\includegraphics[width=.45\textwidth]{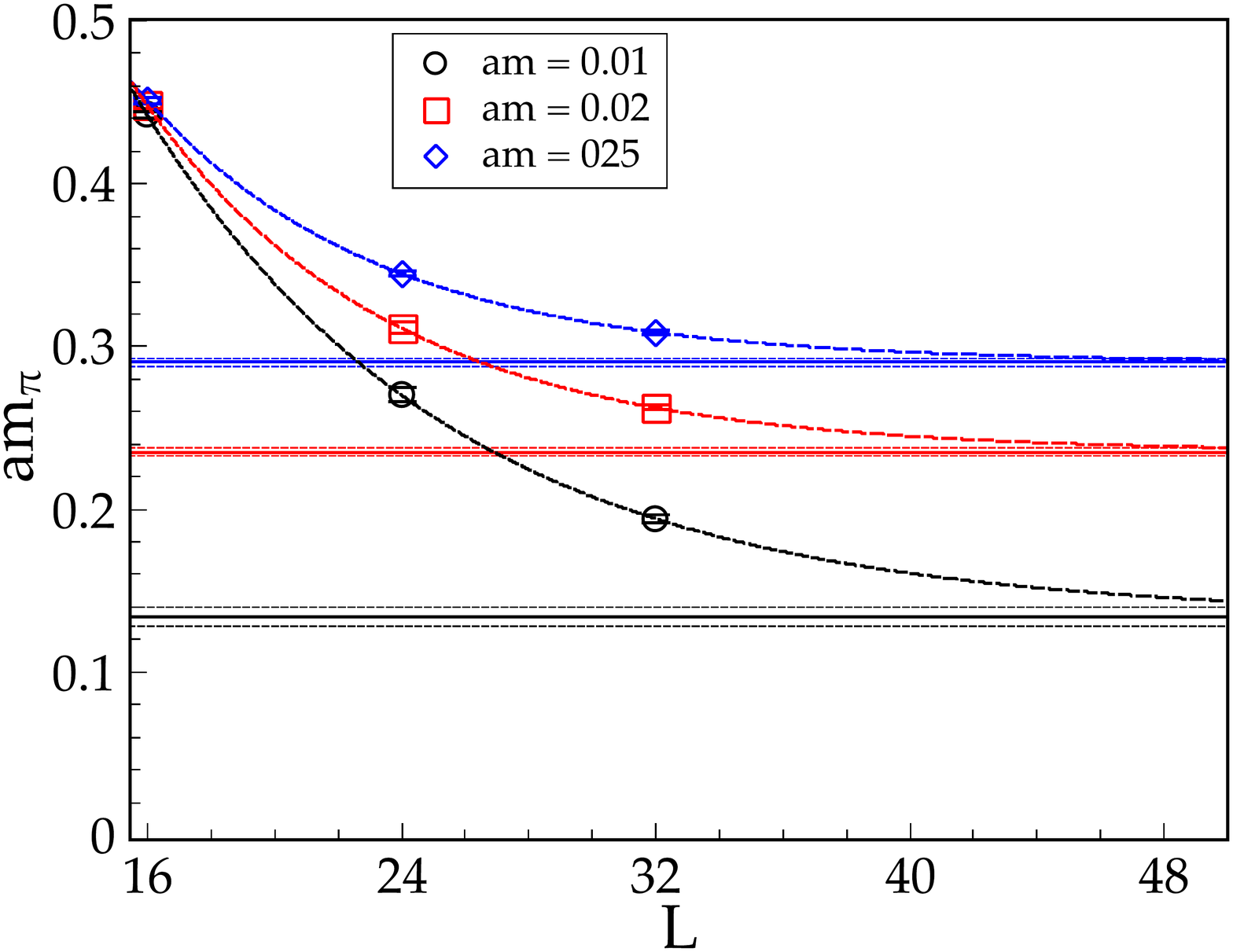}
\includegraphics[width=.45\textwidth,origin=c]{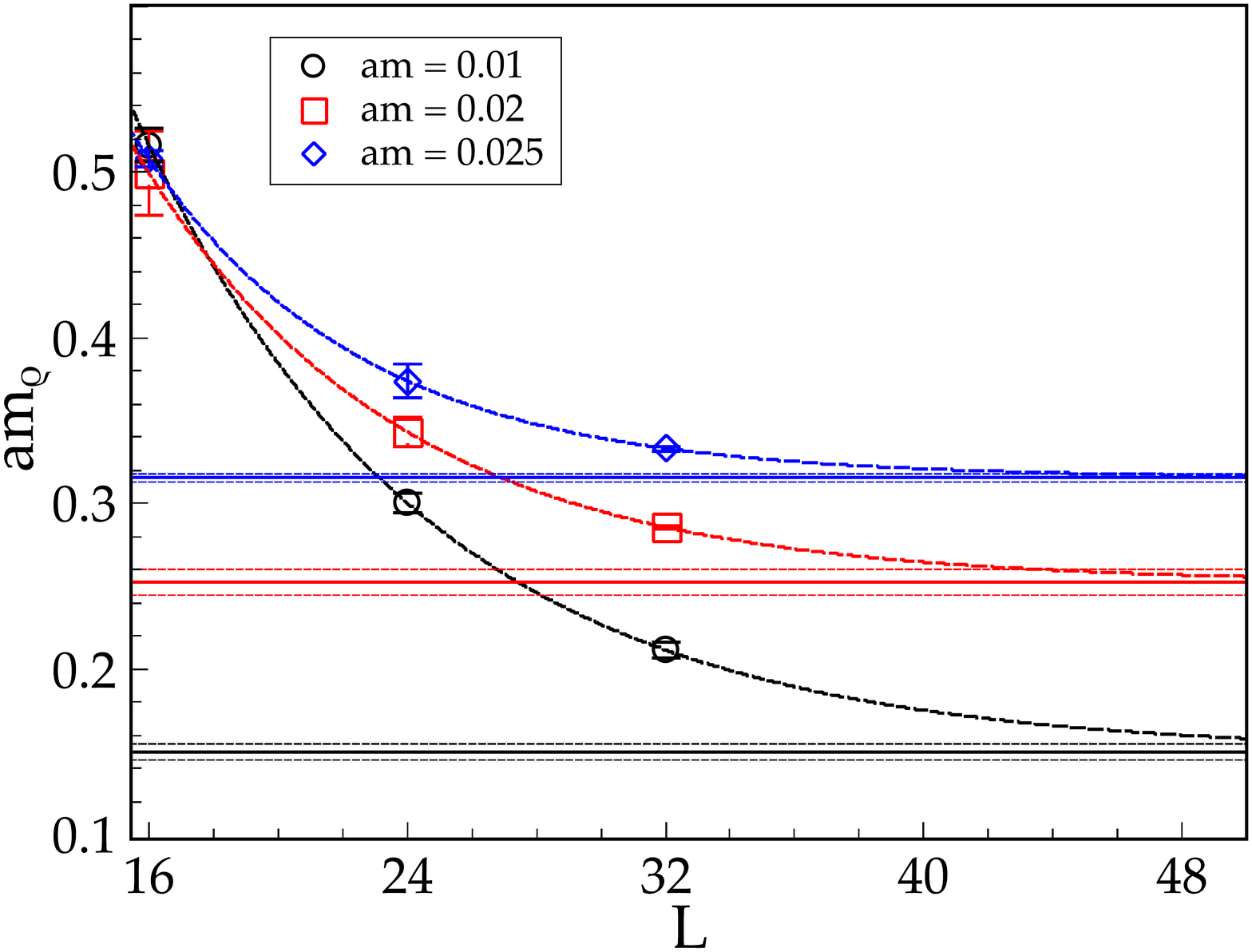}
\caption{ \label{fig:mHversusL} Spatial $L$ dependence of the pseudoscalar mass (left) and the vector mass (right) and  extrapolation to infinite volume according to eq.~(\ref{eq:FV_QED}). From top to bottom, $am=0.025$ (blue diamonds), $am=0.02$ (red squares) and $am=0.01$ (black circles). 
The extrapolated value and its uncertainty is indicated by the horizontal bands and reported in table~\ref{tab:FVextrap}. The channels $H=S,PV,N$ are in figure~\ref{fig:mS_PV_versusL} in the appendix. } 
\end{figure}

A volume dependence milder than QCD and milder than the universal scaling form can be traced back to the Coulomb dynamics in the QED-like phase  and the absence of a confining potential. It is favoured by the combination of data in figure~\ref{fig:smallx_fit}, figure~\ref{fig:mHversusL} and the infinite volume study of the heavier masses $am >0.025$.
Having only three volumes for each bare fermion mass, we have performed the extrapolation of the lightest would-be hadron masses to infinite volume with the simplest ansatz
\begin{equation}
\label{eq:FV_QED}
m_H (L) = m_H + c\, {e^{-\tilde{k} m_H L}}~~~~~~~{\mbox{H=}}\pi ,\sigma , \rho , a_1, N
\end{equation}
with parameters $c$,  $\tilde{k}$ and the infinite volume mass $m_H$ for the channel $H$.  
The results of the extrapolation are summarized in figure~\ref{fig:mHversusL}, table~\ref{tab:FVextrap} and in the appendix. 
\begin{table}[tbp]
\centering
{\small 
\begin{tabular}{|c|c|c|c|c|c|c|}
\hline
$am$ &  $am_\pi$ & $am_\rho$ & $am_N$ & $am_\sigma$ & $am_{a1}$\\
\hline 
$0.01$&   $0.1343(58)({}^{+599}_{-321} )$ & $0.1496(49)({}^{+618}_{- 277})$   & $0.228(16)({}^{+80}_{-17})$  & $0.1696(51)({}^{+361}_{-149} )$    & $0.1842(64)({}^{+453}_{-176})$   \\
\hline
$0.02$&   $0.2353(26)({}^{+271}_{-77}  )$ & $0.2522(79)({}^{+335}_{-136} )$ &   $0.382(11)({}^{+57}_{-3} )$ & $0.3084(27)({}^{+32}_{-0} )$   & $0.3205(40)({}^{+259}_{-0})$  \\
\hline
$0.025$&   $0.2903(27)({}^{+184}_{-60}  )$  & $0.3155(24)({}^{+177}_{-55} )$ &   $0.515(18)({}^{+5}_{-0})$ & $0.3755(76)({}^{+30}_{-0}$ )   & $0.4043(66)({}^{+5}_{-0})$  \\
\hline
\end{tabular}
}
\caption{\label{tab:FVextrap} Values of the would-be hadron masses extrapolated to infinite volume for $am=0.01,\,0.02,\,0.025$. The first uncertainty is given by the best-fit to eq.~(\ref{eq:FV_QED}). The second uncertainty accounts for the lack of complete knowledge of $F(x,L)$, see text.}
\end{table} 
In order to account for the uncertainty induced by the lack of a complete knowledge of the function $F(x,L)$ we add a second uncertainty to each extrapolated mass obtained as follows.  The simple parameterization $g(L)=c(1-bL)$  provides an explicit expression for $F(x,L)$, which in turn gives the volume dependence
\begin{equation}
\label{eq:FSS_FV}
m_H(L)=m_H + c\left(\frac{1}{L}-b\right )\,e^{-\frac{k}{c_H} m_H L}\, .
\end{equation}
Here, we used $x=Lm^\delta$ and the infinite volume mass relation $m_H = c_H m^\delta$  inside $F(x,L)$, where $c_H$ is nothing but the parameter $a$ in the scaling study of figure~\ref{fig:smallx_fit} and table~\ref{tab:smallx_fit}. 
 For each channel $H$, the second asymmetric uncertainty on the infinite volume mass in table~\ref{tab:FVextrap} has the $L=32$ mass value as upper bound, and as a lower bound we take the infinite volume mass given by a fit to   eq.~(\ref{eq:FSS_FV}), with free parameters $c, b$ and $m_H$ and fixed $k/c_H$ equal to its $L=24$ value in table~\ref{tab:smallx_fit} and table~\ref{tab:smallx_fit_SN} in the appendix. 
\subsection{The spectrum at infinite volume}
\label{sec:delta}

\begin{figure}[tbp]
\centering
\includegraphics[width=.45\textwidth]{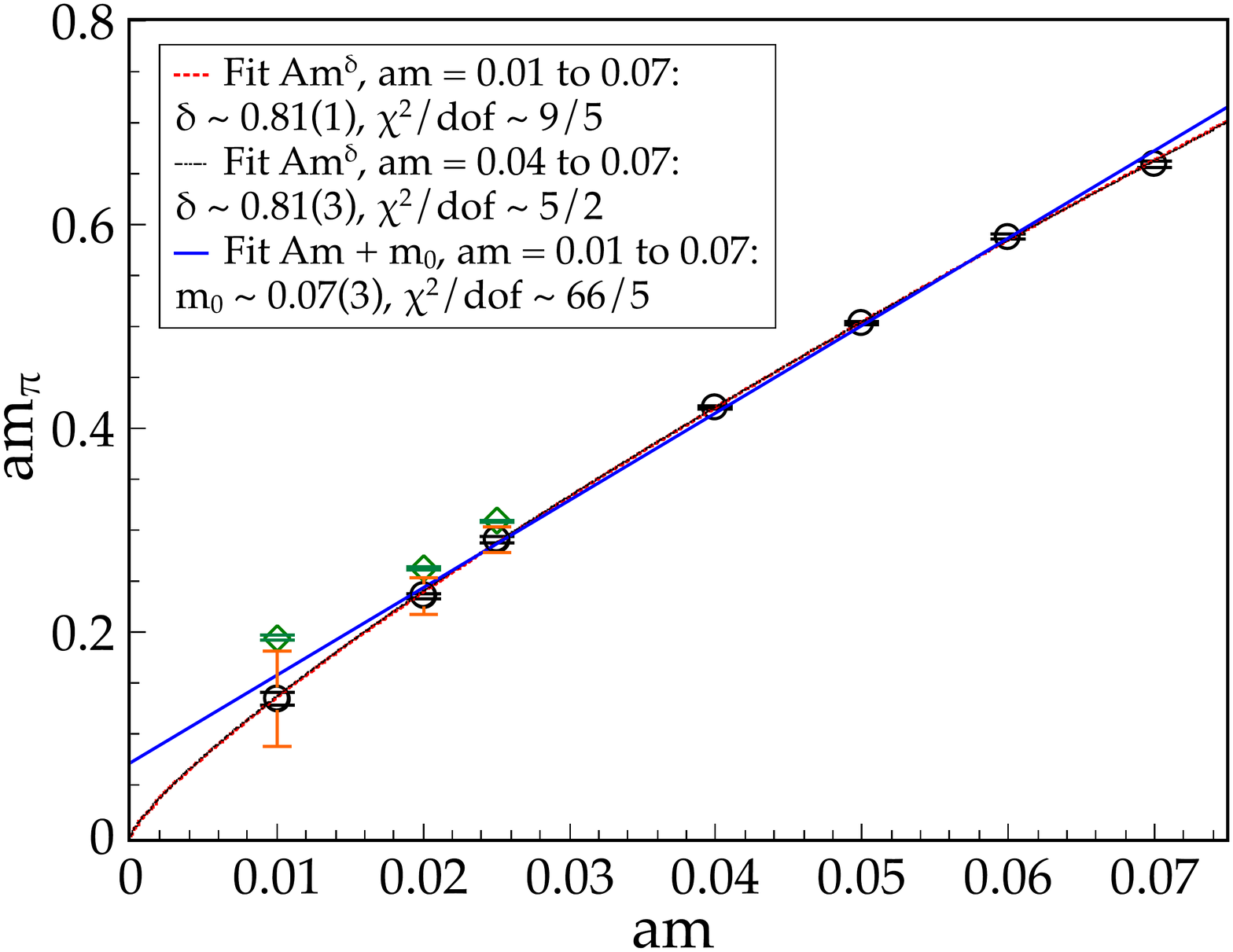}
\includegraphics[width=.45\textwidth,origin=c]{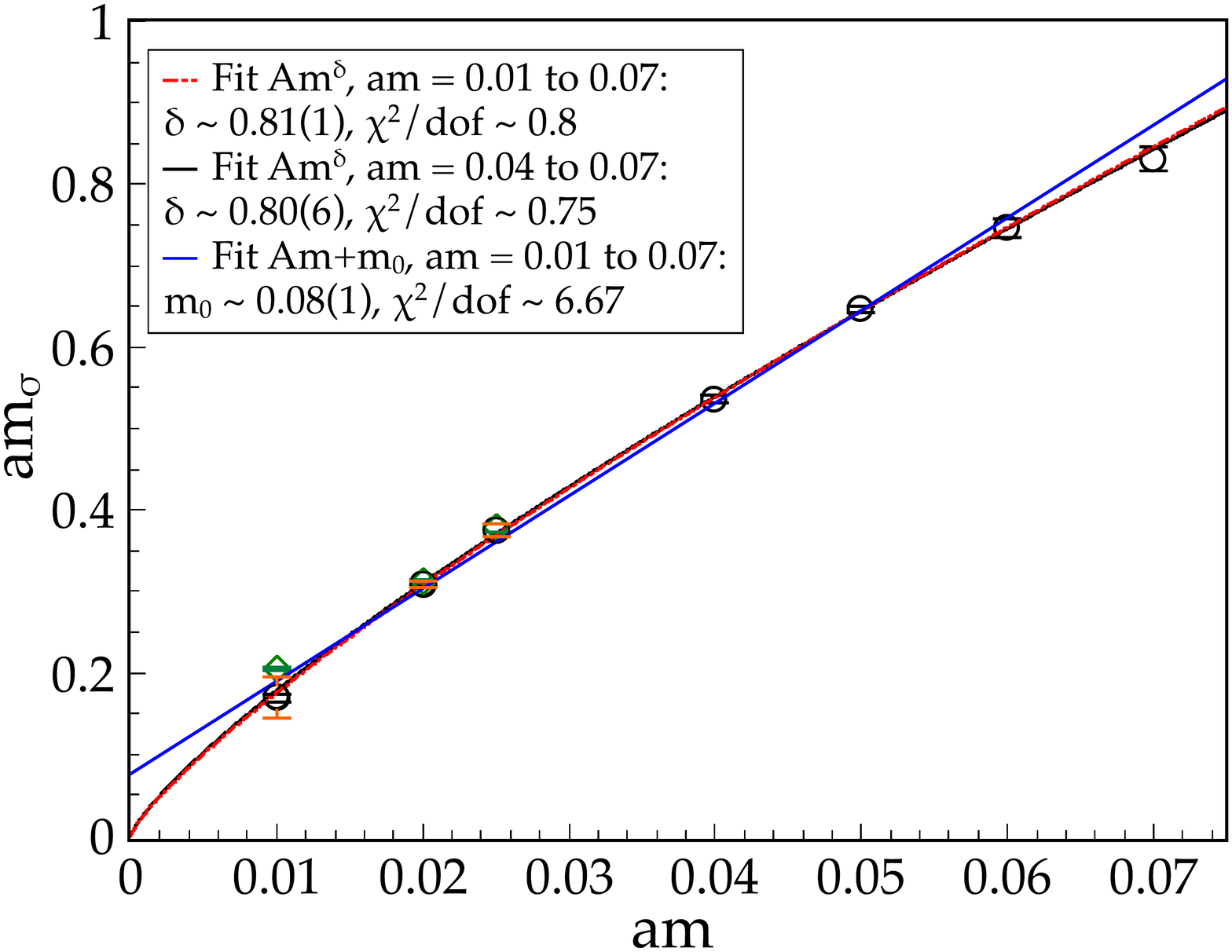}
\vspace{1.cm}
\includegraphics[width=.45\textwidth]{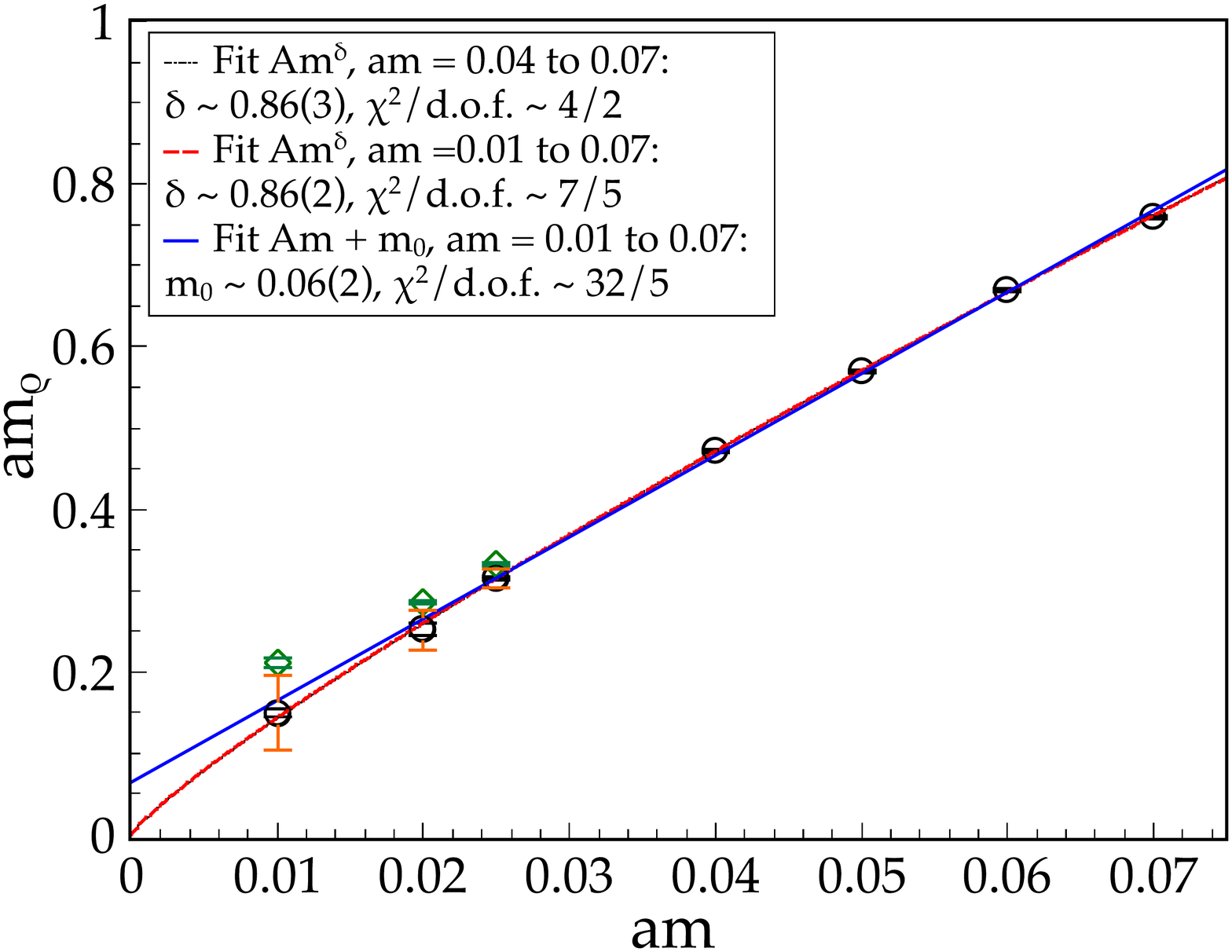}
\includegraphics[width=.45\textwidth,origin=c]{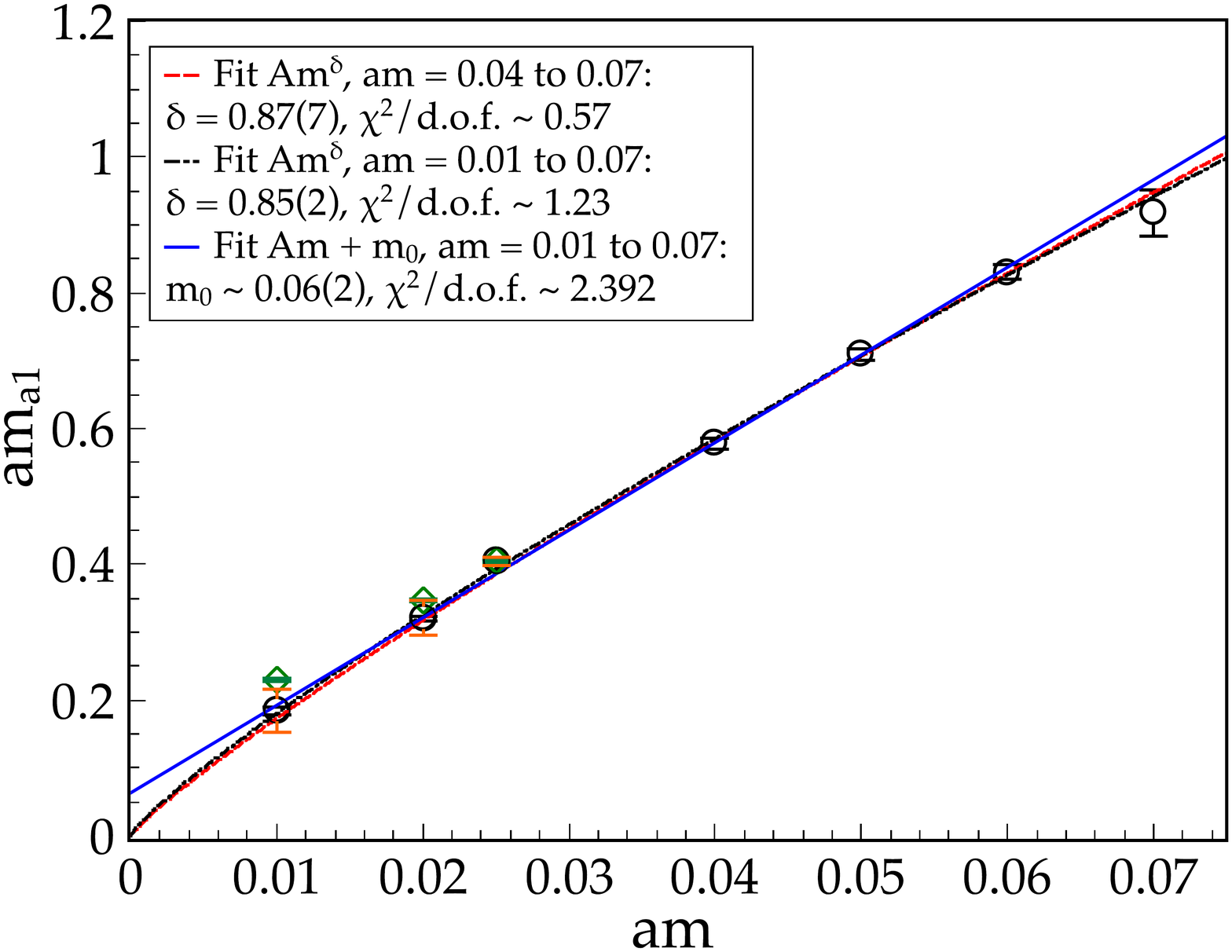}
\vspace{1.cm}
\includegraphics[width=.45\textwidth]{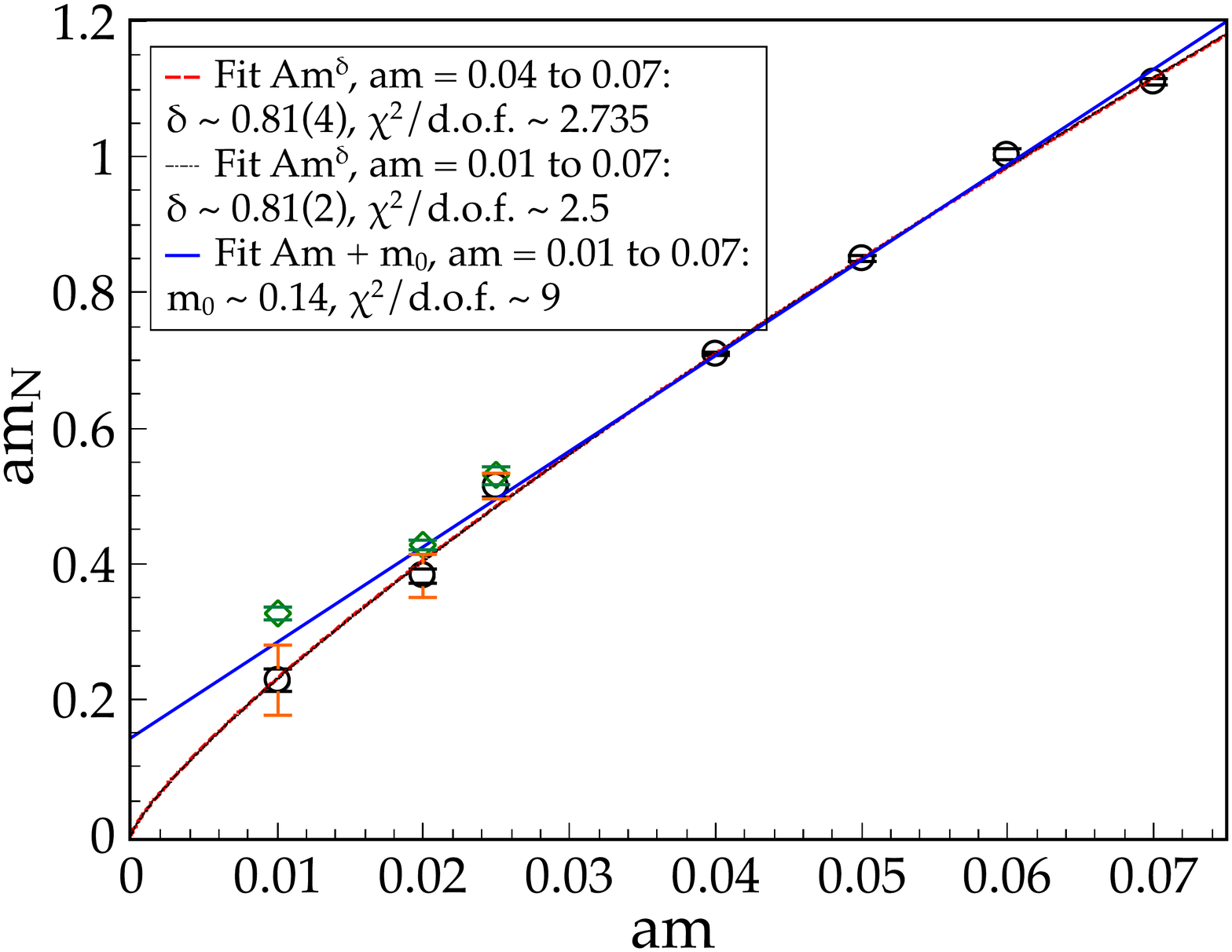}
\caption{\label{fig:fitPS-S}  Would-be hadron masses at $L=\infty$ in the pseudoscalar (top left), scalar (top right), vector (centre left), axial (centre right) and the nucleon (bottom) channels as a function of the bare fermion masses. Three fits are shown: Fit I (solid black) is a power-law on the range $am=0.04$ to $0.07$, Fit II (dashed red) is a power-law on the range $am=0.01$ to $0.07$, and Fit III (solid blue) is a linear fit with free intercept on the range $am=0.01$ to $0.07$. The total uncertainty used in Fit IIb is shown (red bar) for $am=0.01,\,0.02,\,0.025$.
Largest volume data are also shown for the same points (green diamonds). Fit results are in table~\ref{tab:delta} and \ref{tab:FitIIb}.  }  
\end{figure}

Figure~\ref{fig:fitPS-S} shows the bare fermion mass dependence of the would-be hadron masses at infinite volume, taken from tables~\ref{tab:masses} and \ref{tab:FVextrap}. The results of a single power-law fit on the heavier mass range (Fit I), the full mass range (Fit II), and a linear fit with free intercept (Fit III) are summarized in table~\ref{tab:delta} and reproduced in 
figure~\ref{fig:fitPS-S}. 
The linear fit turns out to be significantly worse than the power-law fits in all cases. This confirms, once again, that chiral symmetry is restored.
Fit IIb, reported in table~\ref{tab:FitIIb}, is a single power-law fit on the full mass range where the symmetrized second uncertainty in table~\ref{tab:FVextrap} has been added in quadrature to the first uncertainty. In almost all cases in table~\ref{tab:FitIIb} we obtain a $\chi^2/dof \lesssim 1$, likely indicating that the uncertainties on the lightest points are in this case slightly overestimated.   
\begin{table}[tbp]
\centering
\begin{tabular}{|c|c|c|c|}
\hline
Ch. & Fit I & Fit II  & Fit III \\
\hline 
 $\pi$   & $\delta = 0.81(3), c_\pi = 5.7(5)$  & $\delta = 0.81(1), c_\pi=5.8(2)$  &$m_0 = 0.07(3), c_\pi = 8.6(3)$  \\
 $\rho$   & $\delta = 0.86(3), c_\rho = 7.5(6)$  & $\delta = 0.86(2), c_\rho = 7.4(2)$  &$m_0 = 0.06(2), c_\rho = 10.0(2) $ \\
 $\sigma$   & $\delta = 0.80(6), c_\sigma = 7(1) $  & $\delta = 0.81(1), c_\sigma = 7.2(2) $  &$m_0 = 0.08(1), c_\sigma = 11.4(4)$  \\
 a1   & $\delta = 0.87(7), c_{a1} = 10(2)$  & $\delta = 0.85(2), c_{a1} = 9.0(5)$  &$m_0 = 0.06(2), c_{a1} = 12.9(4)$  \\
 N   & $\delta = 0.81(4), c_N = 10(1)$  & $\delta = 0.81(2), c_N=9.7(6)$  &$m_0 = 0.14(3), c_N = 14.1(6)$  \\
\hline
\end{tabular}
\caption{\label{tab:delta} Best-fit results for the fermion mass dependence of the would-be hadrons at $L=\infty$. Fit I is a power-law $c_H\,m^\delta$ on the range
$am=0.04$ to $0.07$, Fit II is a power-law on the range $am=0.01$ to $0.07$ that includes only the first uncertainty for the three lightest masses, and Fit III is a linear fit with free intercept $m_0 + c_H m$ on the range $am=0.01$ to $0.07$. Values of the $\chi^2/d.o.f.$ are reported in the figures.  }
\end{table} 
\begin{table}[tbp]
\centering
\begin{tabular}{|c|ccc|}
\hline
Ch. & & Fit IIb & \\
\hline 
 $\pi$      & $\delta = 0.81(1)$  & $c_\pi = 5.71(19)$ &   $\chi^2/dof =0.95$ \\
 $\rho$   & $\delta =  0.86(1)$             &  $ c_\rho = 7.47(23) $         &    $\chi^2/dof = 0.82$ \\
 $\sigma$ & $\delta =0.80(1)$  & $ c_\sigma = 7.08(24) $ & $\chi^2/dof =0.43 $   \\
 a1          & $\delta = 0.83(2)$ &  $c_{a1} = 8.43(57) $          & $\chi^2/dof = 0.64$  \\
 N           & $\delta = 0.81(2)$  & $ c_N = 9.58(46)$              &  $\chi^2/dof = 1.72$  \\
\hline
\end{tabular}
\caption{\label{tab:FitIIb} Fit IIb is Fit II of table~\ref{tab:delta} where the second (symmetrized) uncertainty for $am=0.01$, $0.02$ and $0.025$ is added in quadrature to the first one in table~\ref{tab:FVextrap}. }
\end{table} 
What is most interesting is the value of the exponent $\delta$ and its dependence, or lack thereof, on the different quantum numbers $H$. 
A value $\delta\neq 1/2$ for the pseudoscalar state says that it is not a Goldstone boson and chiral symmetry is exact. A value $\delta < 1$ says that we are away from the heavy quark limit where $m_H\sim m$. 
We observe a common $\delta = 0.81$ within errors in the channels $H=PS,S, N$\,---\,a sign of universality\,---\,and a slightly larger value $\delta = 0.86$ within errors for the vector states $H=V, PV$; this could be attributed to the different pattern of spin-spin interactions for spin-1 and spin-0 or 1/2 states. 
Noticeably, the value $\delta = 0.81$ agrees with the four-loop prediction \cite{vanRitbergen:1997va,Vermaseren:1997fq} at the IRFP and it agrees with the best-fit result of \cite{Cheng:2013xha}. We conclude that the lattice results for the pseudoscalar, scalar and the nucleon states are in the universal scaling regime, i.e., at masses sufficiently light to be insensitive to perturbative mass corrections to universal scaling arising for $g\neq g^*$. 
For this reason, we take $\delta=0.81$ and these results for the pseudoscalar (scalar and nucleon) state as reference in the combined analysis with other lattice results, i.e., $R=1$ and $b=0$ in table~\ref{tab:noietal}. 

\begin{figure}[tbp]
\centering
\includegraphics[width=.45\textwidth]{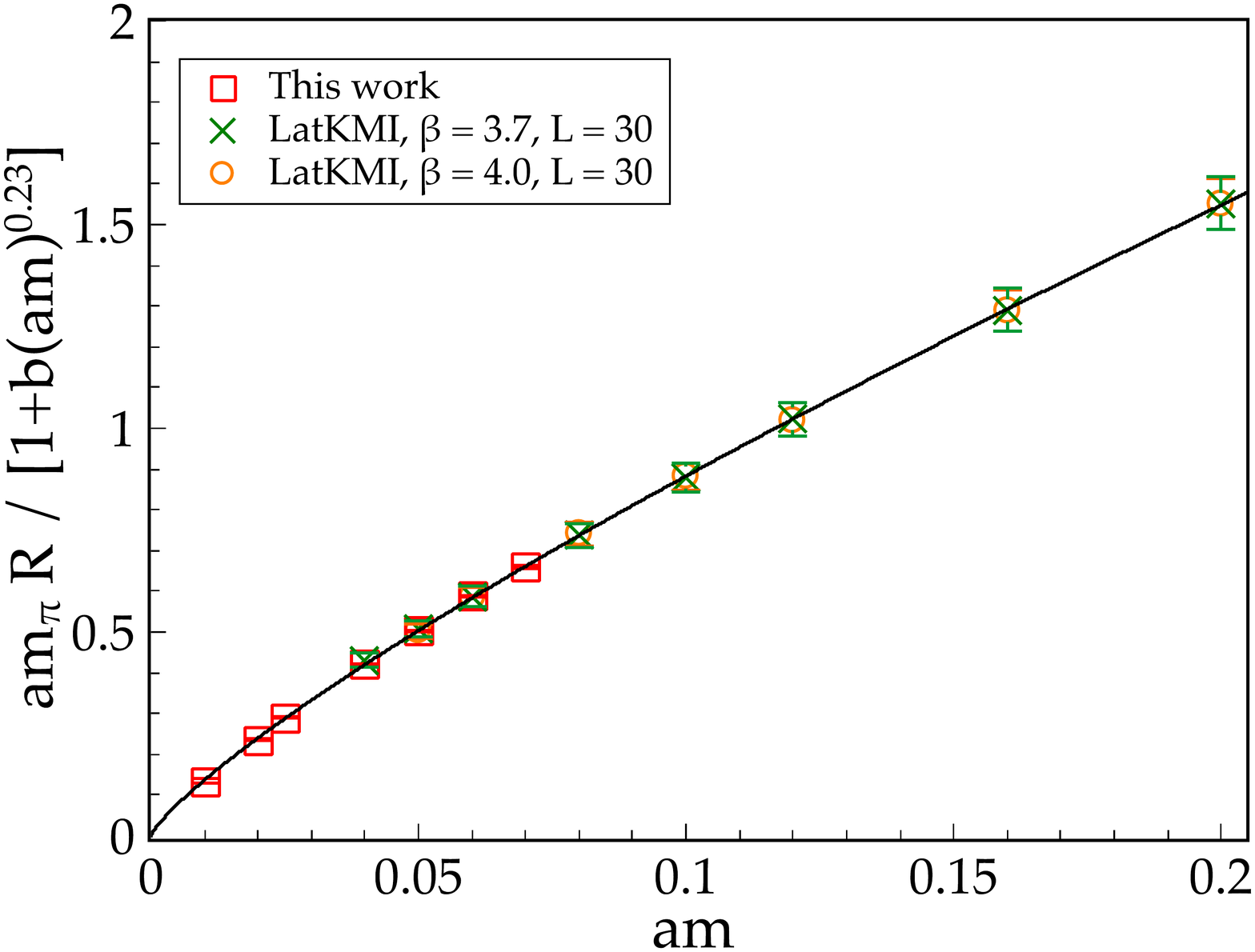}
\includegraphics[width=.45\textwidth,origin=c]{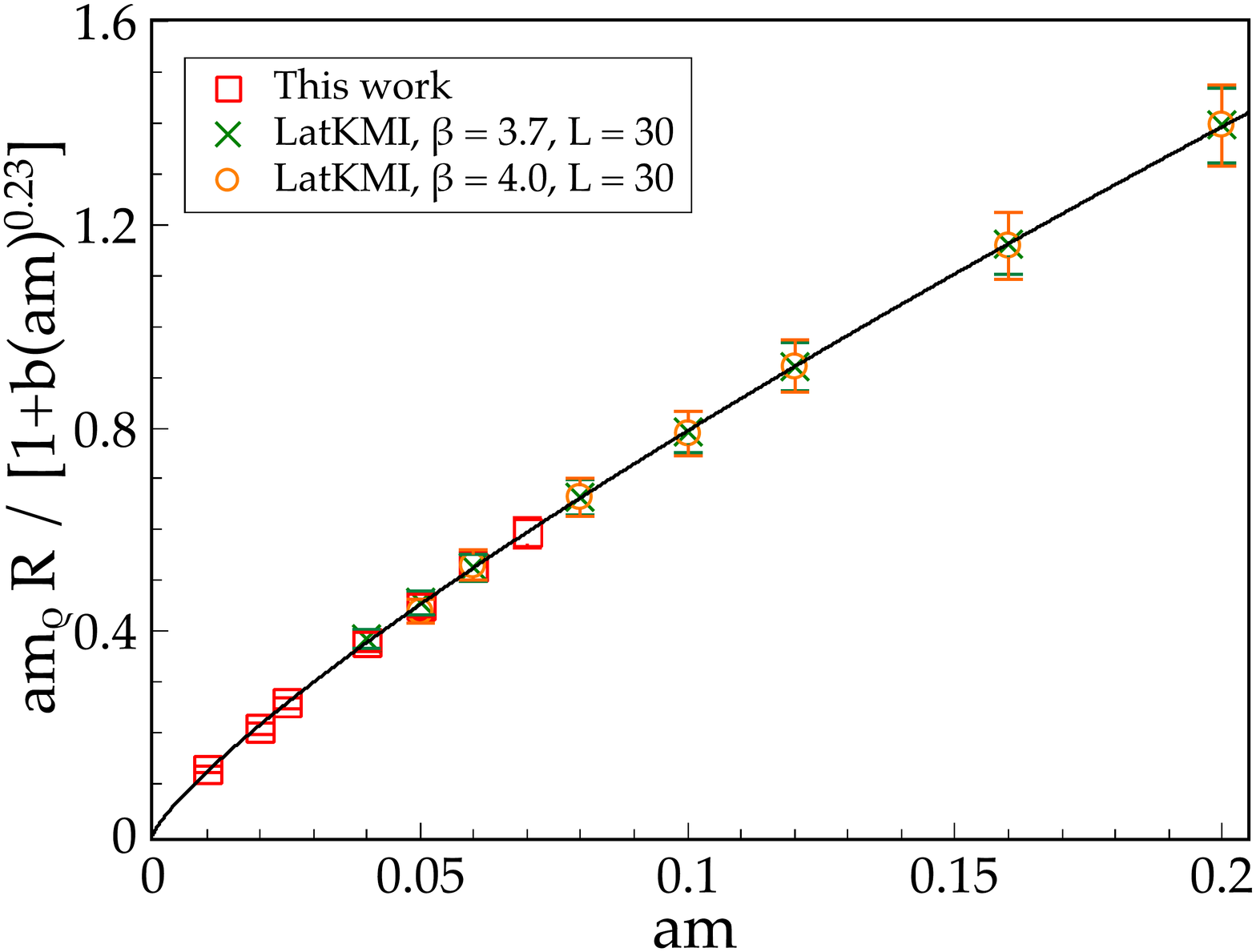}
\caption{ \label{fig:compKMI}Collapse of the rescaled infinite volume masses $am_H R/(1+b_H(am)^\omega )$, $\omega =0.23$,  for the  pseudoscalar ($\pi$) and vector ($\rho$) states in this work (red squares), and in \cite{Aoki:2012eq} at $\beta = 4.0$ (orange circles) and $\beta =3.7$ (green crosses). The values of $R=c_H/c_H^{KMI}$ and $b_H$, $H=\pi ,\rho$, are from table~\ref{tab:usKMI_PSV}.
} 
\end{figure}

Instead, $\delta = 0.86$ for the vector states suggests the presence of perturbative corrections to scaling. The results of a fit with two power laws, according to the parameterization of the perturbative corrections to scaling discussed in section~\ref{sec:scalingv}
\begin{equation}
m_H =c_H\, m^\delta \left (1+b_H m^\omega \right )
\label{eq:2powlaw}
\end{equation}
with $\delta =0.81$ and $\omega =0.23$ for the vector state are in table~\ref{tab:usKMI_PSV} (This work). 
\begin{table}[tbp]
\centering
\begin{tabular}{|c|c|c|c|c|}
\hline
 &  $c_\pi$  & $b_\pi$ & $c_\rho$  & $b_\rho$\\
\hline 
This work   &   5.7(5)       &     0       & $5.13(26)$   & $0.52(12)$ \\
\hline
LatKMI 3.7  &  $5.408(86)$   & $-0.544(16)$ & $6.899(68)$   & $-0.594(10)$  \\
\hline
LatKMI 4.0 & $4.778(64)$    & $-0.493(14)$  & $5.96(11)$    & $-0.560(18)$   \\
\hline
\end{tabular}
\caption{\label{tab:usKMI_PSV} Best-fit values for two power laws eq.~(\ref{eq:2powlaw}) for the infinite volume pseudoscalar and vector masses from this work and \cite{Aoki:2012eq}. The exponents are fixed to $\delta =0.81$ and $\omega =0.23$. The value of $c_\pi$ from this work is from Fit I in table~\ref{tab:delta}. }
\end{table} 
Note that $b_\rho >0$, as expected, consistently with the fact that our system is on the strong coupling side of the IRFP. To further corroborate this statement we combine the data for the vector and the pseudoscalar with those of \cite{Aoki:2012eq}, all at infinite volume within uncertainties. The best-fit values for eq.~(\ref{eq:2powlaw}) are in table~\ref{tab:usKMI_PSV}. While $b_H>0$ on the strong coupling side of the IRFP (This work), $b_H<0$ on the weak coupling side of the IRFP (LatKMI), and we expect $b_H\to 0$ for $\beta_L\to\infty$. 
Finally, figure~\ref{fig:compKMI} shows the collapse of the infinite volume pseudoscalar and vector states of this work and \cite{Aoki:2012eq}, after rescaling. 
The rescaling factor $R=c_H/c_H^{KMI}$ is the ratio of the leading power-law coefficients for the channel $H$ in table~\ref{tab:usKMI_PSV}.
This analysis leads to the determination of the mass anomalous dimension $\gamma^*$ at the IRFP. We quote the value obtained from Fit I in the pseudoscalar channel from table~\ref{tab:delta} 
\begin{equation}
\label{eq:gamma}
\delta = 0.81(3)~~~~\gamma^* = 0.235(46)  
\end{equation}
This value is in agreement with the perturbative four-loop prediction, with the best-fit result of \cite{Cheng:2013xha} and not far from the first lattice determination of the fermion mass anomalous dimension for the $N_f=12$ system in \cite{Deuzeman:2009mh}, though the latter was affected by rather large uncertainties. 
The value of $\gamma^*$ in eq.~(\ref{eq:gamma}) suggests a rather weakly coupled $N_f=12$ system at the IRFP, so that  perturbation theory should be expected to hold. Conversely, four-loop perturbation theory seems to fail for $N_f \sim 8$, where it predicts an IRFP at rather strong coupling $g^*$ and, even worse, a change of sign of the mass anomalous dimension for $g<g^*$, see end of section~\ref{sec:23}\footnote{ This effect is not encountered at two loops.}. 
This reinforces the idea that nonperturbative dynamics, known to be chiral dynamics in this case, has to play a role at the opening of the conformal window, for $8\lesssim N_f\lesssim 12$. 
Also, if $\gamma^* =1$ has to be realized at the lower endpoint of the IRFP line, where the conformal window disappears, a rapid variation of the mass anomalous dimension for $N_f^c\lesssim N_f\lesssim 12$ should be expected in a lattice (or any nonperturbative) determination of the IRFP line, where nonperturbative dynamics is fully encompassed. 
The present study also corroborates the view that the IRFP of these theories is not associated to a physical singularity, no discontinuity happens there and estimates of physical observables including the anomalous dimensions can be attempted on either side of the fixed point.

\begin{figure}[tbp]
\centering
\includegraphics[width=.60\textwidth]{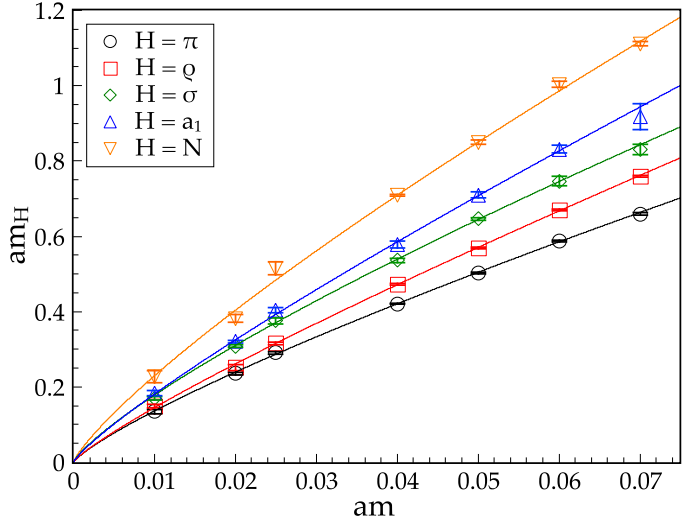}
\caption{\label{fig:orderstates} Ordering of the would-be hadrons in the QED-like phase of the $N_f=12$ system. From bottom to top, the pseudoscalar ($\pi$), the vector ($\rho$), the scalar ($\sigma$), the axial ($a_1$) and the nucleon $N$. } 
\end{figure}

We conclude this section with showing the ordering of the would-be hadrons in figure~\ref{fig:orderstates}. 
 To summarize, a universal power law with exponent $\delta = 0.81$ describes all would-be hadrons, with additional perturbative mass corrections of the type $1+\Delta g m^{\delta\gamma_g^*}$ in the vector and axial channels.  The pseudoscalar is the lightest state, but it is not a Goldstone boson. The vector, the scalar, the axial, and finally the nucleon follow. It is worth noting that the scalar state\footnote{We remind the reader that $\sigma$ of this work is the state extracted from the connected scalar two-point function.} is heavier than the vector state. Their ordering becomes phenomenologically relevant when the theory is just below the conformal window\,---\,it remains, however, difficult to identify a broad scalar resonance, such as $f_0(500)$ of QCD\footnote{This state could in addition be an admixture of ordinary $\bar{q}q$ states and tetraquarks.}, on the lattice.  
\subsection{Mass ratios and degeneracies }
\label{sec:mass-deg}

\begin{figure}[tbp]
\centering
\includegraphics[width=.45\textwidth]{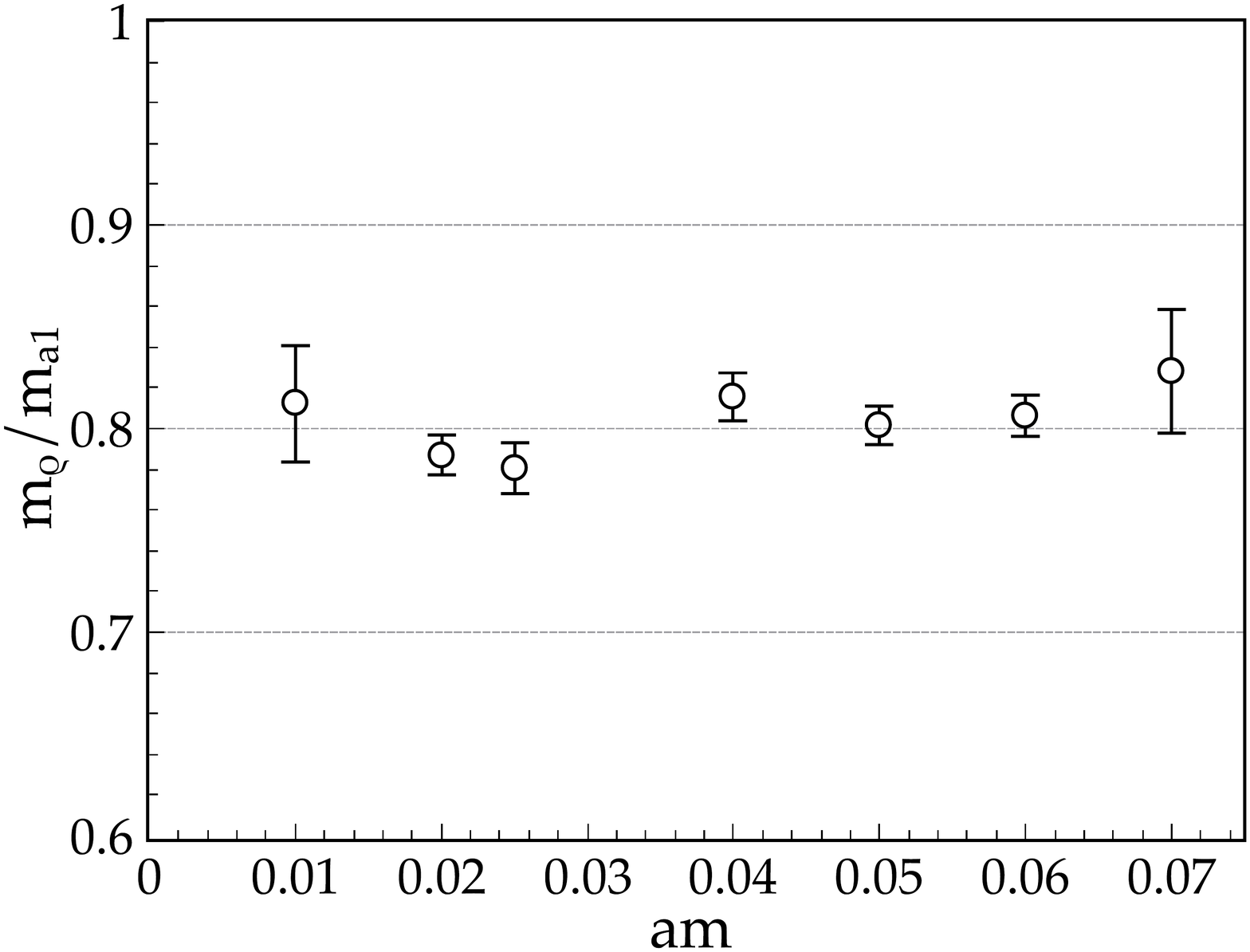}
\includegraphics[width=.45\textwidth,origin=c]{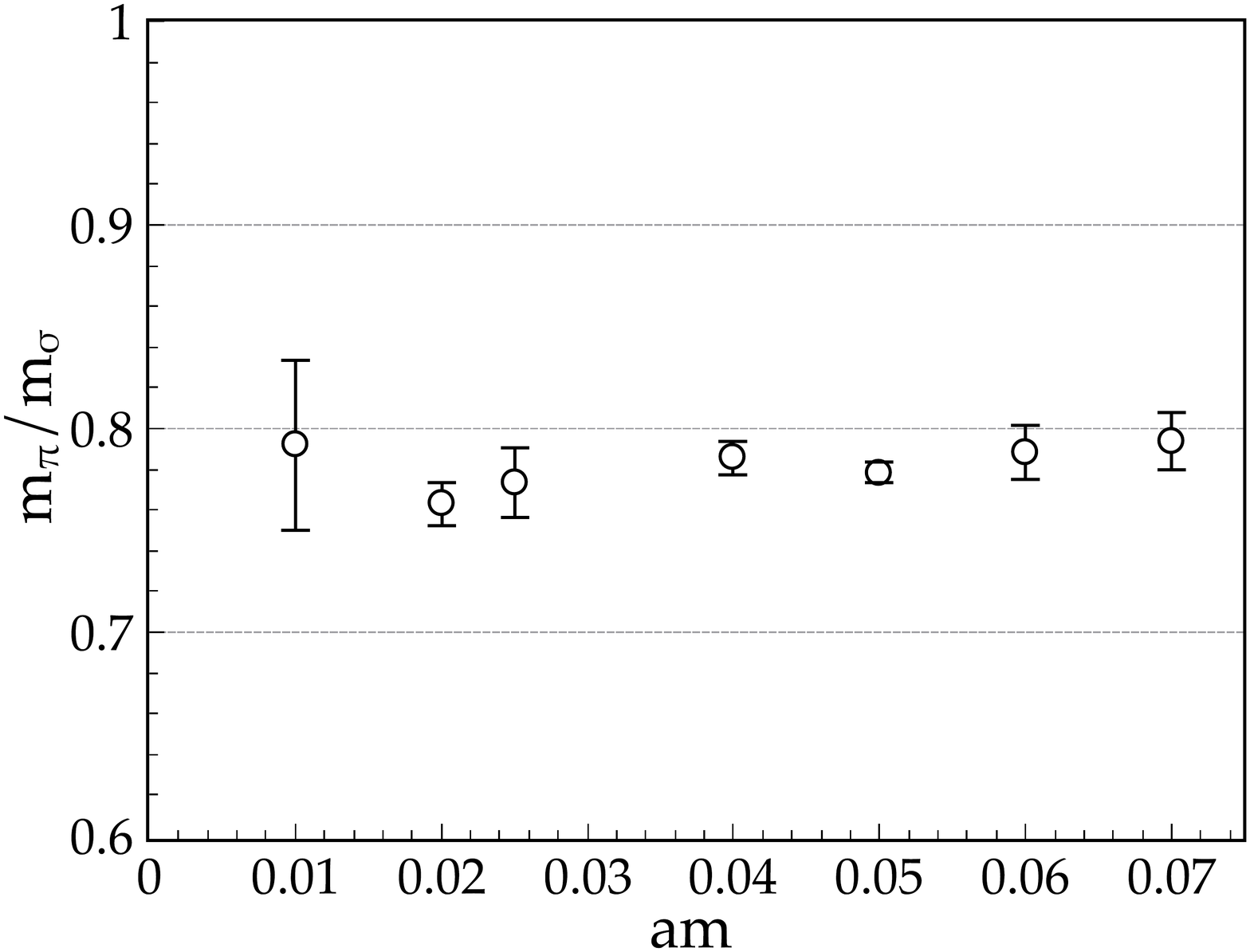}
\caption{\label{fig:ratios_summary} The ratio of the vector and axial masses (left) and the ratio  of the pseudoscalar and scalar masses (right) as a function of the bare fermion mass.   } 
\end{figure}

\begin{figure}[tbp]
\centering
\includegraphics[width=.45\textwidth]{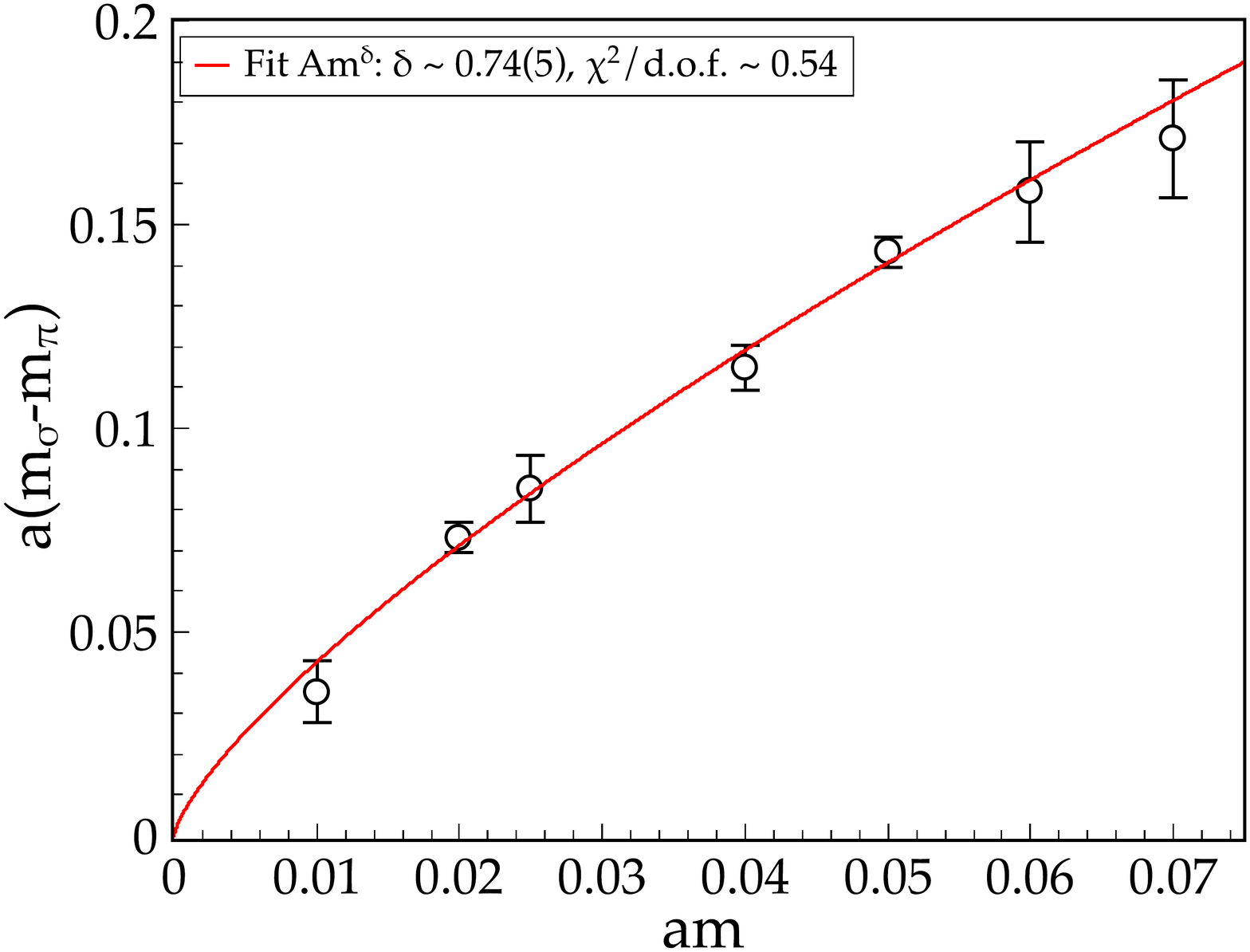}
\includegraphics[width=.45\textwidth,origin=c]{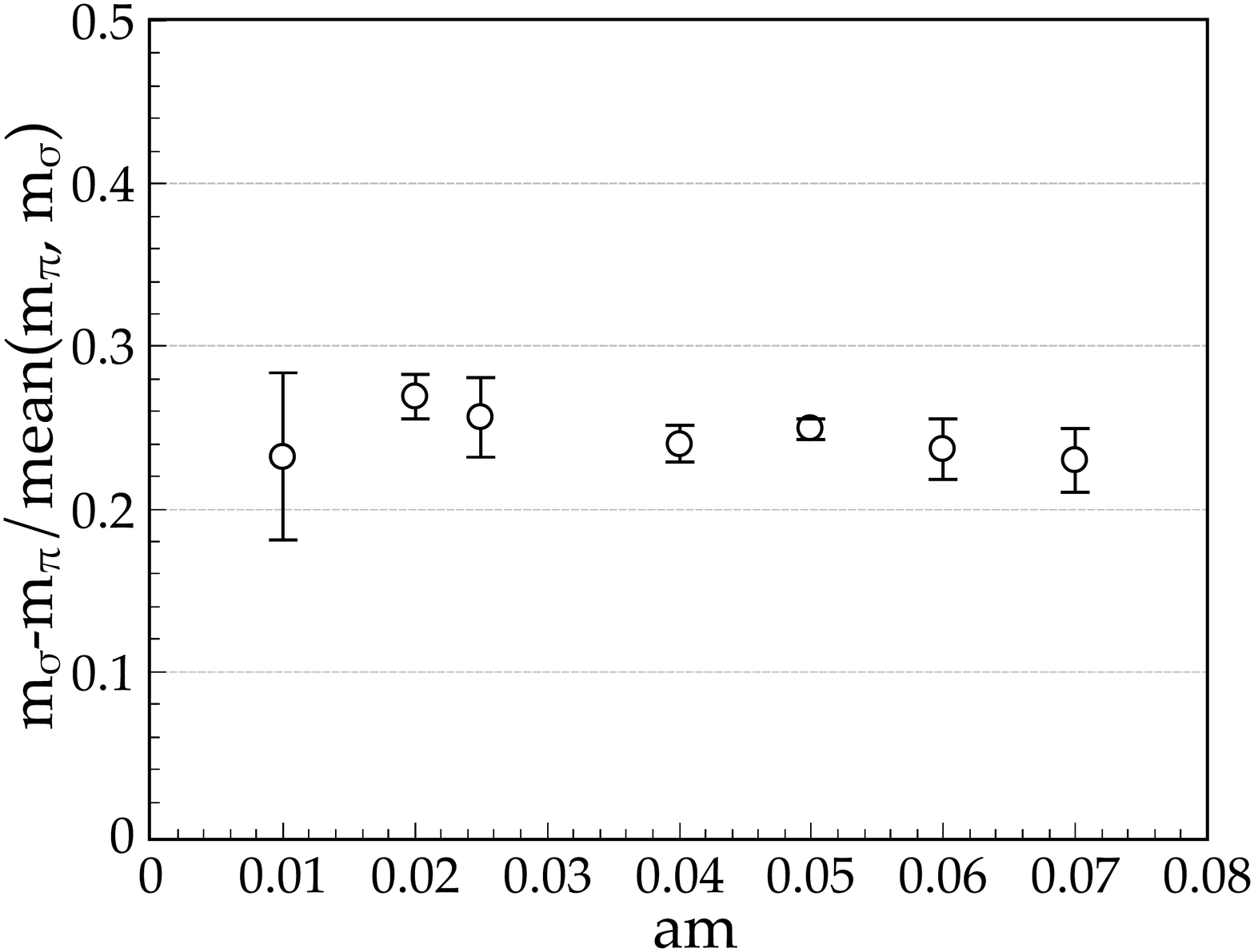}
\vspace{1.cm}
\includegraphics[width=.45\textwidth]{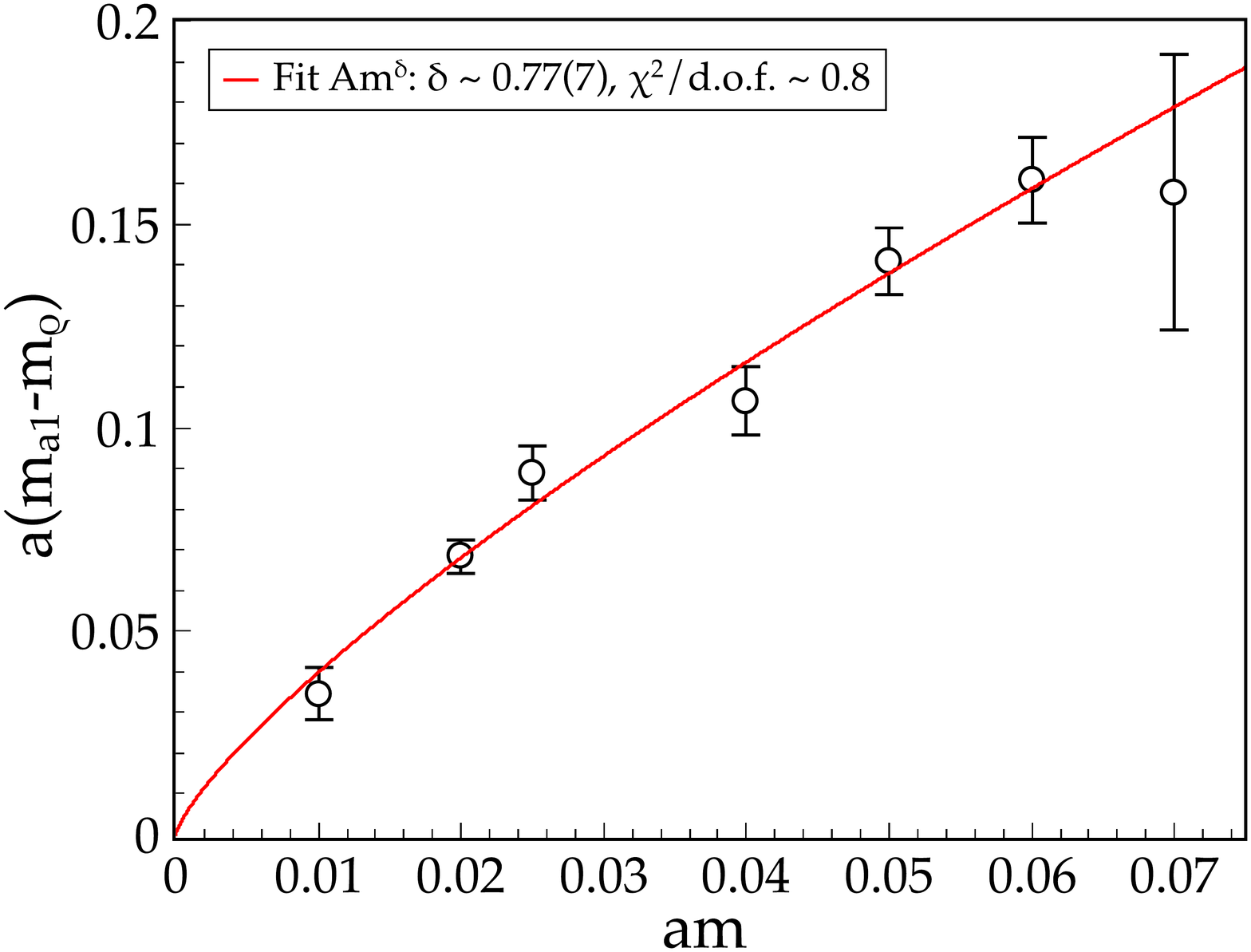}
\includegraphics[width=.45\textwidth,origin=c]{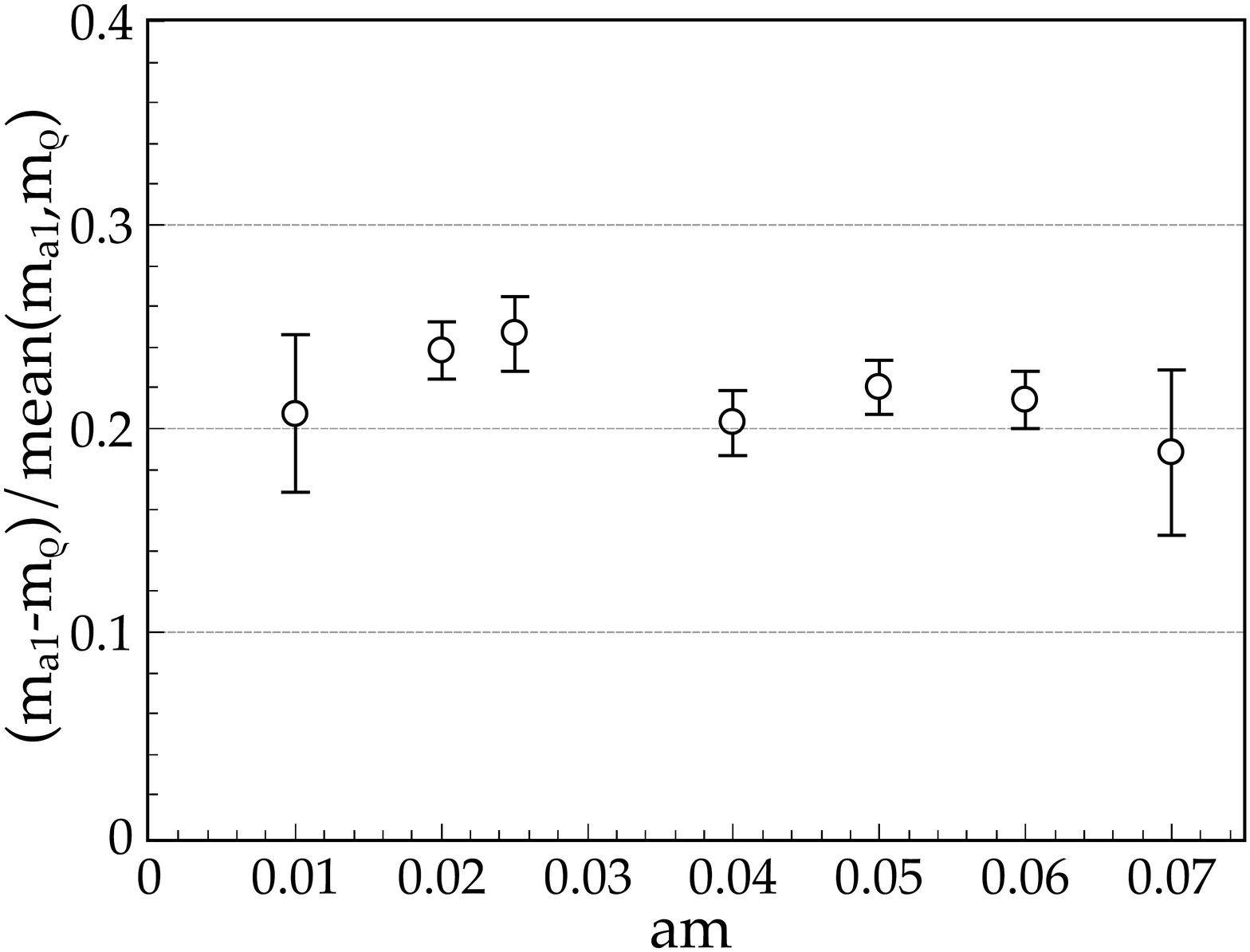}
\caption{\label{fig:degeneracy} Degeneracy pattern and best-fit curve of the pseudoscalar $\pi$ and the scalar $\sigma_c$ (connected) (top left) states. The ratio of the mass difference to the mass average (top right). Analogous figures for the vector $\rho$ and the axial $a_1$ states (bottom). } 
\end{figure}

We conclude this work with some remarks on the interplay of conformal and chiral symmetry inside the conformal window. Ratios and degeneracies of would-be hadron masses are a combined probe of both symmetries, and, as shortly discussed below, the $U(1)$ axial symmetry. At the IRFP conformal symmetry implies exact chiral symmetry. Away from the IRFP, inside the conformal window, restored chiral symmetry implies the degeneracy of chiral partners in the chiral limit.  
Figure~\ref{fig:ratios_summary} shows the mass ratios pseudoscalar-scalar and vector-axial. The two ratios are essentially constant $\sim 0.8$ over the explored mass range, 
as it can be deduced from the best-fit values for the power-law exponent $\delta$.  Due to the presence of perturbative corrections to universal scaling for the vector states, deviations from a constant ratio will instead be observed in all cases that mix the vector (or axial) channel with the other ones, one example is figure~\ref{fig:mpimrhoRatio}. 
Before discussing the degeneracy patterns, it is important to recall that the scalar "$\sigma$" studied here is extracted from the quark-line connected piece of the scalar-isoscalar two-point function; for clarity, we then call this state $\sigma_c$ in the following discussion and with $\sigma$ we refer to the lowest-lying state of the complete (connected plus disconnected) scalar-isoscalar correlator. The pseudoscalar $\pi$ and the scalar $\sigma$, not $\sigma_c$, belong to the same chiral multiplet of $SU(N_f)_L\times SU(N_f)_R$, and they must be degenerate in the chiral limit $m\to 0$ when chiral symmetry is not spontaneously broken. 
The vector $\rho$ and the axial $a_1$ are also chiral partners and show the same degeneracy pattern.
In other words, the mass degeneracy of the chiral partners $\rho$ and $a_1$, $\pi$ and $\sigma$, can be used as an indicator of the restoration of chiral symmetry.  What about the degeneracy of $\pi$ and $\sigma_c$? For degenerate fermion masses, as in our case, the connected contribution to the scalar-isoscalar, $\sigma_c$, equals the connected contribution to the scalar-isovector $\delta$\,---\,the latter  has no disconnected contributions. The $\delta$ is the $U(1)_A$ partner of $\pi$. We should thus conclude that the degeneracy of $\pi$ and $\sigma_c$ is probing the effective restoration of $U(1)_A$, at least at the level of the two-point functions\footnote{A complete probe of $U(1)$ axial and chiral restoration obviously includes the direct observation of the disconnected contributions and the degeneracy patterns of all states, including $\eta^\prime$ and $\sigma$.}.  
Figure~\ref{fig:degeneracy} shows  the mass difference and the ratio $(m_i-m_j)/(\overline{m_i+m_j})$  for the spin-0 $U(1)_A$ partners $\pi -\sigma_c$ and the spin-1 chiral partners $\rho - a_1$.  
The  degeneracies in the chiral limit are evident from the best-fit curves of the mass difference on the left of figure~\ref{fig:degeneracy}, thus confirming once again exact chiral symmetry and the effective restoration of $U(1)_A$. 
This also implies that  the disconnected contributions to the scalar-isoscalar correlator are at least $O(m)$. We defer to future work the question to which degree $U(1)_A$ is exact inside the conformal window beyond the two-point functions and how its restoration pattern compares with high-temperature QCD. 
The approximately constant ratios on the right of figure~\ref{fig:degeneracy}  highlight, once again,  the realization of the scaling  form $c_H\,m^\delta$, modulo small perturbative corrections $(1+b_H m^\omega )$ in the spin-1 case, with universal exponents $\delta$ and $\omega$ and mass independent nonuniversal coefficients $c_H$ and $b_H$. 

\section{Conclusions}
\label{sec:conclusions}

We have studied the $SU(3)$ gauge theory with twelve fundamental fermions as a prototype of theories inside the 
conformal window, with emphasis on the two-point functions and their properties when the IRFP is perturbed by a fermion mass. 
In order to disentangle the imprint of the IRFP in the dynamics of the system, we have analyzed the complete would-be hadron spectrum, the would-be mesons and the nucleon, and performed a universal scaling study at finite and infinite volume. 
The identification of universal contributions dictated by the conformal invariance at the fixed point and deviations from universal scaling induced in the surroundings of the fixed point has allowed for the nonperturbative determination of  the fermion mass anomalous dimension $\gamma^*=0.235(46)$ and a unified description of all lattice results for the would-be hadron spectrum of the $N_f=12$ theory. 

This analysis shows that the lattice system retains all signatures of the underlying conformal symmetry of the fixed point, in addition to the restored chiral symmetry, that the pattern of symmetries can be followed across the IRFP and the critical exponents\,---\,or any other physical observable\,---\,can be determined on either side of the fixed point, be it the asymptotically free phase of the lattice system or the QED-like phase. In other words, one should conclude that no singularity is associated to such a fixed point. 

The obtained nonperturbative value of $\gamma^*$ hints at a rather weakly coupled $N_f=12$ system at the IRFP. It is thus amusing, and perhaps not unexpected, to observe the agreement with the four-loop perturbative prediction at the fixed point. Based on this agreement, it is tempting to infer that the perturbative expansion is working well in this range of $N_f$, and that the missing nonperturbative contributions and the effects of a truncation of the perturbative series amount to a negligible correction for this system.

It also reinforces the idea that nonperturbative dynamics, known to be chiral dynamics in this case, has to play a role at the opening of the conformal window, for $8\lesssim N_f\lesssim 12$. 
Should $\gamma^* =1$ be realized at the lower endpoint of the IRFP line, where the conformal window disappears, a rapid variation of the mass anomalous dimension on the interval $N_f^c\lesssim N_f\lesssim 12$ should be expected. Plausibly, nonperturbative contributions would become increasingly important towards $N_f^c$, making mandatory the use of a nonperturbative strategy, lattice or else, in the study of the infrared dynamics of systems close to the lower endpoint of the conformal window. 

As a byproduct, we have confirmed the restoration of chiral symmetry through the degeneracy of chiral partners and the effective restoration of the $U(1)$ axial symmetry at the level of the two-point functions.

\newpage

\appendix
\section{Volume dependence and extrapolation to infinite volume}

\begin{figure}[h]
\centering
\includegraphics[width=.45\textwidth]{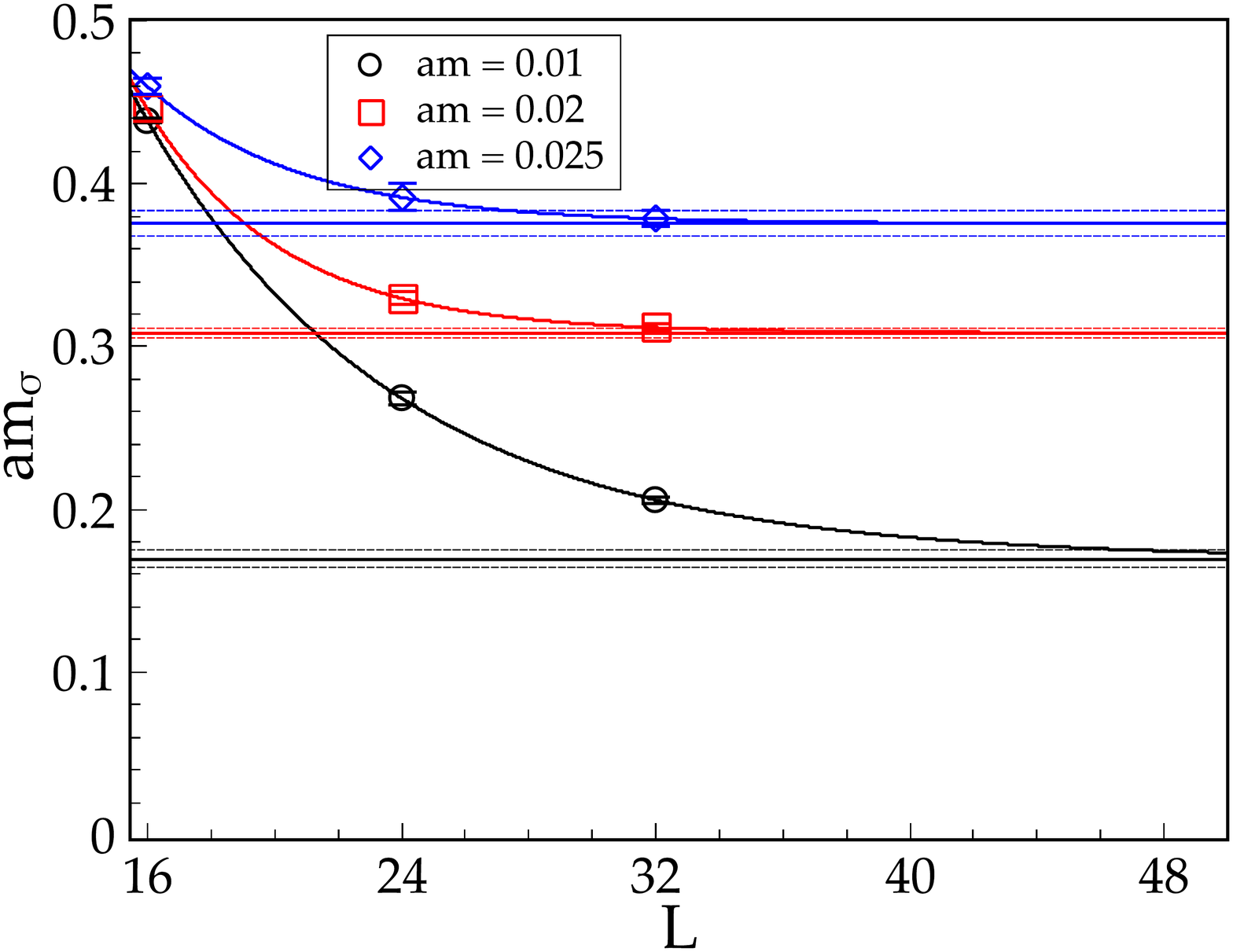}
\includegraphics[width=.45\textwidth,origin=c]{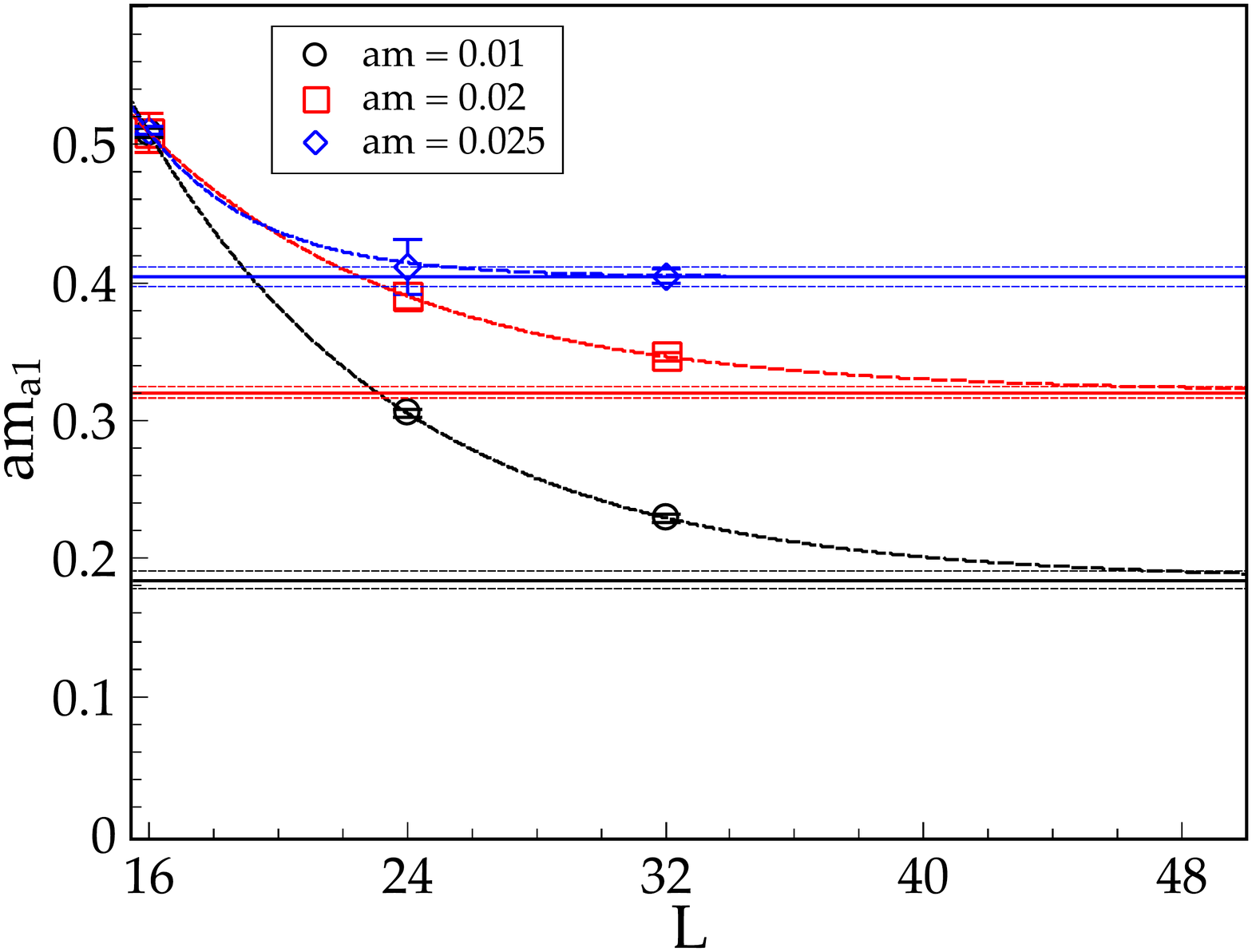}
\vspace{1.cm}
\includegraphics[width=.45\textwidth]{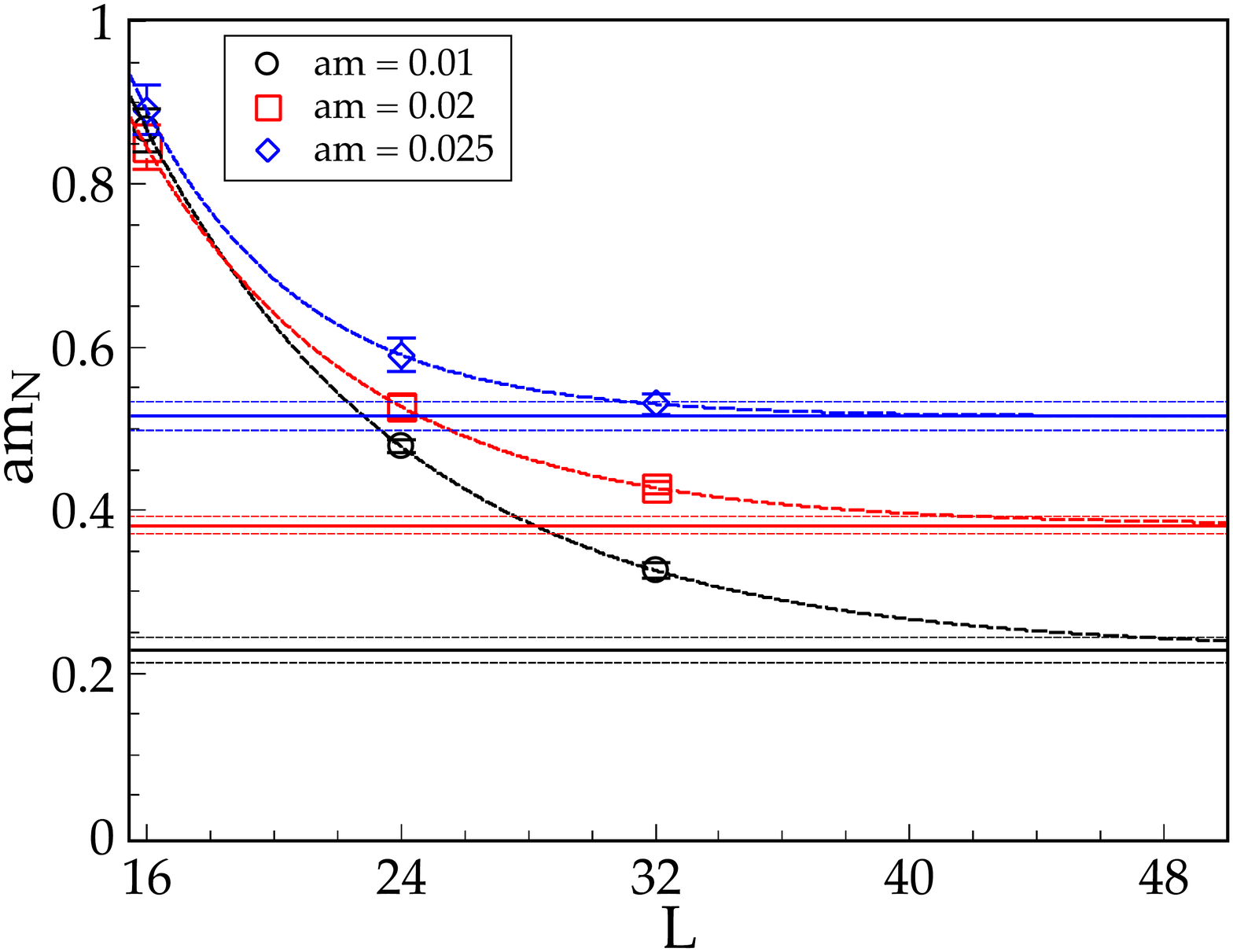}
\caption{ \label{fig:mS_PV_versusL} Spatial volume dependence and  extrapolation to infinite volume with functional form eq.~(\ref{eq:FV_QED})
for the masses of the scalar (top left), axial (top right) and the nucleon (bottom) states, and bare fermion masses $am=0.01$, $0.02$ and $0.025$. The extrapolated value and its uncertainty is indicated by horizontal lines and can be read from table~\ref{tab:FVextrap}. } 
\end{figure}

\begin{figure}[h]
\centering
\includegraphics[width=.45\textwidth]{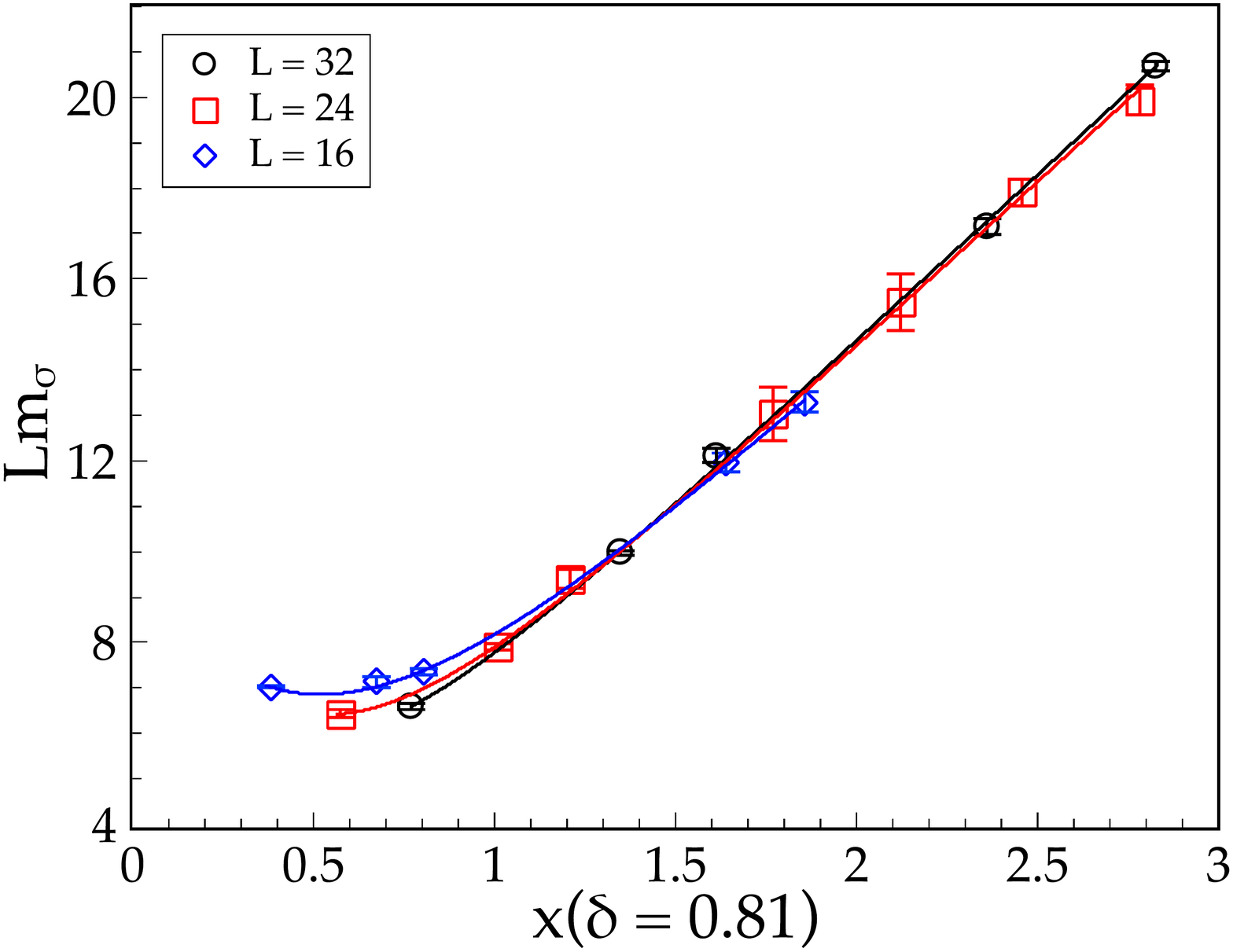}
\includegraphics[width=.45\textwidth,origin=c]{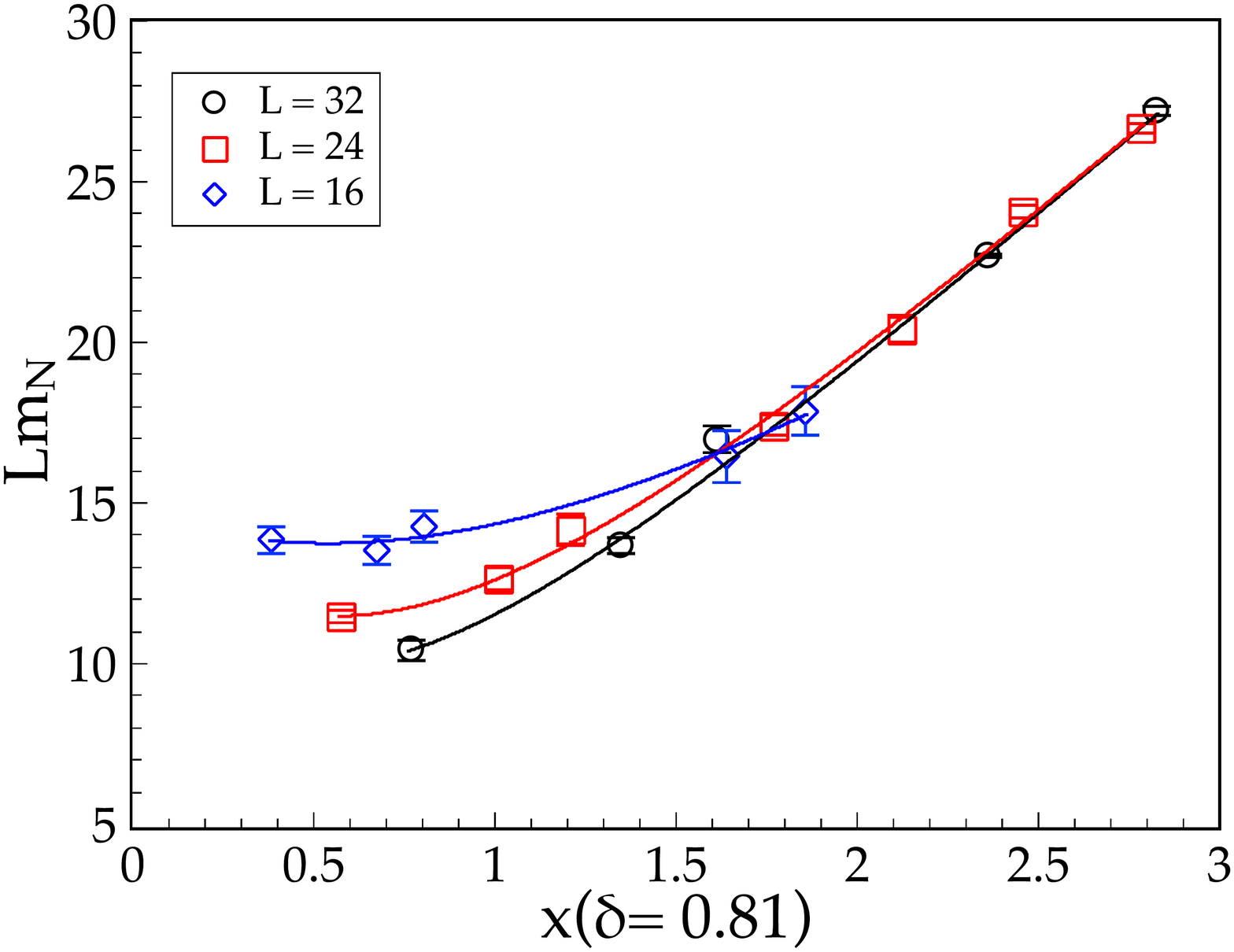}
\vspace{1.cm}
\includegraphics[width=.45\textwidth]{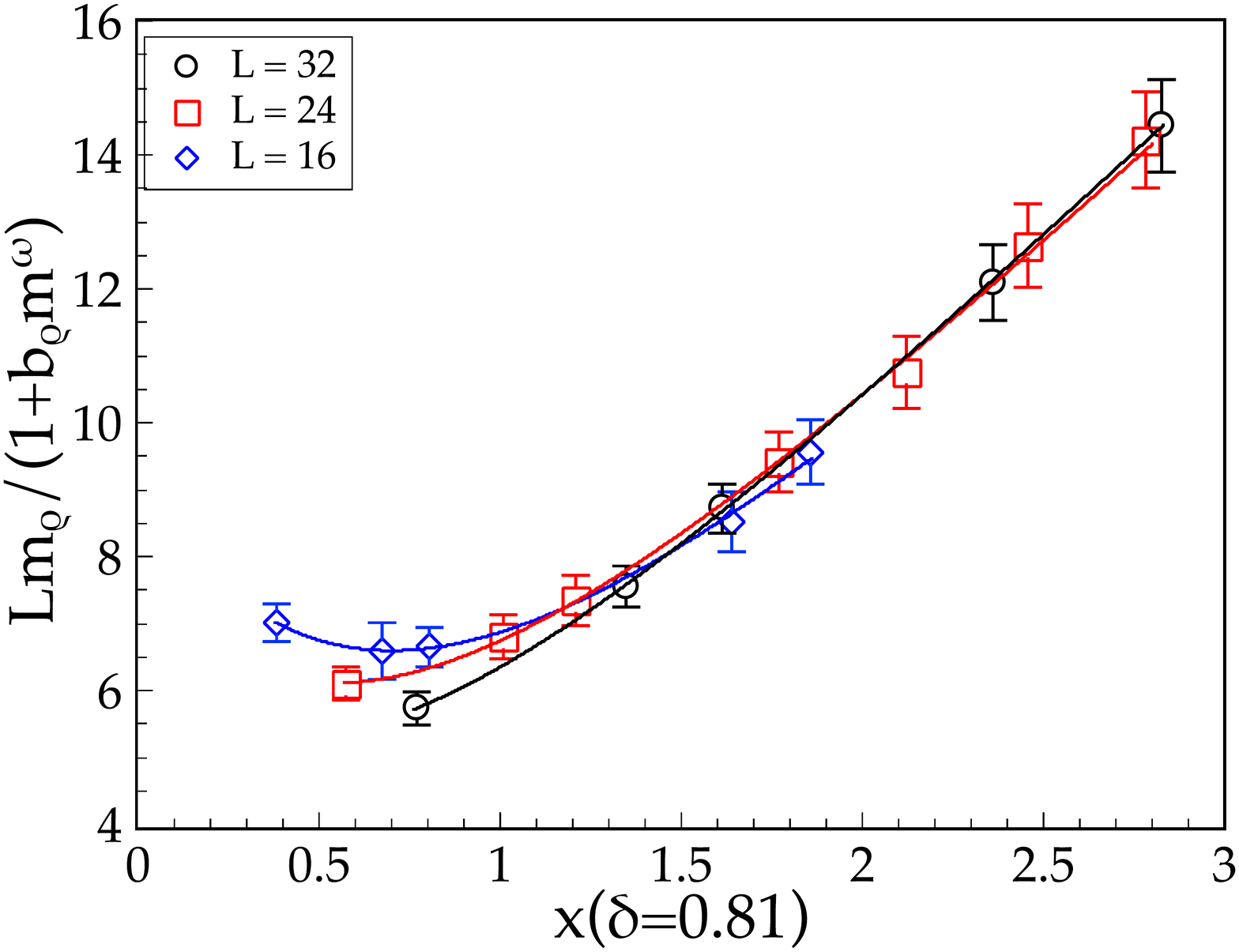}
\includegraphics[width=.45\textwidth,origin=c]{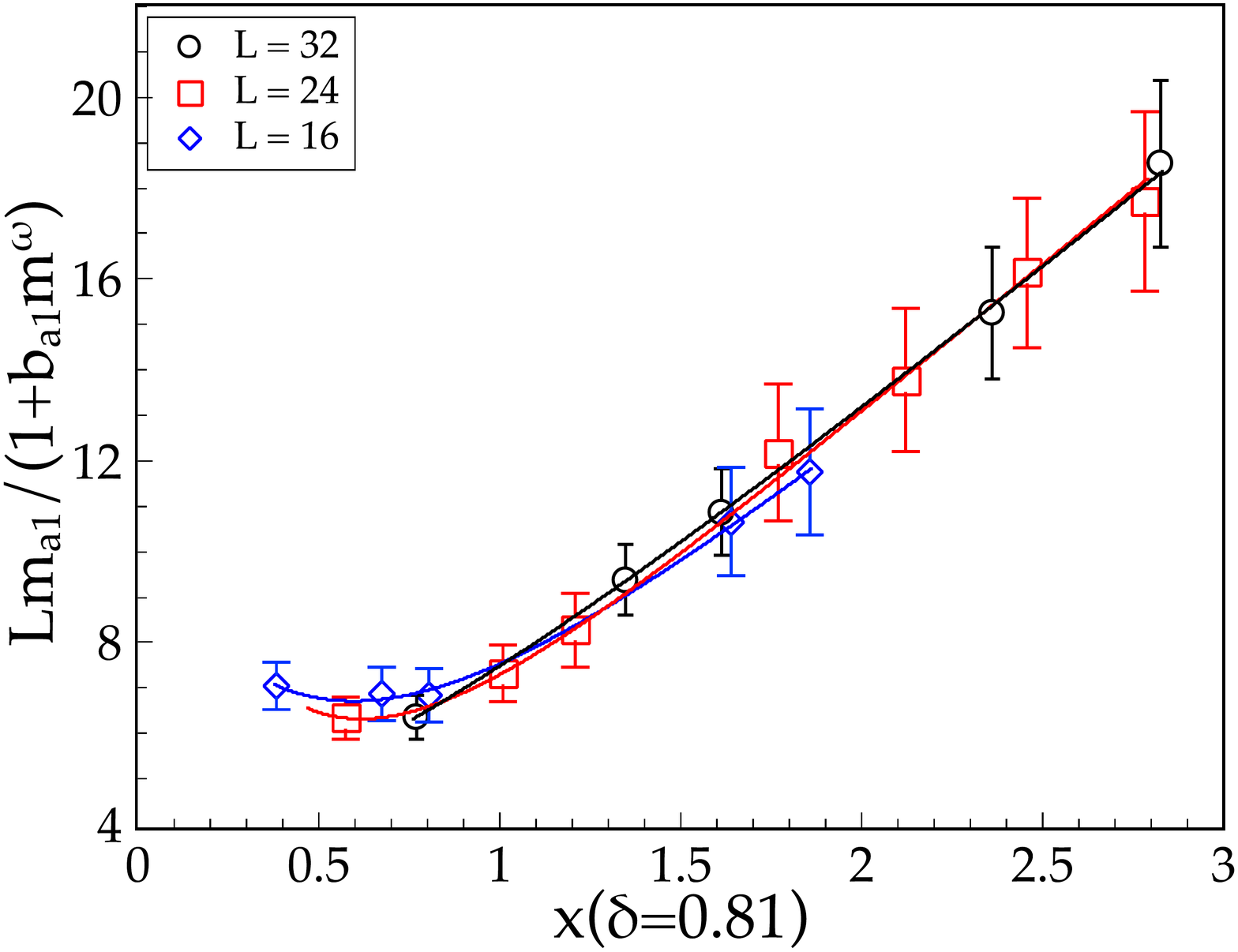}
\caption{ \label{fig:smallx_fit_app} 
$Lm_H/(1+b_H m^\omega )$ for $H= \sigma, N, \rho, a_1$ as a function of the scaling variable $x=Lm^\delta$ with $\delta =0.81$, for $L=16,\,24,\,32$. The coefficient $b_H=0$ for $H=S, N$, $b_\rho = 0.52(12)$ and $b_{a_1}=0.45(24)$. 
The curves are best fits to the functional form $F(x,L)=ax+c\exp{(-kx)}$, see table~\ref{tab:smallx_fit_SN}.  } 
\end{figure}

This appendix completes the study in sections~\ref{sec:results_scaling}, \ref{sec:FV} and \ref{sec:delta} for some of the would-be hadrons.  The extrapolation to infinite volume for the scalar, axial and the nucleon are analogous to the ones presented in section~\ref{sec:FV} and are reported in figure~\ref{fig:mS_PV_versusL}.
The nonperturbative $L$ dependent violations of scaling at small $x$ have been studied for all states. We report in figure~\ref{fig:smallx_fit_app} and table~\ref{tab:smallx_fit_SN} the analogous of figure~\ref{fig:smallx_fit} and table~\ref{tab:smallx_fit} for the channels $H=S, V, PV, N$.  
The best-fit values of $a,c,k$ for the simplified ansatz $F(x,L)=ax+c\exp{(-kx)}$ are reported in table~\ref{tab:smallx_fit_SN}.
\begin{table}[h]
\centering
\begin{tabular}{|c|c|c|c||c|c|c|}
\hline
\multicolumn{4}{|c}{$H=\sigma$ } & \multicolumn{3}{c|}{$H=N$} \\
\hline
 &  $L=16$  & $L=24$ & $L=32$  &  $L=16$  & $L=24$ & $L=32$\\
\hline 
$a$ & $7.10(17) $  & $7.25(10) $ & $7.31(9)  $  
& $7.8(1.4)   $  & $9.54(10)  $ & $ 9.56(19) $    \\
\hline
$c$ & $10.12(80)  $  & $12.2(3.4) $  & $ 10.84(19.58) $   
& $14.83(2.26)  $  & $ 14.84(3.20) $  & $13.49(21.02) $   \\
\hline
$k$ & $2.23(23)  $  & $ 2.93(48)  $  &  $ 3.15(2.35) $  
& $ 0.82(41) $  & $1.57(35)  $  &  $ 1.93(1.84) $    \\
\hline
\multicolumn{4}{|c}{$H=\rho$} & \multicolumn{3}{c|}{$H=a_1$} \\
\hline
 &  $L=16$  & $L=24$ & $L=32$  &  $L=16$  & $L=24$ & $L=32$\\
\hline 
$a$ & $4.77(17)  $  & $5.01(06)    $ & $5.08(06)   $  
& $6.24(30)$  & $6.53(12)  $ & $6.45(16) $    \\
\hline
$c$ & $9.077(76) $  & $7.52(70)    $  & $  6.1(1.1)   $   
& $10.4(1.7) $  & $13.3(5.0) $  & $3.8(1.9) $   \\
\hline
$k$ & $1.46(20)   $  & $ 1.47(15)  $  &  $ 1.57(25)    $  
& $2.08(40) $  & $2.84(66) $  &  $1.30(70) $    \\
\hline
\end{tabular}
\caption{\label{tab:smallx_fit_SN} Best-fit values of the parameters $a,c,k$  for the fits of $Lm_H$, $H=\sigma , N, \rho , a_1$ to the functional form $F(x,L)=ax+c\exp{(-kx)}$, with $x=Lm^\delta$ and $\delta =0.81$. }
\end{table} 
\newpage

\acknowledgments

We thank Dries Coone for his participation in the early stage of this work. EP and MpL  acknowledge the hospitality of the Aspen Center for Physics,  which  is supported by the National Science Foundation Grant No.  PHY-1066293,  as well as the many interesting discussions with 
the participants during their visit. 
KM and MpL thank Yasumichi Aoki and Hiroshi Ohki for fruitful discussions. 
This work was in part based on the MILC collaboration's public lattice gauge theory code,  see

\noindent http://physics.utah.edu/$\sim$detar/milc.html. Simulations were performed on the IBM BG/P at the University of Groningen, the Huygens IBM Power6+ system at Stichting Academisch Rekencentrum Amsterdam (SARA) and the IBM BG/Q Fermi at CINECA. 
This work is part of the research programme of the Foundation for Fundamental Research on Matter (FOM), which is part of the Netherlands Organisation for Scientific Research (NWO).

\end{document}